\documentclass[]{tADP2e}
\def\be{\begin{equation}}
\def\ee{\end{equation}}
\def\bea{\begin{eqnarray}}
\def\eea{\end{eqnarray}}
\def\la{\langle}
\def\ra{\rangle}
\def\om{\omega}
\def\nn{\nonumber}
\def\p{\partial} 
\def\bx{{\bf{x}}}
\def\bp{{\bf{p}}}
\def\bv{{\bf{v}}}

\def\f{\frac}
\def\g{\gamma}
\def\etal{et al.}
\def\ha{\hat{A}}

\def\a{\alpha}

\def\d{\delta}
\def\e{\epsilon}
\def\g{\gamma}

\def\G{\Gamma}

\def\l{\lambda}
\def\om{\omega}

\def\S{\Sigma}

\def\p{\partial}

\def\n{\eta }
\def\tn{\tilde{\eta} }
\def\tX{\tilde{X}}

\def\ie{\emph{i.e. }}
\def\mT{\mathcal{T}}
\def\mJ{\mathcal{J}}
\def\bl{{\bf{n}}}
\def\hM{\hat{M}}
\def\hP{\hat{\Phi}}
\def\hD{\hat{D}}
\def\hE{\hat{E}}
\def\hb{\hat{B}}
\def\hGam{\hat{\Gamma}}
\def\x{\hat{B}_x}
\def\y{\hat{B}_p}
\def\z{{\hat{B}_{xp}}}
\def\lsim{\mathrel{\rlap{\lower4pt\hbox{\hskip1pt$\sim$}}
    \raise1pt\hbox{$<$}}}                
\def\gsim{\mathrel{\rlap{\lower4pt\hbox{\hskip1pt$\sim$}}
    \raise1pt\hbox{$>$}}}  
\begin{document}
\doi{10.1080/0001873YYxxxxxxxx}
\issnp{0001-8732}  \jvol{00} \jnum{00} \jyear{2008} \jmonth{Oct}

\title{Heat Transport in low-dimensional systems}

\author{Abhishek Dhar$^{\ast}$ \thanks{$^\ast$Corresponding
    author. Email: dabhi@rri.res.in}
\vspace{6pt}  
{\em{Raman Research Institute, Bangalore 560080, India}}
}

\maketitle

\begin{abstract}
Recent results on theoretical studies of heat conduction in
low-dimensional systems are presented.
These studies are on simple, yet nontrivial, models. Most of these are
classical systems, but some quantum-mechanical work is also
reported. Much of the work has been on lattice  
models corresponding to phononic  systems, and some on hard particle and hard
disc systems. A recently developed  approach, using generalized
Langevin equations and phonon Green's functions, is explained and
several applications to harmonic systems are given. For interacting
systems, various analytic approaches based on the Green-Kubo
formula are described, and their predictions are compared with the
latest results from simulation. These results indicate that for
momentum-conserving systems, transport is anomalous in one and two
dimensions, and the thermal conductivity $\kappa$, diverges with system
size $L$, as $\kappa \sim L^\alpha$. 
For one dimensional interacting systems there is strong numerical
evidence for a universal exponent 
$\alpha =1/3$, but there is no exact proof for this so far.
A brief discussion of some of the
experiments on  heat conduction in nanowires and nanotubes is also given.
\bigskip

\hbox to \textwidth{\hsize\textwidth\vbox{\hsize30pc
\hspace*{-12pt} 
{1.} Introduction\\
{2.} Methods \\
\hspace*{10pt}{2.1.} Heat bath models, definitions of current,
temperature and conductivity  \\
\hspace*{10pt}{2.2.} Green-Kubo formula  \\
\hspace*{10pt}{2.3.} Nonequilibrium Green's function method \\
{3.} Heat conduction in  harmonic lattices\\
\hspace*{10pt}{3.1.} The Rieder-Lebowitz-Lieb method \\
\hspace*{10pt}{3.2.}  Langevin equations and Greens function (LEGF)
formalism \\
\hspace*{10pt}{3.3.}  Ordered  harmonic lattices \\
\hspace*{24pt} {3.3.1.}  One dimensional case\\
\hspace*{24pt} {3.3.2.}  Higher dimensions \\
\hspace*{10pt}{3.4.}  Disordered harmonic lattices \\
\hspace*{24pt} {3.3.1.}  One dimensional disordered lattice \\
\hspace*{24pt} {3.3.2.}  Two dimensional disordered lattice \\
\hspace*{10pt}{3.5} Harmonic lattices with self-consistent reservoirs \\
{4.}    Interacting systems in one dimension\\
\hspace*{10pt}{4.1.}  Analytic results \\
\hspace*{24pt} {4.1.1.}   Hydrodynamic equations and renormalization
group theory \\
\hspace*{24pt} {4.1.2.}  Mode coupling theory\\
\hspace*{24pt} {4.1.3.}  Kinetic and Peierls-Boltzmann theory \\
\hspace*{24pt} {4.1.4.}  Exactly solvable model \\
\hspace*{10pt}{4.2.} Results from simulations  \\
\hspace*{24pt} {4.2.1.} Momentum conserving models \\
\hspace*{24pt} {4.2.2.}  Momentum non-conserving models \\
{5.} Systems with disorder and interactions \\
{6.} Interacting systems in two dimensions \\
{7.} Non-interacting non-integrable systems \\
{8.} Experiments \\
{9.} Concluding remarks \\
Acknowledgements \\
References
}}
\end{abstract}

\section{Introduction}

It is now about two hundred years since Fourier first proposed the
law of heat conduction that goes by his name. Consider a macroscopic
system subjected to different temperatures at its boundaries. One
assumes that it is possible to have a coarse-grained description with
a clear separation between microscopic and macroscopic scales.  
If this is achieved, 
it is then possible to define, at any spatial point $\bx$ in the system
and at time $t$, a local temperature field $T(\bx,t)$ which varies
slowly both in space and time (compared to microscopic scales).
One then expects heat currents to flow inside
the system and Fourier argued that the local heat current density
${\bf{J}}(\bx,t)$   is given by
\bea
{\bf{J}}(\bx,t)= - \kappa \bf{\nabla} T(\bx,t)~,
\eea    
where $\kappa$ is the thermal conductivity of the system.
If $u(\bx,t)$ represents the local energy density then this
satisfies the continuity equation $\p u/\p t + \bf{\nabla}. \bf{J} = 0$.
Using the relation $\p{u}/\p T = c $, where $c$ is the specific heat
per unit volume, leads to the heat diffusion equation:
\bea
\f{\p T(\bx,t)}{\p t}=\f{1}{c} \nabla. [~{\kappa} \nabla T(\bx,t)~]~.
\eea 
Thus, Fourier's law implies diffusive transfer of energy. In
terms of a microscopic picture, this can be understood in terms of the
motion of 
the heat carriers, \ie, molecules, electrons, lattice
vibrations(phonons), etc., which suffer random collisions and hence move
diffusively. Fourier's law is a phenomological law and has been
enormously succesful in  providing an accurate description of heat
transport phenomena as observed in experimental systems. However there is 
no rigorous derivation of this law starting from a microscopic
Hamiltonian description and this basic question has motivated a large
number of studies on heat conduction in model systems.  
One important and somewhat surprising conclusion
that emerges from these studies is that Fourier's law is probably not
valid in one and two dimensional systems, except when the system is
attached to an external substrate potential. 
For three dimensional systems, one expects that
Fourier's law is true in generic models, but it is not yet known as to
what are the neccessary conditions.

Since one is addressing a conceptual issue it makes sense to
start by looking at the simplest models which incorporate the
important features that one believes are necessary to see normal
transport. For example, one expects that for a solid, anharmonicity
and disorder play important roles in determining heat transport properties. 
Thus most of the theoretical studies have been on these simple models, 
rather than on detailed models including realistic interparticle potentials,
etc. The hope is that the simple models capture the important physics,
and understanding them in detail is the first step towards
understanding more realistic models. This review almost exclusively
will talk about simple models of heat conduction in low dimensional
systems, mostly one dimensional ($1D$) and some two dimensional ($2D$).
Also a lot of the models that have been studied are lattice models,
where heat is transported by phonons, 
and are relevant for understanding heat conduction in electrically
insulating materials.  Some work on hard particle and hard disc
systems will also be reviewed.

There are two very good earlier review articles on this topic,
including those by Bonetto \etal~ \cite{BLR00} and Lepri
\etal~\cite{LLP03}. Some areas that have not been covered in much detail here can be found
in those reviews.  Another good review, which also gives some  historic perspective,
is that by Jackson \cite{jackson78}.   Apart from being an update on
the older reviews, one area which has been covered extensively in this
review is the use of the nonequilibrium Green's function
approach for harmonic systems. This approach nicely shows the
connection between  results from various studies on heat transport in
classical harmonic chain models, and results obtained 
from methods such as the Landauer formalism, which is widely used in
mesoscopic physics.  As we will see, this is one of the few methods
where explicit results can be obtained for the quantum case also.

The article is organized as follows. In Sec.~(\ref{sec:methods}), 
some basic definitions and a description of some of the methods used
in transport studies is given.  
In Sec.~(\ref{sec:harmlat}),  results for the harmonic
lattice are given. The nonequilibrium Green's function theory will be developed
using the Langevin equation approach and various applications of this
method are described. The case of interacting particles ((non-harmonic
inter-particle interactions)  in one
dimension is treated in Sec.~(\ref{sec:INT1D}). This section briefly
summarizes the analytic approaches, and then gives results of
the latest simulations in momentum-conserving and momentum
non-conserving one dimensional systems. The next section
[Sec.~(\ref{sec:disint})] looks at the joint effect of disorder and
interactions in one dimensional systems. In Sec.~(\ref{sec:intsys2D})
results on two dimensional interacting systems are presented while
Sec.~(\ref{sec:nonint}) gives results for billiard like systems of
noninteracting particles. Some of the recent experimental results on
nanowires and nanotubes are discussed in
Sec.~(\ref{sec:expts}). Finally the conclusions of the review are
summarized in Sec.~(\ref{sec:conclusions}) and a list of some
interesting open problems is provided.  

\section{Methods} 
\label{sec:methods}
The most commonly used approaches in heat transport studies have been:
(i) those which look at the {\it nonequilibrium steady state} obtained
by connecting a system to reservoirs at different temperatures, and
(ii) those based on the {\it Green-Kubo relation} between conductivity and
{ equilibrium} correlation functions.  In this section we will
introduce some of the definitions  and concepts necessary in using   these
methods [secs.~(\ref{sec:defs},\ref{sec:GK})]. Apart from these two
methods, an approach that has been especially 
useful in understanding  ballistic transport in mesoscopic systems, is
the nonequilibrium  Green's function method and we will describe this
method in sec.~(\ref{sec:negf}). Ballistic transport of electrons
refers to the case where electron-electron interactions are negligible. In the
present context ballistic transport means that phonon-phonon
interactions can be neglected.

\subsection{Heat bath models, definitions of current, temperature and
  conductivity} 
\label{sec:defs}
To study steady state heat transport in a Hamiltonian
system, one has to connect it to heat reservoirs. In this
section we will first discuss some commonly used models of reservoirs,
and  give the definitions of heat current,
temperature and thermal conductivity. It turns out that there are
some subtle points involved here  and we will try to explain these. 

First let us discuss a few models of heat baths that have been
used in the literature. For simplicity we here discuss the $1D$ case
since the generalization to higher dimension is straight-forward. We
consider a classical $1D$ system of 
particles interacting through a nearest neighbour interaction potential
$U$ and which are in an external potential $V$. The Hamiltonian is thus:
\bea
H=\sum_{l=1}^N [\f{p_l^2}{2 m_l} + V(x_l)~] + \sum_{l=1}^{N-1}
U(x_l-x_{l+1}) \label{sec2:ham1D}
\eea
where $\{m_l,x_l,p_l=m_l \dot{x}_l \}$ for $l=1,2,...N$  denotes the masses,
positions and momenta of the $N$ particles.  
For the moment we will assume that the interparticle
potential is such that the particles do not cross each other and so
their ordering on the line is maintained. 

To drive a heat current in the above Hamiltonian system, one needs to
connect it to heat reservoirs. Various models of baths have been used
in the literature and here we discuss three popular ones.

(i) {\bf Langevin baths}:  These are defined by  adding additional force
terms in the equation of motion of the particles in contact with
baths. In the simplest form, the additional forces consist of a
dissipative term, and a stochastic term, which is taken to be Gaussian
white noise.   
Thus with Langevin reservoirs connected to particles $l=1$ and $l=N$,
the equations of motion are given by:
\bea
 \dot{p}_1&=&f_1-\f{\g_L}{m_1}
{p}_1+\eta_L(t) \nn \\
\dot{p}_l&=&f_l~~~{\rm for }~~l=2,3...N-1 \nn \\
\dot{p}_N&=&f_N-\f{\g_R}{m_N}
{p}_N+\eta_R(t)~~\label{sec2:eqmot1D} \\
{\rm where}~~~~~~~f_l&=&-\f{\p H}{\p x_l} \nn
\eea 
is the usual Newtonian force on the $l^{\rm th}$ particle.
The noise terms given by $\eta_{L,R}$ are Gaussian, with zero
mean, and related to the dissipation
coefficients $\g_{L,R}$ by the usual fluctuation dissipation relations
\bea
\la \eta_L(t) \eta_L(t') \ra&=&2 k_B T_L \gamma_L \d(t-t') \nn \\
\la \eta_R(t) \eta_R(t') \ra&=&2 k_B T_R \gamma_R \d(t-t') \nn \\
\la \eta_L(t) \eta_R(t') \ra&=&0~, \nn
\eea
 where $T_{L},T_{R}$ are the temperatures of the left and right
 reservoirs respectively. 

More general Langevin baths where the noise
 is correlated will be described in sec.~(\ref{sec:legf}). Here we
 briefly discuss one particular example of a correlated bath, namely the
 Rubin model. This model is obtained by connecting our system of
 interest to two reservoirs which are each described by semi-infinite harmonic
 oscillator chains with Hamiltonian of the form $H_b=\sum_{l=1}^{\infty} P_l^2/2
 + \sum_{l=0}^{\infty}  (X_l-X_{l+1})^2/2$, where $\{X_l,P_l\}$ denote
 reservoir degrees of freedom and $X_0=0$. One assumes that the
 reservoirs are initially in thermal equilibrium at different
 temperatures and are then linearly coupled, at time $t=-\infty$, to the two ends of
 the system. Let us assume the coupling of system with left reservoir to
 be of the form $-x_1 X_1$. Then, following the methods to be discussed
 in sec.~(\ref{sec:legf}), one finds that the effective equation of
 motion of the left-most particle is a generalized Langevin equation
 of the form:
\bea
\dot{p}_1=f_1+\int_{-\infty}^t dt' \Sigma_L (t-t') x_1(t') + \eta_L (t)~,
\eea
where the fourier transform
of the kernel $\Sigma_L (t)$ is given by:
\bea
\tilde{\Sigma}_L(\om) =  \int_{0}^\infty dt \Sigma_L
 (t) e^{i \om t}  
&=& e^{i q} ~~~~{\rm for}~~~ |\om| < 2 \nn \\
&=& -e^{-\nu}~~~~{\rm for}~~~ |\om | > 2 \nn~,
\eea
and $q,\nu$ are defined through $\cos (q) =1-\om^2/2$,
 $\cosh(\nu)=\om^2/2-1$ respectively.   
The noise correlations are now given by:
\bea
\la \tilde{\eta}_L (\om) \tilde {\eta}_L (\om') \ra = \f{k_B T_L}{\pi
 \om} Im[\tilde{\Sigma}_L (\om)]~\delta (\om+\om'), 
\eea
where $\tilde{\eta}_L(\om) = (1/2\pi) \int_{-\infty}^\infty dt \eta_L
 (t) e^{i \om t}$. Similar equations of motion are obtained for the
 particle coupled to the right reservoir.

(ii) {\bf Nos\'e-Hoover baths}: These are deterministic baths with {\it
   time-reversible} dynamics which however, surprisingly, have the  ability to
   give rise to {\it irreversible} dissipative behaviour.  In
   its simplest form,  Nos\'e-Hoover baths attached to  the end particles
   of the system described by the Hamiltonian Eq.~(\ref{sec2:ham1D}),
are defined through the following equations of motion for the set of
   particles:
\bea
\dot{p}_1&=&f_1-\zeta_L
{p}_1 \nn \\
\dot{p}_l&=&f_l~~{\rm for }~~l=2,3...N-1
   \nn \\ 
\dot{p}_N&=&f_N-\zeta_R
{p}_N ~, \label{sec2:nheqmot1D}
\eea 
where $\zeta_L$ and $\zeta_R$ are also dynamical variables which satisfy
   the following equations of motion:
\bea
\dot{\zeta}_L &=& \f{1}{\theta_L}\left( \f{p_1^2}{m_1 k_B T_L}-1 \right)
   \nn  \\  
\dot{\zeta}_R &=& \f{1}{\theta_R}\left( \f{p_N^2}{m_N k_B T_R}-1 \right)~, \nn 
\eea
with $\theta_L$ and $\theta_R$ as parameters which control the
strength of    coupling to reservoirs.

Note that in both models (i) and (ii) of baths, we have described
situations where baths are connected to { \emph{ particular particles}}
and not located at fixed positions in space. These are 
particularly suited for simulations of lattice models, where particles
make small displacements about equilibrium positions. Of course one
could  modify the dynamics by saying that particles experience the bath
forces (Langevin or Nos\'e-Hoover type) whenever they are in a given region of
space, and then these baths can be applied to fluids too. Another
dynamics   where the heat bath is located at  a fixed 
position, and is particularly suitable for simulation of fluid
systems, is the following: 

(iii) {\bf Maxwell baths}: Here we take  particles described by
the Hamiltonian Eq.~(\ref{sec2:ham1D}),and moving within a closed box
extending from $x=0$ to $x=L$. The particles execute usual Hamiltonian
dynamics except when any of the end particles hit the walls. 
Thus when  particle $l=1$ at the left end ($x=0$) hits the wall at temperature
$T_L$, it is reflected with a random velocity chosen from the distribution:  
\bea
P(v)=\f{m_1 v}{k_B T_L} ~\theta (v)~e^{-m_1 v^2/(2k_B T_L)}~,
\eea
where $\theta(v)$ is the Heaviside step function. A similar rule is
applied at the right end. 

There are two ways of defining a current variable depending on whether
one is using a discrete or a continuum description. For  lattice
models, where every particle moves around specified lattice points,
the discrete definition is appropriate. In a fluid system, where the
motion of particles is unrestricted, one has to use the continuum
definition. 
For the $1D$ hard particle gas, the ordering of particles is
maintained, and in fact both definitions
have been used in simulations  to calculate  the steady state
current. We will show 
here explicitly that they are equivalent. Let us first discuss the discrete
definition of heat current.

For the Langevin and Nos\'e-Hoover baths, we note that the equation of
motion has the form $\dot{p}_l=f_l+\delta_{l,1}f_L+\delta_{l,N} f_R$
where $f_L$ and $f_R$ are forces from the bath. The instantaneous rate
at which work is done by the left and right reservoirs on the system
are respectively given by:
\bea
j_{1,L}&=&f_L v_1 \nn \\
{\rm and}~~~j_{N,R}&=&f_R v_N~, \nn 
\eea  
and these give the instantaneous energy currents from the reservoirs
into the system.
To define the local energy current inside the wire we first define the
local energy density 
associated with the $l^{\rm th}$ particle (or energy at the lattice site $l$) as follows:
\bea
\e_1 &=& \f{p_1^2}{2 m_1} +V(x_1)+ \f{1}{2} U(x_1-x_2)~, \nn \\
\e_l &=& \f{p_l^2}{2 m_l}+ V(x_l) + \f{1}{2}
  [~U(x_{l-1}-x_l)+U(x_l-x_{l+1})~]~, ~~~{\rm for }~~l=2,3...N-1 \nn \\
 \e_N &=& \f{p_N^2}{2 m_N} +V(x_N)+ \f{1}{2} U(x_{N-1}-x_N)~. \label{sec2:enerdef1} 
\eea
Taking a time derivative of these equations, and after some
straightforward manipulations, we get the discrete continuity
equations given by:
\bea
\dot{\e}_1&=&-j_{2,1}+j_{1,L} \nn \\
\dot{\e}_l&=&-j_{l+1,l}+j_{l,l-1} ~~~{\rm for }~~l=2,3...N-1 \nn \\
\dot{\e}_N&=&j_{N,R}+j_{N,N-1}~, \label{sec2:conteq1} \\
{\rm with}~~ j_{l,l-1}&=&\f{1}{2} (v_{l-1} + v_{l}) f_{l, l-1}
\label{sec2:jdef1} \\
{\rm and ~ where}~~~~f_{l,l+1}&=& -f_{l+1,l}=-\p_{x_l} U(x_l-x_{l+1})~ \nn
\eea 
is the force that the $(l+1)^{\rm th}$ particle exerts on the $l^{\rm
  th}$ particle and $v_l=\dot{x}_l$. 
From the above equations one can  identify
$j_{l,l-1}$ to be the energy current  from site $l-1$ to $l$.  
The steady state average of this current can be written in a slightly
different form which has a clearer physical meaning. We will denote
steady state average of any physical quantity $A$ by $\la A \ra$. Using the fact
that $\la dU(x_{l-1}-x_l)/dt \ra =0$ it follows that $\la v_{l-1} f_{l,l-1}
\ra = \la v_l f_{l,l-1} \ra$ and hence:
\bea
\la j_{l,l-1} \ra = \la \f{1}{2} (v_l +v_{l-1}) f_{l, l-1}
\ra =\la v_l f_{l,l-1} \ra~, \label{sec2:sscur}
\eea
and this has the simple interpretation as the average rate at which the
$(l-1)^{\rm th}$ particle does work on the $l^{\rm th}$ particle. 
In the steady state, from Eq.~(\ref{sec2:conteq1}), we get the
equality of current flowing between any neighbouring pair of particles: 
\bea
J=\la j_{1,L} \ra=\la j_{2,1} \ra=\la j_{3,2} \ra=...\la j_{N,N-1}\ra=
-\la j_{N,R} \ra~,\label{sec2:sscur2}
\eea  
where we have used the notation $J$ for the steady state energy current per
bond. In simulations one can use the above definition, which involves
no approximations, and  a good check of convergence to steady state is to verify the
above equality on all bonds.
In the case where interaction is in the form of hard particle
collisions we can write the expression for steady state current in a
different form. Replacing the steady state average by a time average we get:
\bea
\la j_{l,l-1}\ra= \la  v_l f_{l,l-1} \ra = \lim_{\tau \to \infty} \f{1}{\tau}
\int_0^\tau dt  v_l(t) f_{l,l-1}(t)= \lim_{\tau \to \infty}
\f{1}{\tau} \sum_{t_c} \Delta K_{l,l-1}~,\label{hpcur1}
\eea
where $t_c$ denotes time instances at which particles $l$ and $(l-1)$ collide
and $\Delta K_{l,l-1}$ is the change in energy of the $l^{\rm th}$
particle as a result of the collision.

Next we discuss the continuum definition of current which is more
appropriate for fluids but is of general validity. We will discuss it
for the case of Maxwell boundary conditions with the system confined
in a box of length $L$. 
Let us define the local energy density at position
$x$ and at time $t$ as:
\bea
\e(x,t)=\sum_{l=1}^N \e_l~ \d [x-x_l(t)]~,
\eea
where $\e_l$ is as defined in Eq.~(\ref{sec2:enerdef1}).
Taking a time derivative we get the required continuum continuity
equation in the form (for $ 0<x<L$ ):
\bea
&&~~~~~~~~~\f{\p \e(x,t)}{\p t} + \f{\p j(x,t)}{\p x} = j_{1,L} \delta
(x) +j_{N,R} \delta (x-L) \label{sec2:conteq} \\
&&{\rm where}~~~ j(x,t)= j_K(x,t)+j_I(x,t) \nn \\
&&{\rm with}~~~ j_K(x,t)=\sum_{l=1}^N \e_l(t) v_l(t) \delta
  [x-x_l(t)] \nn \\
&&{\rm and}~~~ j_I(x,t)= \sum_{l=2}^{N-1} (j_{l+1,l}-j_{l,l-1})~ \theta
  [x-x_l(t)] + j_{2,1} \theta [x-x_1(t)]-j_{N,N-1}\theta[x-x_N(t)]~.  \nn
\eea
Here $j_{l,l-1},~j_{1,L},~j_{N,R}$ are as defined earlier in the discrete case, and we have written
the current as a sum of two parts, $j_K$ and $j_I$, whose physical meaning 
we now discuss. To see this, consider a particle
configuration with 
$x_1,x_2,...x_k < x < x_{k+1},x_{k+2},...x_N$. Then we get
\bea
j_I(x,t)=j_{k+1,k} \nn
\eea
which is thus simply the rate at which the particles on the left of
$x$ do work on the particles on the right. Hence we can interpret
$j_I(x,t)$ as the contribution to the current density coming from  
interparticle interactions. The other part $j_K(x,t)$
arises from the physical flow of particles carrying energy across the
point $x$. Note, however, that even in the absence of any net
convection particle flow, both $j_K$ and $j_I$ can contribute to the  
energy flow. In fact for point particles interacting purely by hard
elastic collisions $j_{k+1,k}$ is zero whenever the $k^{\rm th}$ and
$(k+1)^{\rm th}$ particles are on the two sides of the point $x$ and
hence $j_I$ is exactly zero. The only contribution to the energy
current then comes from the part $j_K$ and we thus for the steady-
state current we obtain
\bea
\la j(x,t) \ra = \sum_{l=1}^N \la \f{m_l v_l^3}{2}~\d(x-x_l) \ra~. \label{hpcur2}
\eea
In simulations either of expressions Eq.~(\ref{hpcur1}) or
Eq.~(\ref{hpcur2}) can be used to evaluate the steady state current
and will give identical results. For hard particle simulations one
often uses a simulation which updates between collisions and in this
case it is more efficient to evaluate the current using
Eq.~(\ref{hpcur1}). 
We now show that, in the nonequilibrium steady state, the average
current from the discrete and continuum definitions are the same, $\ie$,
\bea
\la j_{l,l-1} \ra =\la j(x,t) \ra=J~.
\eea
Note that the steady state current is independent of $l$ or $x$. 
To show this we first define the total current as:
\bea
\mJ (t) &=&\int_0^L dx j(x,t)=\sum_{l=1,N} \e_l v_l- \sum_{l=2,N-1} x_l
(j_{l+1,l}-j_{l,l-1} ) -x_1j_{2,1}+x_N j_{N,N-1} \nn \\
&=& \sum_{l=1,N} \e_l v_l+ \sum_{l=1,N-1} (x_{l+1}-x_l) j_{l+1,l}
\label{sec2:totcur1}~.   
\eea
Taking the steady-state average of the above equation and using the
fact that $\la \e_l v_l \ra= -\la \dot{\e}_l x_l \ra = \la (
j_{l+1,l}-j_{l,l-1}) x_l \ra$, where $j_{1,0}=j_{1,L},
j_{N+1,N}=-j_{N,R}$ we get:
\bea
\la \mJ \ra= -\la x_1 j_{1L}+x_N j_{NR} \ra~. \nn
 \eea
Since the Maxwell baths are located at $x=0$ and $x=L$, the above
then gives
$\la \mJ \ra=L \la j(x,t) \ra= -L \la j_{NR} \ra$ and hence from 
Eq.~(\ref{sec2:sscur2}), we get $\la j(x,t) \ra = J = \la
j_{l,l-1}\ra$, which proves the equivalence of the two definitions.

The extensions of the current definitions, both the discrete and
continuum versions, to higher dimensions is
straightforward. Here, for reference,  we outline the derivation for the continuum
case since it is not easy to see a discussion of this in the literature.
Consider a system in $d$-dimensions with Hamiltonian given by:
\bea
H=\sum_l \left[~\f{ \bp_l^2}{2 m_l}+V(\bx_l)~\right] +\f{1}{2} \sum_{l \neq n}
\sum_n U(r_{ln})~, 
\eea
where $\bx_l=(x_l^1,x_l^2,...x_l^d)$ and
$\bp_l=(p_l^1,p_l^2,...p_l^d)$ are the vectors denoting the position
and momentum of the $l^{\rm th}$ particle and
$r_{ln}=|\bx_l-\bx_n|$. The particles are assumed to be inside a
hypercubic box of volume $L^d$.   
As before we define the local energy density as: 
\bea
\e(\bx,t)&=&\sum_l \d (\bx-\bx_l) \e_l {\rm~~~where} \nn \\ 
\e_l &=& \f{ \bp_l^2}{2m_l}+V(\bx_l) +\f{1}{2} \sum_{n \neq l} U(r_{ln})~. \nn
\eea
Taking a derivative with respect to time (and suppressing the source
terms arising from the baths) gives:
\bea
\f{\p \e(\bx,t)}{\p t} &=& -\sum_{\alpha =1}^d \left[ \f{\p}{\p x_\a}
\sum_l \d (\bx - \bx_l)~ \e_l v_l^\a \right] ~+~ 
\sum_l \d (\bx-\bx_l) \dot{\e}_l   \\
&=& -\sum_\alpha \f{\p}{\p x^\a} [~j_{\a}^K + j_{\a}^I~]~,  \nn 
\eea
where 
\bea
j^\a_K (\bx,t) &=& \sum_i \d (\bx - \bx_l)~ \e_l~ v^\a_l  \nn \\
{\rm and}~~
j^\a_I(\bx,t) &=& - \sum_l \sum_{n \neq l} \theta (x^\a-x_l^\a) \prod_{\nu
  \neq \a} \d (x^\nu-x_l^\nu)~ j_{l,n} \\
{\rm where}~~j_{l,n}&=& \f{1}{2} \sum_{\nu}
(~v_l^\nu+v_n^\nu~)~f_{l,n}^\nu~, \nn 
\eea
and $f^\a_{l,n}=-\p U(r_{l,n})/\p x^\a_l$ is the force, in the
$\alpha^{\rm th}$ direction, on the $l^{\rm th}$
particle due to the $n^{\rm th}$ particle. We have  defined $j_{l,n}$ as the
current, from particle $n$ to particle $l$, analogously to the discrete
$1D$ current.  
The  part $j^\a_I$ gives the energy flow as a result of
physical motion of particles across $x^\a$.  
The part $j^\a_I$ also  has a simple physical
interpretation, as in the $1D$ case. First note that we need to sum over only those $n$ for
which $x_n^\a < x^\a$. Then the formula basically gives us the net rate, at
which work is done, by particles on the left of $x^\a$, on the particles
to the right. This is thus the rate at which energy flows from left to
right. 
By integrating the current density over the full volume of the system, we
get the total current:
\bea
\mJ^\a(t) = \sum_l  \e_l v_l^\a  +\f{1}{2} \sum_{l \neq
  n}  \sum_n (x_l^\a-x_n^\a)~ j_{l,n}~. \label{totcur2}
\eea

Thus we get an expression similar to that in $1D$ given by
Eq.~(\ref{sec2:totcur1}).  
In simulations making nonequilibrium measurements, any of the various
definitions for current can be used to find the steady state current.
However, it is not clear whether other quantities, such as correlation
functions obtained from the discrete and continuum definitions, 
will be the same.   

The local temperature can also be defined using
either a discrete approach (giving $T_l$) or  a continuum approach
(giving $T(\bx,t)$). In the  steady state these are
respectively given by (in the $1D$ case):
\bea
k_B T_l &=& \left\la \f{p_l^2}{m_l} \right\ra \nn \\
k_B T(x) &=& \f{\la \sum_l \f{p_l^2}{m_l} \delta (x-x_l)\ra }{\la \sum_l
  \delta (x-x_l) \ra}
\eea
Again it is not obvious that these  two definitions will always agree.
Lattice simulations usually use the discrete definition while hard
particle simulations use the continuum definition.  

The precise definition of thermal conductivity would be:
\bea
\kappa= \lim_{L \to \infty} \lim_{\Delta T \to 0} \f{ J L}{\Delta T}~,
\eea
where $\Delta T= T_L-T_R$. In general $\kappa$ would depend on
temperature $T$. The finiteness of $\kappa (T)$, along with Fourier's law,
implies that even for arbitrary fixed values of $T_L, T_R$ the current
$J$ would scale as $\sim L^{-1}$ (or $N^{-1}$). What is of real interest is this
scaling property of $J$ with system size. 
We will typically be interested in the large $N$ behaviour of the
conductivity defined as:
\bea
\kappa_N = \f{JN}{\Delta T}~,
\eea
which will usually be denoted by $\kappa$.
For large $N$, systems with normal diffusive transport give a finite
$\kappa$ while anomalous transport refers to the scaling
\bea
\kappa \sim N^\alpha~ ~~~~\alpha \neq 0~,
\eea
and the value of the heat conduction exponent $\alpha$ is one of the
main objects of interest. For the current, this implies $J \sim
N^{\alpha-1}$. 

The only examples where the steady state current can  be 
analytically evaluated, and exact results are available for the
exponent $\alpha$, corresponds to harmonic lattices, using very
specific methods that will be discussed in sec.~(\ref{sec:harmlat}).

{\bf{ Coupling to baths and contact resistance}}: In the various
models of heat baths that we have discussed, the
efficiency with which heat exchange takes place between reservoirs and
system depends on the  strength
of coupling constants. For example, for the Langevin and Nos\'e-Hoover
baths, the parameters $\gamma$ and $\theta$  respectively determine
the strength of  coupling ( for the Maxwell bath one could introduce a
 parameter which gives the probability that after  a collision
the particle's speed changes and this can be used to tune the
coupling between system and reservoir). From simulations it is found
that typically there is an optimum value of the coupling parameter for
which energy exchange takes place most efficiently, and at this value
one gets the maximum current for given system and fixed bath
temperatures. For too high or too small values of the coupling
strength the current is small. The coupling to bath can be thought of
as giving rise to a contact resistance. An effect of this resistance
is to give rise to  boundary jumps  in the temperature profile measured
in simulations. One expects that these jumps will be  present
as long as the contact resistance is comparable to the systems resistance. 
We will later see that in order to be sure that one is measuring the
true resistance of the system, it is necessary to be in parameter
regimes where the contact resistances can be  neglected.

\subsection{Green-Kubo Formula}
\label{sec:GK}
The Green-Kubo formula provides a relation between transport
coefficients, such as the thermal conductivity $\kappa$ or the
electrical conductivity $\sigma$, and {\it equilibrium} time correlation functions of the
corresponding current. For the thermal conductivity in a
classical $1D$ system, the Green-Kubo formula gives:
\bea
\kappa= \f{1}{k_B T^2} \lim_{\tau \to \infty} \lim_{L \to \infty}
\f{1}{ L} \int_{0}^\tau
dt \la \mJ(0) \mJ(t) \ra~, \label{sec2:kubo}
\eea
where $\mJ$ is the total current as defined in Eq.~(\ref{sec2:totcur1})
 and $\la...\ra$ denotes an average over initial conditions
chosen from either a micro-canonical ensemble or a canonical one at
temperature $T$. 
Two important points to be remembered with regard to use of the above
Green-Kubo formula are the following: 

(i) It is often necessary to subtract a convective part from the current definition or,
alternatively, in the microcanonical case one can work with initial
conditions chosen such that the centre of mass velocity is zero (see
also discussions in \cite{green60,kadanoff63}). 
To understand this point, let us consider the case with
$V(\bx)=0$. Then for a closed system that is not in contact 
with reservoirs, we expect the time average of  the total current to
vanish. But this is true  only if we  are in the centre of mass
frame. If the centre of mass is moving with velocity $\bv$ then the
average velocity of any particle $\la \bv_l \ra= \bv$. Transforming to
the moving frame let us write $\bv_l=\bv'_l+\bv$. Then the average total current in
the rest frame is given by (in $d$-dimensions):
\bea
\la \mJ^\a \ra &=& \left[ \f{M}{2} \bv^2 + \sum_l \la \e_l' \ra~\right] v^\a +
\sum_\nu ~\left[ ~\sum_l \la m_l v^{'\a}_l v^{'\nu}_l \ra +
  \f{1}{2} \sum_{\stackrel{l,n}{l \neq n}} \la 
~(x_l^\a-x_n^\a) ~ f^\nu_{l,n}~\ra~\right] ~ v^\nu \nn \\ 
&=& (E+P V)~ v^\a~,
\eea
where $M=\sum m_l$ and $E$ is the average total energy of the system
as measured in the 
rest frame and $V=L^d$ . In deriving the above result  we have used 
the standard 
expression for equilibrium stress-tensor  given by:
\bea
V \sigma_{\a \nu}= \sum \la m_l v^{'\a}_l v^{'\nu}_l \ra + \f{1}{2}
\sum_{l \neq n} \la ~(x_l^\a-x_n^\a) ~ f^\nu_{l,n}~\ra~ ~,
\eea
and assumed an isotropic medium. 
Thus, in general, to get the true energy current in an arbitrary
equilibrium ensemble one should use the expression:
\bea
\mJ^\a_c=\mJ^\a-(E+PV)v^\a ~. \label{JC}
\eea
The corresponding form in $1D$ should be used to replace $\mJ$ in  Eq.~(\ref{sec2:kubo}).

(ii) The second point to note is that in Eq.~(\ref{sec2:kubo}) the order of
limits $L \to \infty$  {\it and then} $\tau \to \infty$ has to be strictly maintained.  
In fact for a system of particles inside a finite box of length $L$ it
can be shown exactly that:
\bea
\int_0^\infty dt  \la \mJ_c(0) \mJ_c(t) \ra =0 ~.  \label{GKfinite}
\eea 
To prove this, let us consider a microcanonical ensemble (with $\la \mJ
\ra=0$, so that $\mJ_c=\mJ$), in which case from
Eq.~(\ref{sec2:totcur1}) we get:
\bea
\mJ (t) = \f{d}{dt} \left[\sum_{l=1}^N \e_l(t) x_l(t)~\right]~.
\eea
Multiplying both sides of the above equation by $\mJ (0)$, integrating
over $t$ and noting that both the boundary terms on the right hand
side vanish, we get the required result in Eq.~(\ref{GKfinite}). 
With the correct order of limits in Eq.~(\ref{sec2:kubo}), one can calculate the correlation
functions with arbitrary boundary conditions and apply the formula to obtain the
response of an open system with reservoirs at the ends.

{\bf{Derivation of the Green-Kubo formula for thermal
    conductivity}}: There have been a number of derivations of this formula
 by various authors including  Green, Kubo, Mori, McLennan, Kadanoff and
Martin, Luttinger, and Visscher  
 \cite{green54,kubo57b,mori58,green60,kadanoff63,luttinger64,visscher74,forster75,mclennan88}.
 None of these derivations are rigorous and all require certain assumptions. 
Luttinger's derivation was an attempt at a mechanical derivation and
involves introducing a fictitious 'gravitational field' which couples
to the energy density and drives an energy current. However one now
has to relate the response of the system to the field and its response
to imposed temperature gradients. This requires additional inputs such
as use of the  Einstein relation relating diffusion
coefficient (or thermal conductivity) to the response to the
gravitational field. We believe that this derivation, as is also the
case for most other derivations 
, implicitly assumes local thermal equilibrium. 
Although these derivations are not rigorous, they are 
quite plausible, and it is likely that the assumptions made are
satisfied in a large number of cases of practical interest. 
Thus the wide use of the Green-Kubo formula in calculating thermal
conductivity and transport properties of different systems is possibly
justified in many situations. 

Here we give an outline of a non-mechanical derivation of the
Green-Kubo formula, one in which the  assumptions can be somewhat
clearly stated and their physical basis understood. The
assumptions we will make here are:

(a) The nonequilibrium state with energy flowing in the system  can be
described by coarse-grained variables and the condition of local
thermal equilibrium is satisfied.

(b) There is no particle flow, and energy current is equal to
heat current. The energy current satisfies Fourier's law which we
write in the form $J(x,t)=-D \p u(x,t)/\p x$ where $D=\kappa/c$,  
$c$ is the specific heat capacity, and $u(x,t)=\bar{ \e} (x,t)
,~J(x,t)=\bar{ j} (x,t)$ are macroscopic variables obtained by a
coarse-graining (indicated by bars) of the microscopic fields.   

(c) Finally, we assume that regression of equilibrium energy
fluctuations occurs in the same way as nonequilibrium flow of energy. 

We consider a macroscopic system of size $L$. Fluctuations in energy density in equilibrium can be described by the
correlation function $S(x,t)= \la \e(x,t) \e(0,0) \ra-\la \e(x,t)
\ra~ \la \e (0,0) \ra$. Assumption (c) above means that the decay of
these fluctuations  is determined by the heat diffusion equation and
given by:
\bea
\f{\p S(x,t)}{\p t}=D \f{\p^2 S(x,t)}{\p x^2}~~~~~{\rm for}~~~t>0 \nn~,
\eea
where we have also assumed temperature fluctuations to be small enough
so that the temperature dependence of $D$ can be neglected. 
From time reversal invariance we have $S(x,t)=S(x,-t)$. Using this and
the above equation we get:
\bea
\tilde{S}(k,\om)=\int_{-\infty}^\infty dt
\hat{S} (k,t) e^{-i \om t} = \f{2 D k^2 \hat{S}(k,t=0)}{D^2 k^4 + \om^2} \nn~,
\eea
where $\hat{S} (k,t)=\int_{-\infty}^\infty dx S(x,t) e^{ik x}$. 
Now, from equilibrium statistical mechanics we have $\hat{S}(k=0,t=0)=c k_B
T^2$ and  using this in the  above equation we obtain: 
\bea
\kappa = c D =\f{1}{2  k_B T^2} \lim_{\om \to 0} \lim_{k \to 0}
\f{\om^2}{k^2}\tilde{S}(k,\om)~. \label{kubder1}
\eea 
One can relate the energy correlator to the current correlator
using the continuity equation $\p \e (x,t)/\p t + \p j (x,t)/\p x=0$
and this gives:
\bea
\la \tilde{j}(k,\om) \tilde{j}(k',\om') \ra=(2\pi)^2 \f{\om^2}{k^2} \tilde{S}(k,\om)~
\delta (k+k') \delta (\om+\om') \nn~,
\eea
where $\tilde{j} (k,\om)=\int_{-\infty}^\infty d x d t j (x,t) e^{i
  (kx -\om t)}$. Integrating the above equation over $k'$ gives:
\bea
\int_{-\infty}^\infty dx \int_{-\infty}^\infty  dt~ \la j(x,t) j(0,0) \ra~e^{i (kx -\om t)} =
\f{\om^2}{k^2} \tilde{S}(k,\om)~. \label{kubder2}
\eea
From Eq.~(\ref{kubder1}) and Eq.~(\ref{kubder2}) we then get:
\bea
\kappa=\lim_{\om \to 0} \lim_{k \to 0} \f{1}{2 k_B T^2} \int_{-\infty}^\infty  dx~ \int_{-\infty}^\infty  dt~ \la j(x,t) j(0,0) \ra~e^{i (kx
  -\om t)}~.  
\eea
Finally, using time-reversal and translational invariance and
interpreting the ${\om \to 0},~ {k \to 0}$ limits in the
alternative way ( as ${\tau \to \infty},~ {L \to \infty}$) we
recover the Green-Kubo formula in Eq.~(\ref{sec2:kubo})~.

{\bf{Limitations on use of the Green-Kubo formula}}: There are
several situations where the Green-Kubo formula in 
Eq.~(\ref{sec2:kubo}) is not applicable. 
For example, for the small structures that are studied in mesoscopic
physics, the thermodynamic limit is meaningless, and one is interested
in the conductance of a specific finite system. 
Secondly, in  many low dimensional systems, heat transport is anomalous
and the thermal conductivity diverges. 
In such cases it is impossible to take the limits as
in Eq.~(\ref{sec2:kubo}); one is there interested in the thermal
conductance as a function of $L$ instead of an $L$-independent thermal
conductivity. 
The usual procedure that has been followed in the heat conduction
literature is to put a cut-off at $t_c \sim L$, in the  upper limit in
the Green-Kubo integral \cite{LLP03}. 
The argument is that for a finite system
connected to reservoirs, sound waves traveling to the boundaries at a
finite speed, say  $v$, lead to a decay  of correlations in a time
$\sim L/v$. However 
there is no rigorous justification of this assumption.
A  related case is that of integrable systems, where the infinite time
limit of the correlation function in Eq.~(\ref{sec2:kubo}) is non-zero.

Another  way of using the Green-Kubo formula for finite systems
is to include the infinite reservoirs also while applying 
the formula and this was done,  
for example, by Allen and Ford \cite{allen68} for heat transport and by Fisher and
Lee \cite{fisher81} for electron transport. 
Both these cases are for non-interacting systems and the final expression 
for conductance (which is more relevant than conductivity in such systems) is basically what one 
also obtains from the Landauer formalism \cite{landauer70}, or the
nonequilibrium Green's function approach [see
  sec.~(\ref{sec:negf})]. 

It has been shown that  
Green-Kubo like expressions for the linear response heat current for finite  open systems can be 
derived rigorously by using the steady state fluctuation theorem
\cite{gallavotti96a,gallavotti96b,lebowitz99,bellet02,bellet03,gaspard07,dharsaito07}.
This has been done for lattice models coupled to stochastic Markovian
baths and the expression for linear 
response conductance of a one dimensional chain is given  as: 
\bea
\lim_{\Delta T \to 0} \f{J}{\Delta T}=\f{1}{k_B T^2} \int_0^\infty \la j_l(0) j_l(t) \ra~, \label{sec2:LRopen}
\eea
where $j_l$ is the discrete current defined in
sec.~(\ref{sec:defs}). 
Some of the important differences of this formula with the usual
Green-Kubo formula are 
worth keeping in mind: (i) the dynamics of the system here is
non-Hamiltonian  since they are for a system coupled to reservoirs,
(ii) one does not need to take the limit $N \to \infty$ first, the
formula being valid for a finite system, (iii) the discrete bond current
appears here, unlike the continuum one in the usual Green-Kubo formula.
Recently, an exact linear response result similar to Eq.~(\ref{sec2:LRopen}),
for the conductance of a finite open system has been derived using a
different approach  \cite{kundu08}. This has been done for quite general classical
Hamiltonian systems and for a number of implementations of heat
baths.

It appears that the linear response formula given by
Eq.~(\ref{sec2:LRopen}) is the correct one to use to evaluate the
conductance in systems where there is a problem with the usual
Green-Kubo formula, {\emph{e.g.}} in finite systems or low dimensional
systems showing anomalous transport (because of slow decay of $\la
\mJ(0) \mJ(t)\ra$. 
We note here the important point that the current-current correlation can
have very different scaling properties, for a purely Hamiltonian
dynamics, as compared to a heat bath dynamics. This has been seen in
simulations by Deutsche and Narayan \cite{deutsch03} for the
random collision model [defined in sec.~(\ref{sec:momcons})].   
Of course this makes things somewhat complicated since the usefulness
of the Green-Kubo formalism arises from the fact that it allows a
calculation of transport properties from equilibrium properties of the
system, and without any reference to heat baths, etc.  In fact, as we
will see in sec.~(\ref{sec:THEORY}), all the  analytic results for heat conduction
in interacting one dimensional systems rely on Eq.~(\ref{sec2:kubo})
and involve a calculation of the current-current correlator for a
closed system.   
Some of the simulation results discussed later suggest that, for interacting (nonlinearly)
systems, in the limit of large system size the heat current is
independent of details of the heat baths. This means that, in the  linear
response regime and the limit of large system sizes,  a description
which does not take into account  bath 
properties may still be possible.

\subsection{Nonequilibrium Green's function method} 
\label{sec:negf}
The nonequilibrium Green's function method (NEGF) is a method, first
invented in the context of electron transport, to calculate steady
state properties of a finite system connected to reservoirs which are
themselves modeled by noninteracting Hamilitonians with infinite
degrees of freedom \cite{caroli71,meir92,datta95}. Using the Keldysh
formalism, one can obtain formal 
expressions for the current and other observables such as electron
density, in models of electrons such as those described by
tight-binding type Hamiltonians. Recently this approach has been
applied both to phonon  \cite{yamamoto06,wang06} and photon
\cite{ojanen07,ojanen08} transport. 

The main idea in the formalism is as
follows. One 
starts with an initial density matrix describing the decoupled system, and two
infinite reservoirs which are in thermal equilibrium, at different
temperatures and chemical potentials. The system and reservoirs are
then coupled together and the density matrix is 
evolved with the full Hamiltonian for an infinite time so that one
eventually reaches a
nonequilibrium steady state. 
Various quantitites of interest such as currents and local
densities, etc can be obtained using the steady state density matrix
and can be written in terms of the so called Keldysh Green's
functions. 
An alternative and equivalent formulation is the Langevin approach
where, instead of dealing with the density matrix, or in the classical
case with the phase space density, one works with equations of motion
of the phase space variables of the full Hamiltonian (system plus
reservoirs). Again the reservoirs, which are initially in thermal
equilibrium, are coupled to the system in the remote past. It is then
possible to integrate out the reservoir degrees of freedom, and these
give rise to generalized Langevin terms in the equation of motion.
For non-interacting systems, one can show that it is possible to
recover all the results of NEGF exactly, both for electrons and
phonons. Here we will discuss this approach for the case of phonons,
and  describe the main results  that have been obtained [sec.~(\ref{sec:legf})]. 
  
For non-interacting systems, the formal expressions for current
obtained from the NEGF approach is in terms of transmission
coefficients of the heat carriers (electrons, phonons or photons)
across the system, with appropriate weight factors corresponding to
the population of modes in the reservoirs. These expressions are
basically what one also obtains from the Landauer formalism
\cite{landauer70}. We note that in the Landauer approach one simply thinks
of transport as a transmission problem and the current across the
system is obtained directly using this picture. In the simplest set-up
one thinks of one-dimensional reservoirs (leads) filled with
non-interacting electrons at different chemical potentials. On
connecting the system in between the reservoirs, electrons are transmitted
through the system from one reservoir to the other. The net current in
the system is then the sum of the currents from left-moving and
right-moving electron states from the two reservoirs respectively.  

\section{Heat conduction in harmonic lattices}
\label{sec:harmlat}
The harmonic crystal is a good starting point for understanding
heat transport in solids. Indeed in the equilibrium case we know that
studying the harmonic crystal already gives a good understanding of,
for example, the specific heat of an (electrically) insulating solid. 
In the nonequilibrium case, the problem of heat conduction in a
classical one dimensional
harmonic crystal was studied for the first time by Rieder, Lebowitz
and Lieb (RLL) \cite{rieder67}. They considered the case of stochastic Markovian baths and
were able to obtain the steady state exactly. 
The main results of this paper were: (i) the temperature in the bulk
of the system was  constant and 
equal to the mean of the two bath temperatures, (ii) the heat
current approaches a constant value for large system sizes and an
exact expression for this was obtained. These results can be
understood physically when one realizes that in the ordered crystal,
heat is carried by freely propagating phonons.  RLL considered the case where
only nearest neighbour interparticle interactions were
present. Nakazawa (NK) \cite{nakazawa68}
extended these results to the case with a constant onsite harmonic
potential at all 
sites and also to higher dimensions. 
The approach followed in both the RLL and NK papers was to obtain the
exact nonequilibrium stationary  state measure which, for this
quadratic problem, is a Gaussian distribution. A complete solution for
the correlation matrix was obtained and from this one could obtain both
the steady state temperature profile and the heat current.  

In sec.~(\ref{sec:rll}) we will briefly describe the RLL
formalism.  In sec.~(\ref{sec:legf}) we will describe a different and
more powerful formalism. This is the 
Langevin equation and nonequilibrium Green's function method and
various applications of this will be discussed in secs.~(\ref{sec:ordharmlat},\ref{sec:disharmlat},\ref{sec:selfcons}).

\subsection{The Rieder-Lebowitz-Lieb method}
\label{sec:rll}
Let us consider a classical harmonic system of $N$ particles with
displacements about the equilibrium positions described by the vector
$X=\{x_1,x_2,...x_N\}^T$, where $T$ denotes transpose. The particles
can have arbitrary masses and 
we define a diagonal matrix $\hat{M}$ whose diagonal elements 
$M_{ll}=m_l$, for $l=1,2...N$, give the masses of the particles. The
momenta of the particles are given by the vector $P=(M \dot{X})=\{p_1,p_2,...p_N\}^T$. We
consider the following harmonic  Hamiltonian for the system:
\bea
H=\f{1}{2}P^T \hM^{-1} P +\f{1}{2}X^T \hP X~,
\eea
where $\hP$ is the force matrix. Let us consider the general case where
the $l^{\rm th}$ particle is coupled to a white noise heat reservoir at
temperature $T^B_l$ with a coupling constant $\gamma_l$. The equations
of motion are given by:
\bea
\dot{x}_l&=&\f{p_l}{m_l} \nn \\
\dot{p}_l&=&-\sum_n \Phi_{ln} x_n -\f{\gamma_l}{m_l} p_l +\eta_l  ~~~{\rm
  for}~~l=1,2...N  \label{3.1:eqmot}
\eea
with the noise terms satisfying the usual fluctuation dissipation
relations
\bea
\la \n_l(t) \n_n(t') \ra =2 \g_l k_B T^B_l \d_{l,n} \delta (t-t')~.
\label{3.1:nncor} 
\eea 
Defining  new variables $q=\{ q_1,q_2...q_{2N} \}^T= \{
x_1,x_2...x_N,p_1,p_2...p_N \}^T$ we can rewrite 
Eqs.~(\ref{3.1:eqmot},\ref{3.1:nncor}) in the form: 
\bea
\dot{q}&=&-\ha~q+\n \nn \\
\la \n(t) \n^T(t') \ra &=& \hD \delta (t-t') 
\label{3.1:neqmot}
\eea
where the $2N$ dimensional vector $\n^T=(0,0,..0,\eta_1,\n_2...\n_N)$
and the $2N \times 2N$ matrices $\ha, \hD$ are given by:
\bea
 \ha= \left( \begin{array}{cc}
0 & -\hM^{-1} \\  \hP & \hM^{-1} \hGam \end{array} \right) ~~~~~~~
 \hD= \left( \begin{array}{cc}
0 & 0 \\  0 & \hE  \end{array} \right) \nn 
\eea
and $\hGam$ and $\hE$ are $N\times N$ matrices with elements
$\Gamma_{ln}=\g_l 
\d_{l,n}$,~ $E_{ln}=2 k_B T_l^B \gamma_l \delta_{ln}$ respectively. 
In the steady state, time averages of total time derivatives vanish,
hence we
have $\la d(qq^T)/dt \ra =0$. From Eq.~(\ref{3.1:neqmot}) we get 
\bea
\la \f{d}{dt} (q q^T) \ra &=& \la (-\ha q + \n) q^T  + q (-q^T \ha^T +\n^T)
\ra \nn \\
&=& -\ha.\hb -\hb.\ha^T + \la \n q^T + q \n^T \ra = 0  \label{3.1:derivsst}
\eea  
where $\hat{B}$ is the correlation matrix $\la q q^T \ra$. To find the
average of the term involving noise we first write the formal solution
of Eq.~\ref{3.1:neqmot}:
\bea
q(t)= \hat{G}(t-t_0) q(t_0) + \int_{t_0}^t dt' \hat{G}(t-t') \n(t')~~~~{\rm
  where}~~~\hat{G}(t)=e^{-\ha t}.  
\eea
Setting $t_0 \to -\infty$ and assuming $\hat{G}(\infty)=0$ (so that a
unique steady state exists), we get $q(t)=\int_{-\infty}^t dt'
\hat{G}(t-t') \n(t')$ and hence 
\bea
\la  q \n^T \ra =  \int_{-\infty}^t dt'
\hat{G}(t-t') \la \n(t') \n^T(t) \ra \nn \\   
=\int_{-\infty}^t dt' \hat{G}(t-t') \hD \d (t-t') = \f{1}{2} \hD \label{3.1:qeta}
\eea
where we have used the noise correlations given by Eq.~(\ref{3.1:nncor}),
and the  fact that $\hat{G}(0)= \hat{I}$, a unit matrix.
Using Eq.~(\ref{3.1:qeta}) in Eq.~(\ref{3.1:derivsst}), we finally get 
\bea
\ha.\hb+\hb.\ha^T= \hD  
\label{3.1:mateq}
\eea
The solution of this equation gives the steady state correlation
matrix $\hb$ which completely determines the steady state since we are
dealing with  a Gaussian process. 
In fact the steady state is given by the Gaussian distribution
\bea
P(\{ q_l \} ) =(2 \pi)^{-N} Det[\hb]^{-1/2} e^{-\frac{1}{2} q^T \hb^{-1} q}~.
\eea
Some of the components of the matrix equation Eq.~(\ref{3.1:mateq}) have
simple physical interpretations. To see this we first write $\hb$ in the
form
\bea
\hb=\left( \begin{array}{cc} \x & \z \\ \z^T & \y \end{array} \right)   \nn
\eea
where $\x,~\y$ and $\z$ are $N \times N$ matrices with elements $(\x)_{ln}=
\la x_l x_n \ra$, $(\y)_{ln}=\la p_l p_n \ra$ and $({\z})_{ln}= \la x_l p_n
\ra$. From Eq.~(\ref{3.1:mateq}) we then get the set of equations:
\bea
&&\hM^{-1} \z^T + \z \hM^{-1} = 0 \label{3.1:eq1}\\ 
&&\x \hP^T - \hM^{-1} \y + \z \hM^{-1} \hGam = 0 \label{3.1:eq2}\\
&&\hM^{-1} \hGam \y + \y \hGam \hM^{-1} + \hP \z + \z^T \hP = \hE \label{3.1:eq3}  
\eea   
From Eq.~\ref{3.1:eq1} we get the identity $ \la x_l p_n/m_n\ra= -\la x_n
p_l/m_l \ra$. Thus $\la x_l p_l \ra = 0$. The diagonal terms in
Eq.~\ref{3.1:eq2} give
\bea
&&\sum_n \la x_l (\Phi_{ln} x_n) \ra-\f{1}{m_l} \la p^2_l \ra +\la x_l p_l \ra
\f{\g_l}{m_l}= 0 \nn \\   
&&\Rightarrow \la x_l \f{\p H}{\p x_l} \ra = \la \f{p_l^2}{m_l} \ra=
k_B T_l \nn 
\eea
where we have defined the local temperature $k_B T_{l}=\la p_{l}^2/m_{l}
\ra$. This equation has the form of the `equipartition' theorem of 
equilibrium physics. It is in fact valid quite generally for any Hamiltonian
system at all bulk points and simply follows from the fact that $\la
(d/dt) x_l p_l \ra =0$.
Finally let us look at the diagonal elements of Eq.~\ref{3.1:eq3}. 
This gives the equation
\bea
\sum_n \la (\Phi_{ln} x_n) \f{p_l}{m_l} \ra = \f{\g_l}{m_l}k_B
(T^B_l-T_l) \label{3.1:enercons}
\eea
which again has a simple interpretation. The right hand can be seen to
be equal to $\la  (-\g_l {p}_l/m_l +\eta_l) p_l/m_l \ra$ which is
simply the 
work done on the $l$th particle by the heat bath attached to it (and
is thus the 
heat input at this site). 
On the left hand side  $\la (-\Phi_{ln} x_n) {p_l}/{m_l} \ra $ 
is the rate at which
the $n^{\rm th}$ particle does work on the $l^{\rm th}$  particle and is
therefore the energy current from site $n$ to site $l$. The left side
is thus is just the sum of the 
outgoing energy currents from the $l$th site to all the sites connected to
it by $\hP$. 
Thus we can 
interpret Eq.~(\ref{3.1:enercons}) as an energy-conservation
equation. The energy current between two sites is given by
\bea
J_{n \to l}=  -\la (\Phi_{ln}  x_n) \f{p_l}{m_l} \ra~.
\eea
Our main interests are in computing the temperature profile and the
energy current in the steady state. This requires solution of
Eq.~(\ref{3.1:mateq}) and this is quite difficult and has been
achieved only in a few special
cases. For the one-dimensional ordered harmonic lattice, RLL were able to solve
the equation exactly and obtain both the temperature profile and the
current. The  extension of their solution  to  higher dimensional lattices
is straightforward and was done by Nakazawa. More recently
Bonetto et al \cite{bonetto04} have used this approach to solve the case with
self-consistent reservoirs attached at all the bulk sites of a ordered
harmonic lattice.

A numerical solution of Eq.~(\ref{3.1:mateq}) 
requires inversion of a $N(2N+1) \times N(2N+1)$ matrix which
restricts one to rather small system sizes. 
The RLL approach is somewhat restrictive since it is not
easily generalizable to other kinds of heat baths or to the quantum
case. Besides, except for the ordered lattice, it is difficult to
obtain useful analytic results from this approach.   
In the next section we
discuss a different approach which is both analytically more tractable,
as well as numerically more powerful.

\subsection{Langevin equations and Green's function (LEGF) formalism }
\label{sec:legf}
This approach  involves a direct solution of  generalized  Langevin equations. 
Using this solution one can evaluate various quantities of interest such as steady state current
and temperature profiles. 
Compact formal expressions  for various quantities of interest can be
obtained and, as pointed
out earlier, it turns out that
these are identical  to results obtained  by the nonequilibrium
Green's function (NEGF) method described in sec.~(\ref{sec:negf}).
The method can be developed for quantum 
mechanical systems, in which case we deal with quantum Langevin equations (QLE), and
we will see that the classical results follow in 
the high temperature limit.
In all applications we will restrict ourselves to this
approach which we will henceforth refer to as the Langevin equations
and Green's function (LEGF) method. As we will see, for the ordered
case to be discussed in sec.~(\ref{sec:ordharmlat}), one can recover
the standard classical results   
 as well as extend them to the quantum domain using the LEGF method. For the
disordered case too [sec.~(\ref{sec:disharmlat})], one can make significant progress.  Another
important model that has been well studied in the context of harmonic
systems is the case where self-consistent heat reservoirs are attached at all sites of
the lattice.  In sec.~(\ref{sec:selfcons})  we will review results for
this case obtained also from the LEGF approach. All examples in this
section deal with the  case where particle displacements are taken to
be scalars. The generalization to vector displacements is straightforward.

The LEGF formalism  has been developed in the papers by
\cite{zurcher90,saito00,dharshastry03,segal03,dharroy06} and relies on
the approach first proposed  by Ford, Kac and Mazur \cite{ford65} of
modeling a heat bath by an infinite set of oscillators in thermal
equilibrium. Here we will outline 
the basic steps as given in \cite{dharroy06}. 
One again considers a harmonic system which is 
coupled to reservoirs which are of a more general form than the white
noise reservoirs studied in the last section. The reservoirs are now
themselves taken to be a collection of harmonic oscillators, whose
number will  be eventually taken to be infinite. As we will see,
this is equivalent to considering generalized Langevin
equations where the noise is still Gaussian but in general can be
correlated. We will  present the discussion for the quantum case and
obtain the classical result as a limiting case. 

We consider here the case of two reservoirs, labeled as $L$ (for left) and
$R$ (right) , which are at two different temperatures. It is easy to
generalize to the case where there are more than two reservoirs.
For the system let $X=\{x_1,x_2,...x_N\}^T$ now be the set of Heisenberg
operators corresponding to the displacements (assumed to be scalars) 
of the $N$ particles, about equilibrium lattice positions. Similarly
let $X_L$ and $X_R$ refer to position operators of the particles in
the left and right
baths respectively. The left reservoir has $N_L$ particles and the
right has $N_R$ particles. Also let $P, P_L, P_R$ be the corresponding
momentum variables satisfying usual commutation relations with the
position operators (\ie~$[x_l,p_n]=i \hbar \d_{l,n}$, etc.).     
The Hamiltonian of the entire system and reservoirs is taken to be:
\bea
H &=&H_S+H_L+H_R+H^I_{L}+H^I_{R}
\label{fullH} \\ 
{\rm where} \quad H_S &=&\f{1}{2}{P}^T \hM^{-1} {P}+ 
\f{1}{2} X^T \hP X ~, \nn \\
H_L &=&\f{1}{2}{P}_L^T \hM_L^{-1} {P_L}+ \f{1}{2}X^T_L \hP_LX_L ~, \nn \\
H_R &=&\f{1}{2}{P}_R^T \hM_R^{-1} {P}_R+ \f{1}{2}X_R^T \hP_R X_R ~, \nn \\
H^I_{L}&=&X^T \hat{V}_L X_L,~~~~H^I_{R}=X^T \hat{V}_R X_R ~, \nn
\eea
where $M,~M_L,~M_R$ are real diagonal matrices representing masses
of the particles in the  system, left, and right reservoirs
respectively. The quadratic potential energies are given by the real
symmetric matrices $\hP,~ \hP_L, ~\hP_R$ while $\hat{V}_L$ and
$\hat{V}_R$ denote the interaction between the system and the two
reservoirs respectively. It is assumed that at time $t=t_0$, the
system and reservoirs are decoupled and the reservoirs are in thermal
equilibrium at temperatures $T_L$ and $T_R$ respectively.

The Heisenberg equations of motion for the system (for $t > t_0$) are: 
\bea
\hM \ddot{X} = - \hP X -\hat{V}_L X_L -\hat{V}_R X_R ~ \label{3.2:eqmsys},
\eea
and the equations of motion for the two reservoirs are
\bea
\hM_L \ddot{X}_L &=& -\hP_L X_L - \hat{V}_L^T X ~, \label{3.2:res3} \\
\hM_R \ddot{X}_R &=& -\hP_R X_R - \hat{V}_R^T X ~.
\eea
One first solves the reservoir equations by considering them as linear inhomogeneous
equations. 
Thus for the left reservoir the general solution to Eq.~(\ref{3.2:res3})
is (for $t > t_0$):
\bea
X_L(t)&=&\hat{f}^+_L (t-t_0) \hM_L X_L(t_0)+ \hat{g}^+_L (t-t_0) \hM_L \dot{X}_L
(t_0)\nn \\
&&  - \int_{t_0}^t dt'~ \hat{g}^+_L (t-t') \hat{V}_L^T X(t')~,~~~~~~\label{solL} \\ 
{\rm with} \quad \hat{f}^+_L(t)&=& \hat{U}_L\cos{(\hat{\Omega}_L t)} \hat{U}_L^{T}~ \theta(t),~~~
\hat{g}^+_L(t)=\hat{U}_L \f{\sin{(\hat{\Omega}_L t)}}{\hat{\Omega}_L} \hat{U}^{T}_L~ \theta(t) ~, \nn
\eea
where $\theta(t)$ is the Heaviside function, and $\hat{U}_L,~\hat{\Omega}_L $
are the normal mode eigenvector and eigenvalue 
matrices respectively, corresponding to the left reservoir Hamiltonian
$H_L$, and which satisfy the equations:
\bea
\hat{U}_L^T \hP_L \hat{U}_L = \hat{\Omega}^2_L ~, ~~\hat{U}_L^T \hM_L \hat{U}_L =\hat{I}~. \nn
\eea
A similar solution is obtained for the right reservoir.  
Plugging these solutions back into the equation of motion for the
system, Eq.~(\ref{3.2:eqmsys}), one gets the following effective
equations of motion for the system: 
\bea
\hM \ddot{X} &=& - \hP X + \eta_L + \int_{t_0}^t dt' 
\hat{\Sigma}_L (t-t')  X(t')  + \eta_R + \int_{t_0}^t dt' \hat{\Sigma}_R(t-t') X(t'),~~~~~~~
\label{3.2:eqm}  \\
{\rm where}~~\hat{\Sigma}_L(t) &=&\hat{V}_L~ \hat{g}^+_L (t)  ~\hat{V}^T_L,~~ \hat{\Sigma}_R(t)=
\hat{V}_R~ \hat{g}_R^+(t)~  \hat{V}_R^T \nn \\
{\rm and}~~ \eta_L &=&-\hat{V}_L ~\left[~\hat{f}^+_L (t-t_0) \hM_L X_L(t_0) ~+~ \hat{g}_L^+ (t-t_0)
\hM_L \dot{X}_L (t_0)~\right] \nn \\
\eta_R&=&-\hat{V}_R~\left[~\hat{f}_R^+(t-t_0) \hM_R X_R(t_0) ~+~
  \hat{g}_R^+(t-t_0) \hM_R \dot{X}_R(t_0)~\right] ~. \nn
\eea
This equation has the form of a generalized quantum Langevin equation.
The properties of the noise terms $\eta_L $ and $\eta_R$ are 
determined using the condition that, at time $t_0$, the two isolated
reservoirs are described by equilibrium phonon distribution functions.
At time $t_0$, the left reservoir is in equilibrium at temperature
$T_{L}$ and the population of the normal modes (of the isolated left
reservoir) is given by the
distribution function $f_b(\omega,T_L)=1/[e^{\hbar \omega/k_B
 T_L}-1]$. 
One then  gets the following correlations for 
the left reservoir noise:  
\bea
\la \eta_L(t) \eta_L^T(t') \ra &=& \hat{V}_L \hat{U}_L~ \left[\cos{\hat{\Omega}_L (t-t')} 
\f{\hbar}{2 \hat{\Omega}_L} \coth{(\f{\hbar \hat{\Omega}_L}{2 k_B T_L})} \right.\nn \\
 && \left. -i
\sin{\hat{\Omega}_L (t-t')} \f{\hbar}{2 \hat{\Omega}_L}\right]~ \hat{U}_L^{T} \hat{V}_L^T~,~~~
\label{3.2:nono}
\eea
and a similar expression for the right reservoir.
The limits of infinite reservoir sizes ($N_L, N_R \to \infty$) and
$t_0 \to -\infty$ are now taken.  One can then solve Eq.~(\ref{3.2:eqm}) by taking  Fourier
transforms. Let us define the Fourier transforms 
$\tX(\omega) = ({1}/{2 \pi}) \int_{-\infty}^{\infty} dt ~X(t) e^{i \omega t}, ~
\tn_{L,R}(\omega)= ({1}/{2 \pi}) \int_{-\infty}^{\infty} dt~ \n_{L,R}(t) 
e^{i \omega t} ,~ 
\hat{g}_{L,R}^+(\omega)=\int_{-\infty}^{\infty} dt~ \hat{g}_{L,R}^+(t) e^{i \omega t} 
$~. One then gets the following stationary solution to the equations
of motion  Eq.~(\ref{3.2:eqm}): 
\bea
X(t)&=&\int_{-\infty}^\infty d \omega \tX (\omega) e^{-i \omega t}
~, \label{3.2:solSS}  \\
{\rm with} \quad \tX (\omega) &=& \hat{G}^+(\omega)~  {[\tn_L(\omega) + \tn_R (\omega)]}
 ~, \nn \\
{\rm where} \quad \hat{G}^+(\om) &=&\f{1}{[-\omega^2 \hM+  
 \hP -\hat{\S}_L^+ (\omega)-\hat{\S}_R^+(\omega)]}~, \nn \\  
{\rm and} \quad \hat{\S}_L ^+ (\omega)&=& \hat{V}_L \hat{g}_L^+
 (\omega) \hat{V}_L^T ,~~~\hat{\S}_R^+(\omega)= \hat{V}_R \hat{g}_R^+ (\omega)  \hat{V}_R^T~. \nn
\eea
For the reservoirs one obtains [using Eq.~(\ref{solL})]
\bea
-\hat{V}_L \tX_L(\omega) &=& \tn_L (\omega) +\hat{\S}_L^+ \tX(\omega) ~, \nn \\
-\hat{V}_R \tX_R(\omega) &=& \tn_R(\omega) +\hat{\S}_R^+ \tX(\omega) ~.
\label{3.2:solLR}
\eea
The noise correlations, in the frequency domain, can be obtained from
Eq.~(\ref{3.2:nono}) and we 
get (for the left reservoir):
\bea
\la \tn_L (\omega) \tn_L^T (\omega') \ra 
&=& \delta (\omega +\omega')~ \hat{\G}_L(\omega)~ \f{\hbar}{ \pi}
    [1+f_b(\omega,T_L)]~~~ \label{3.2:nncor} \\
{\rm where}~~~ \hat{\G}_L(\omega)&=&Im[\hat{\S}_L^+(\omega)]  \nn 
\eea
which is a fluctuation-dissipation relation. This also leads to the
more commonly used correlation:
\bea
\f{1}{2}\la~ \tn_L (\omega) \tn_L^T (\omega') +\tn_L (\omega') \tn_L^T
(\omega)~ \ra  
= \delta (\omega +\omega') ~\hat{\G}_L(\omega)~ \f{\hbar}{2
  \pi}~{\coth(\f{\hbar \omega}{2k_B
 T_L})}.~ \label{3.2:nncor2}
\eea
Similar relations hold for the noise from the right reservoir. 
The identification of $\hat{G}^+(\omega)$ as a phonon Green function, with
 $\hat{\S}^+_{L,R}(\omega)$ as self energy contributions
coming from the baths, is the main step that
enables a comparison of results derived by the LEGF
 approach with those obtained from the NEGF method. 
This has been demonstrated in refn.~\cite{dharroy06}.

{\bf{Steady state properties}}: The simplest way to evaluate the steady state current is to 
evaluate the following expectation value for current from left
reservoir into the system:
\bea
J &=&-\la~ \dot{X}^T \hat{V}_L X_L~ \ra. 
\eea
This is just the rate at which the left reservoir does work on the
wire. Using the solution in
Eq.~(\ref{3.2:solSS},\ref{3.2:solLR},\ref{3.2:nncor}) one obtains,
after some manipulations:
\bea
J&=& \f{1}{4\pi} \int_{-\infty}^\infty d \omega~\mT (\om) 
 {\hbar \omega}~ [f(\omega,T_L)-f(\omega,T_R)]
 ~.\label{3.2:SSJ} \\
{\rm where}~~\mT (\om) &=& 4~ Tr[~ {\hat{G}^+}(\omega)~  \hat{\G}_L (\omega) 
\hat{G}^-(\omega) ~ \hat{\G}_R (\omega)]~, \nn
\eea
and $\hat{G}^{-}(\om) = {\hat{G}^{+\dagger}}(\om) $. 
This expression for current is of identical form as
the NEGF expression for electron current (see for example
\cite{caroli71,meir92,datta95,dharsen06}) and has also been derived for phonons using the NEGF
approach in refns.\cite{yamamoto06,wang06}. In fact this expression was
first  proposed  by Angelescu \etal \cite{angelescu98} and by Rego
and Kirczenow \cite{rego98} for a $1D$ 
channel and they obtained this using the Landauer approach. Their result was
obtained more systematically later by Blencowe \cite{blencowe99}. 
Note that Eq.~(\ref{3.2:SSJ}) above can also be written as an integral over
only positive frequencies using the fact that the integrand is an even
function of $\omega$. 
The factor $\mT(\om)$ is the transmission
coefficient of phonons at frequency $\om$ through the system, from the
left to right reservoir. The usual Landauer result for a $1D$ channel
precisely corresponds to Rubin's model of bath, to be discussed in
sec.~(\ref{sec:disharmlat1d}).  

For small temperature differences $\Delta T=
T_L-T_R << T$, where $T=(T_L+T_R)/2$, \ie, in the linear response
regime the above expression reduces to:
\bea
J&=& \f{\Delta T}{4 \pi} \int_{-\infty}^\infty d \omega~\mT (\om) 
 {\hbar \omega}~ \f{\p f(\om,T)}{\p T}
 ~.\label{3.2:SSJLR} \\
\eea

The classical limit is obtained by taking
the high temperature limit $\hbar \omega/k_B T \to 0$. 
This gives:
\bea
J=  \f{k_B ~\Delta T}{4 \pi} ~\int_{-\infty}^\infty d \omega~ \mT (\omega)~. \label{3.2:SSJcl}
\eea
One can similarly evaluate various other quantities such as 
velocity-velocity correlations and position-velocity correlations. The
expressions for these in the classical case are respectively:
\bea
K&=& \la \dot{X} \dot{X}^T \ra \nn \\
&=& \f{k_B T_L}{\pi} \int_{-\infty}^\infty d \omega ~\omega ~
G^+(\omega)  \G_L (\omega)
G^-(\omega)  +  \f{k_B T_R}{\pi} \int_{-\infty}^\infty d \omega ~\omega~
G^+(\omega)  \G_R (\omega)  G^-(\omega)~,\nn \\   
C &=&  \la {X} \dot{X}^T \ra \nn \\
&=& \f{i k_B T_L}{\pi} \int_{-\infty}^\infty d \omega ~ G^+(\omega)  \G_L (\omega)
G^-(\omega)  +  \f{i k_B T_R}{\pi} \int_{-\infty}^\infty d \omega ~
G^+(\omega)  \G_R (\omega)  G^-(\omega)~.  \nn
\eea 
The correlation functions $K$ can be used to define the local
energy density which can in turn be used to define the 
temperature profile in the non-equilibrium steady state of the
wire. Also we note that the correlations $C$ give the local heat
current density. Sometimes it is more 
convenient to evaluate the total steady state current from this
expression  rather than the one in 
Eq.~(\ref{3.2:SSJ}).

For one-dimensional wires the above results can be shown \cite{dhar01} to lead to 
expressions for current and temperature profiles obtained  in earlier studies of heat conduction in disordered
harmonic chains \cite{rubin71,oconnor74,dhar01}.

In our derivation of the LEGF results we have {\it implicitly assumed that
a unique steady state will be reached}. One of the necessary conditions for
this is that no modes outside the bath spectrum are generated for the
combined model of system and baths. These modes, when they exist, are
localized near the system and any initial excitation of the mode is
unable to decay.  This has been demonstrated and discussed in detail in the 
electronic context  \cite{dharsen06}.

We note that unlike other approaches such as the Green-Kubo formalism
and Boltzmann equation approach, the Langevin equation approach explicitly
includes the reservoirs. The Langevin equation is  physically
appealing since it gives a nice picture of the reservoirs as sources
of noise and dissipation. Also just as the Landauer formalism and NEGF
have been extremely useful in understanding electron transport in
mesoscopic systems it is likely that a similar description will be
useful for the case of heat transport in (electrically) insulating nanotubes,
nanowires, etc. 
The LEGF approach has some advantages over NEGF. 
For example, in the classical case,  
it is easy to write Langevin equations for nonlinear systems and study
them numerically. Unfortunately, in the quantum case, one does not yet
know how to achieve this, and understanding steady state transport in
interacting quantum systems is an important open problem.

\subsection{Ordered harmonic lattices}
\label{sec:ordharmlat}

As mentioned above, heat conduction in the ordered harmonic chain was
first studied in the Rieder, Lebowitz and Lieb (RLL) paper and its higher
dimensional generalization was obtained by Nakazawa.
The approach followed in both the RLL and Nakazawa papers was to obtain the
exact nonequilibrium stationary  state measure which, for this
quadratic problem, is a Gaussian distribution. A complete solution for
the correlation matrix was obtained and from this one could obtain both
the steady state temperature profile and the heat current.  

Here we follow refn.~\cite{roydhar08a} to show how the LEGF  method, discussed in the previous section,
can be used to calculate the heat 
current in ordered harmonic lattices connected to Ohmic reservoirs
(for a classical system this is white noise Langevin dynamics). 
We will see how exact expressions for the asymptotic current ($N\to
\infty$) can be 
obtained  from this approach. We also briefly discuss the model in the quantum
regime and extensions to higher dimensions.

\subsubsection{One dimensional case}
\label{sec:ordharmlat1d}
The model considered in \cite{roydhar08a}  is  a slightly generalized version
of those studied by RLL and Nakazawa. An external potential is present at
all sites and the pinning strength  at the
boundary sites are taken to be different from those at the bulk sites. 
Thus both the RLL and Nakazawa results can be obtained as limiting cases. Also it seems
that this model more closely mimics the experimental situation. In experiments
the boundary sites would be interacting  
with fixed reservoirs, and the coupling to those can be modeled by an effective spring
constant that is expected to be different from the interparticle
spring constant in the bulk. We also note here that the constant
onsite potential 
present along the wire relates to experimental situations such as that
of heat transport in a nanowire attached to a substrate  or, in the
two-dimensional case, a monolayer film on a substrate. Another example
would be the heat current contribution from the optical modes of a
polar crystal.

Consider $N$ particles
of equal masses $m$  connected to each other by harmonic springs of
equal spring constants $k$.  The
particles are also pinned by onsite  quadratic potentials with 
strengths $k_o$ at all sites except the boundary sites where the
pinning strengths are $k_o+k'$.
The Hamiltonian is thus:
\begin{eqnarray}
H=\sum_{l=1}^N [\f{p_l^2}{2m}  +\f{1}{2}k_o{x}_l^2]+\sum_{l=1}^{N-1}\f{1}{2}k(x_{l+1}-x_l)^2+ \f{1}{2}k'(x_1^2+x_N^2) ~,\label{ham}
\end{eqnarray}
where $x_l$ denotes the displacement of the $l^{\rm th}$ particle from
its equilibrium position. The particles $1$ and $N$ at the two ends are
connected to heat baths at temperature $T_L$ and $T_R$
respectively, assumed to be modeled by Langevin
equations corresponding to Ohmic baths ($\Sigma (\om)=i \gamma \om$). In the classical case the 
steady state heat current from left to right reservoir can be obtained
from Eq.~(\ref{3.2:SSJcl}) and given by
\cite{dhar01,casher71}:
\bea
J&=&\f{k_B(T_L-T_R)}{4 \pi}\int_{-\infty}^\infty d \om
\mT_N(\om),\label{3.3:ocur1}  \\ 
{\rm where}~~
\mT_N(\om)&=&4 ~\G^2(\omega)|\hat{G}_{1N}(\om)|^2,~\hat{G}(\om) =
\hat{Z}^{-1}/k \nn \\ 
{\rm and}~~ \hat{Z}&=&[-m \omega^2 \hat{I} +\hat{\Phi} - \hat{\Sigma}(\omega)]/k~,\nn
\eea
where  $\hat{I}$ is a unit matrix, $\hP$ is the force matrix corresponding
to the Hamiltonian in Eq.~(\ref{ham}). The $N\times N$ matrix
$\hat{\Sigma}$ has mostly zero elements except for  $\Sigma_{11}=\Sigma_{NN}=i
\Gamma (\om)$ where $\Gamma (\om) = \gamma \om$.
The matrix $\hat{Z}$ is  tri-diagonal matrix with 
$Z_{11}=Z_{NN}=(k+k_o+k'-m\om^2-i\g \om)/k$, all other diagonal
elements equal to~ $2+k_o/k-m \om^2/k$ and all off-diagonal elements
equal to $-1$. 
Then it can be shown easily that $|G_{1N}(\om)|=1/(k~|\Delta_N|)$ where
$\Delta_N$ is the determinant of the matrix $\hat{Z}$. This can be
obtained exactly.
For large $N$, only phonons within the spectral band of the system can
transmit, and the integral over $\om$ in Eq.~(\ref{3.3:ocur1}) can be converted to one over
$q$ to give: 
\bea
J&=& \f{2\gamma^2 k_B (T_L-T_R)}{k^2\pi} \int_{0}^\pi dq |\f{d \omega}{dq}| ~ \f{\omega^2_q}{|\Delta_N|^2}~, \label{3.3:JSCR2} 
\eea
with $m \omega_q^2=k_o+2 k [1-\cos{(q)}]$.
Now using the result:
\bea 
\lim_{N \rightarrow \infty}\int_{0}^{\pi} dq \f{g_1(q)}{1+g_2(q)\sin
  Nq}=\int_{0}^{\pi} dq \f{g_1(q)}{[1-g^2_2(q)]^{1/2}}~,\label{iden} 
\eea
where $g_1(q)$ and $g_2(q)$ are any two well-behaved functions, one
can show that in the limit $N \to \infty$, Eq.~(\ref{3.3:JSCR2}) gives   
\bea
J
&=&\f{\g k^2 k_B(T_L-T_R)}{m \Omega^2}(\Lambda-\sqrt{{\Lambda}^2-{\Omega}^2})~,\label{Class} \\
{\rm where}~~\Lambda&=& 2k(k-k')+{k'}^2+\f{(k_o+2k)\g^2}{m}~~{\rm
  and}~~\Omega= 2k(k-k') + \f{2k\g^2}{m}~.\nn
\eea
Two different special cases lead to the RLL and Nakazawa results.
First in the case of
fixed ends and without onsite potentials, {\emph{i.e.}} $k'=k$ and $k_o=0$,
we recover the RLL result \cite{rieder67}:
\bea
J^{RLL}=\f{k k_B(T_L-T_R)}{2\g}\Big[1+\f{\nu}{2}-\f{\nu}{2}\sqrt{1+\f{4}{\nu}}~\Big]~~{\rm where}~~\nu=\f{mk}{\g^2}~.\label{3.3:RLL}
\eea
In the other case of free ends, {\emph{i.e.}} $k'=0$, one gets the
Nakazawa result \cite{nakazawa68}: 
\bea
J^{N}=\f{k\g k_B(T_L-T_R)}{2(mk+\g^2)}\Big[1+\f{\lambda}{2}-\f{\lambda}{2}\sqrt{1+\f{4}{\lambda}}~\Big]~~
{\rm where}~~\lambda=\f{k_o \g^2}{k(mk+\g^2)}~.\label{3.3:NakO}
\eea

In the quantum case, in the linear response regime,
Eq.~(\ref{3.2:SSJLR}) and similar manipulations  made above for the $N \to
\infty$ limit leads to the following final expression for current: 
\bea
J&=&\f{\g k^2 \hbar^2 (T_L-T_R)}{4 \pi k_B m T^2}\int_{0}^\pi dq
\f{\sin^2 q}{\Lambda -\Omega \cos q}~ \om^2_{q}~{\rm{cosech}}^2
\Big(\f{\hbar \om_{q}}{2k_B T}\Big), \label{3.3:ocur2} \\ 
{\rm where}~~\om^2_{q}&=&[k_o+2k(1-\cos q)]/m~.\nn
\eea  
While one cannot perform  this integral exactly, numerically it is easy
to obtain the integral for given parameter values. It is interesting
to examine the temperature dependence of the conductance $J/\Delta T$.
In the classical case this  is independent of temperature while one finds
that at low temperatures the quantum result is completely different. 
For three different cases one finds, in the  low temperature ($T <<
\hbar (k/m)^{1/2}/k_B$) regime, the following behaviour: 
\bea
J \sim  \begin{cases} T^3 ~~~ {\rm for}~~~k'=k, k_o=0
\\ \sim T  ~~~{\rm for}~~~k'=0, k_o=0
\\\sim \f{e^{-\hbar   \omega_o/(k_B T)}}{T^{1/2}} ~~~{\rm for}~~~ k'=0, k_o \neq 0~,
\end{cases}
\eea
where $\omega_o=(k_o/m)^{1/2}$.   
In studies trying to understand  experimental work
on nanosystems [see sec.~(\ref{sec:expts})] ,
the temperature dependence of the conductance is usually derived from
the Landauer formula, which corresponds to the Rubin model of
bath. The temperature dependence will then be different from the above results.

\subsubsection{Higher dimensions}
\label{sec:ordharmlathd}
As shown by Nakazawa \cite{nakazawa68} 
the problem of heat conduction in ordered
harmonic lattices in more than one dimension   can be
reduced to an effectively one-dimensional problem. We will briefly give 
the arguments here and also give the quantum generalization.

Let us consider a $d$-dimensional hypercubic lattice with lattice
sites labeled by the vector $\bl=\{n_\alpha\}, \alpha=1,2...d$, where
each $n_\alpha$ takes values from $1$ to $L_\alpha$. The total number of
lattice sites is thus $N=L_1L_2...L_d$. We assume that heat conduction
takes place in the $\alpha=d$ direction. Periodic boundary conditions
are imposed in the remaining $d-1$ transverse directions. The
Hamiltonian is described by a scalar displacement $X_{\bl}$ and, as in
the $1D$ case, we
consider  nearest neighbour harmonic interactions with a spring
constant $k$ and harmonic onsite pinning at all sites with spring
constant $k_o$. All boundary particles at $n_d=1$ and $n_d=L_d$ are
additionally pinned
by harmonic springs with stiffness $k'$ and follow Langevin dynamics
corresponding to baths at temperatures $T_L$ and $T_R$ respectively. 

Let us write $\bl=(\bl_t,n_d)$ where $\bl_t=(n_1,n_2...n_{d-1})$. Also
let ${\bf q}=(q_1,q_2...q_{d-1})$  with $q_\alpha= 2 \pi s/L_\alpha$
where $s$ goes from $1$ to $L_\alpha$. Then defining variables 
\bea
X_{n_d}({\bf q})=\f{1}{L_1^{1/2}L_2^{1/2}...L_{d-1}^{1/2}}\sum_{\bl_t}
X_{\bl_t,n_d} e^{i {\bf q}.\bl_t}~, 
\eea    
one finds that, for each fixed ${\bf q}$, $X_{n_d}({\bf q})$
($n_d=1,2...L_d$) satisfy  Langevin equations corresponding
to the $1$D Hamiltonian in Eq.~(\ref{ham}) with the  onsite spring
constant $k_o$ replaced by
\bea
\lambda({\bf q})=k_o+2~k~[~d-1-\sum_{\alpha=1,d-1} \cos{(q_\alpha)}~]~.
\eea   
For $L_d\to \infty$, the heat current $J({\bf q})$ for each mode with
given ${\bf q}$ is 
then simply given by  Eq.~(\ref{Class}) with $k_o$ replaced by
$\lambda_{\bf q}$.   In the quantum mechanical case we use
Eq.~(\ref{3.3:ocur2}).  The heat current per bond is then given by:
\bea
J=\f{1}{L_1L_2....L_{d-1}} \sum_{{\bf q}} J({\bf q})~.
\eea 
Note that the result holds for finite lengths in the transverse
direction. For infinite transverse lengths we get $J=\int...\int_0^{2 \pi} d
{\bf q} J({\bf q}) /(2 \pi)^{d-1}$~.

\subsection{Disordered harmonic lattices}
\label{sec:disharmlat}
From the previous section we see that the heat current in an 
ordered harmonic lattice is independent of system size (for large
systems) and hence transport is ballistic. Of course this is expected
since there is no mechanism for scattering of the heat carriers,
namely the phonons. Two ways of introducing scattering of phonons are by
introducing disorder in the system, or by including anharmonicity
which would cause phonon-phonon interactions. In this section we consider the
effect of disorder on heat conduction in a harmonic system.

Disorder can be introduced in various ways, for example by making the
masses of the particles random as would be the case in a isotopically
disordered  solid, or by making the spring constants random. Here we
will discuss the  case of mass-disorder only since the most important features do
not seem to vary much with the type of disorder one is
considering. Specifically, we will consider  harmonic systems where
the mass of each particle is an independent random variable
chosen from some fixed distribution. 

It can be expected that heat conduction in disordered  harmonic
systems will be strongly affected by the physics of Anderson
localization. In fact the problem of finding the normal modes
of the harmonic lattice can be directly mapped to that of
finding  the eigenstates of an electron in a disordered potential
(in a tight-binding model, for example)   
and so we expect the same kind of physics as in electron localization.         
In the electron case the effect of localization is strongest in one
dimensions where it can 
be proved rigorously that all eigenstates are exponentially
localized, hence the current decays exponentially with system size and the
system is an insulator. This is believed to be true in two dimensions also.
In the phonon case the picture is much the same except that, in the
absence of an external potential, the translational invariance of the
problem leads to the fact that {\it low frequency modes are not localized and
are effective in transporting  energy}. Another important difference
between the electron and phonon problems is that {\it electron transport    
is dominated by electrons near the Fermi level while in the case
of phonons, all frequencies participate in transport}. These two
differences lead to the fact that the disordered harmonic crystal in
one and two dimensions is not a heat insulator, unlike its electronic
counterpart. Here we will present
results  using the LEGF approach  to determine the
system size dependence of the current in one dimensional
mass-disordered chains. We note that basically this same approach (NEGF)
is also popular in the electron case and is widely used in mesoscopic
physics. Also earlier treatments by, for example, 
Rubin and Greer \cite{rubin71} and by Casher and Lebowitz
\cite{casher71} of the disordered harmonic chain, can be viewed 
as special examples of the LEGF approach. We will also discuss results of
simulations for the two-dimensional case.

Our main conclusions here will be that Fourier's law is not valid in a
disordered harmonic crystal in one and two dimensions, the current
decays as a power law with system size and the exponent $\alpha$ is
sensitive to boundary conditions (BC) and spectral properties of the
heat baths.

\subsubsection{One dimensional disordered lattice}
\label{sec:disharmlat1d}

We first briefly review earlier work on this problem \cite{LLP03}. 
The thermal conductivity of disordered harmonic lattices was first
investigated by Allen and Ford \cite{allen68} who, using the Kubo formalism,
obtained an exact expression for the thermal conductivity of a finite
chain attached to infinite reservoirs. From this expression they
concluded, erroneously as we now know, that the thermal conductivity
remains finite in the limit of infinite system size. Simulations of
the disordered lattice connected to white noise reservoirs were
carried out by Payton \etal~\cite{payton67}. They were restricted to  small system   sizes
($N \sim 400$) and also obtained a finite thermal conductivity. 

Possibly the first paper to notice anomalous transport was that by Matsuda
and Ishii (MI) \cite{matsuda70}. 
In an  important work  on the localization of normal
modes in the disordered harmonic chain, 
MI showed that all high frequency modes  were exponentially
localized. However, for small $\omega$ the localization length in an
infinite sample was shown to vary as 
$\omega^{-2}$, hence normal modes with frequency
$\om \stackrel{<}{\sim} \om_d$ have localization length greater than
$N$, and cannot be considered as localized. 
For a harmonic chain of length $N$, given the average mass $m=\langle
m_l \rangle$, the variance 
$\sigma^2 =\langle (m_l- m )^2\rangle$ and interparticle spring constant $k$,
it was shown that 
\bea
\om_d \sim \left( \f{k m}{N \sigma^2} \right)^{1/2}
\label{dem}
\eea
They also evaluated expressions for thermal
conductivity of a finite disordered chain connected to two different
bath models, namely:

$\bullet$ model(a): white noise baths and

$\bullet$ model(b): baths modeled by semi-infinite ordered harmonic
chains (Rubin's model of bath). 

In the following we will also consider these two models of baths.
For model(a) MI  used fixed BC (boundary particles in
external potential) and the limit
of weak coupling to baths, while for case (b) they considered free
BC (boundary particles not pinned)  and  
this was treated using the Green-Kubo formalism given by Allen and Ford \cite{allen68}. They
found $\alpha=1/2$ in both 
cases, a conclusion which we will see is incorrect. 
The other two important theoretical papers on heat conduction in the disordered
chain were those by Rubin and Greer  \cite{rubin71} who considered
model(b) and of Casher and Lebowitz \cite{casher71}  who
used model(a) for baths. A lower bound  $[J] \geq 1/N^{1/2}$ was
obtained  for the disorder averaged current $[J]$ in
refn.~\cite{rubin71} who also gave numerical
evidence for an exponent $\alpha = 1/2$.  
This was later proved rigorously by Verheggen \cite{verheggen79}. On the other hand,
for model(a), \cite{casher71} found a rigorous bound $[J] \geq 1/N^{3/2}$ and
simulations by Rich and 
Visscher \cite{rich75} with the same baths supported the corresponding exponent
$\alpha = -1/2$. The  
work in \cite{dhar01} gave a unified
treatment  of the 
problem of heat conduction in disordered harmonic chains connected to baths
modeled by generalized 
Langevin equations and showed that models(a,b)
were two special cases. An efficient numerical scheme was proposed and 
used to obtain the exponent $\alpha$ and it was established that 
$\alpha=-1/2$ for model(a) (with fixed BC) and $\alpha=1/2$ for
model(b) (with free BC).  
It was also pointed out that in general, {\it $\alpha$ depended on the
spectral properties of the baths}.
We will briefly describe  this formulation \cite{roydhar08b}  here 
and  see how one can  understand the effect of boundary conditions  
on heat transport in the disordered chain. 
We will consider both the white noise [model(a)] and Rubin 
baths [model(b)]. One of the main conclusions will be that the
difference in exponents obtained for these two cases arises from use
of different boundary conditions, rather than because of differences in spectral
properties of the baths.

\begin{figure}[t]
\begin{center}
\includegraphics[width=4.5in]{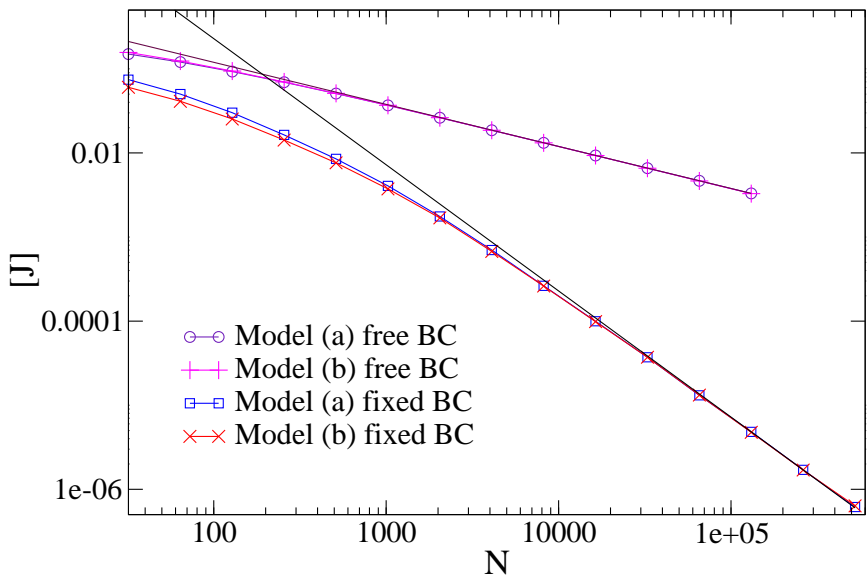}
\end{center}
\caption{Plot of $[J]$ versus $N$
for free  and fixed boundary conditions.
Results are given for both models(a,b) of baths.
The two straight lines correspond to the asymptotic expressions given
in Eqs.~(\ref{3.4:jfr},\ref{3.4:jfi}) and have slopes $-1/2$ and
$-3/2$. Parameters used were $m=1,~\Delta=0.5,~k=1,~ 
\g=1$, $T_L=2$, $T_R=1$ and $k_o=1$ (from \cite{roydhar08b}).}
\label{3.4:fig1}
\end{figure}

The Hamiltonian of the mass-disordered chain is given by
\begin{eqnarray}
H=\sum_{l=1}^N \f{p_l^2}{2m_l}
+\sum_{l=1}^{N-1}\f{1}{2}k(x_{l+1}-x_l)^2
+ \f{1}{2}k'(x_1^2+x_N^2) ~.\label{3.4.1:hamdis}
\end{eqnarray}
The random masses $\{m_l\}$ are chosen from, say,
a uniform distribution between $(m-\Delta)$ to $(m+\Delta)$. The strength
of onsite potentials at the boundaries is 
$k'$. The particles at two ends are connected to  heat baths, at
temperature $T_L$ and $T_R$, and 
modelled by generalized Langevin equations. 
The steady state classical heat current
through the chain is given by:   
\bea
J&=&\f{k_B(T_L-T_R)}{4 \pi}\int_{-\infty}^\infty d \om
\mT_N(\om),\label{ocur1}  \\ 
{\rm where}~~
\mT_N(\om)&=&4 \G^2(\omega)|\hat{G}_{1N}(\om)|^2,~\hat{G}(\om) =
\hat{Z}^{-1}/k \nn \\ 
{\rm and}~~ \hat{Z}&=&[-\omega^2 \hat{M} +\hat{\Phi} -
  \hat{\Sigma}(\omega)]/k~,\nn 
\eea
where $\hat{M}$ and $\hat{\Phi}$ are respectively the mass and force
matrices corresponding to Eq.~(\ref{3.4.1:hamdis}).
As shown in \cite{dhar01}, the non-zero elements of the diagonal
matrix $\hat{\Sigma} (\om)$ for models(a,b) are
$\Sigma_{11}(\omega)=\Sigma_{NN} (\omega) =\Sigma (\omega)$ and 
given by
\bea
\Sigma(\omega)&=&-i\g\om   ~~~~{\rm model (a)} \nn \\ 
\Sigma(\omega) &=& k\{1-m \om^2/2k-i\om (m/k)^{1/2}
 {[1-m \om^2/(4k)]}^{1/2}\}~~~~{\rm model (b)}~, \label{selfener}
\eea
where $\g$ is the 
coupling strength with the white noise  baths, while
in case of Rubin's baths it has been assumed that the Rubin bath 
has spring constant $k$ and equal masses $m$. 
As  noted above,  $\mT_N(\om)$ is 
the transmission coefficient of phonons through the disordered
chain. To extract the asymptotic $N$ dependence of 
the disorder averaged current $[J]$ one needs to determine the  Green's function element
$G_{1N}(\om)$. It is convenient to write the  matrix elements
$Z_{11}=-m_1 \om^2/k +1+k'/k-\Sigma/k=-m_1 \om^2/k +2- \Sigma'$ where
$\Sigma'=\Sigma/k-k'/k+1$ and similarly $Z_{NN}=-m_N \om^2/k +2 
-\Sigma'$. Following the  techniques used in \cite{casher71,dhar01}
one gets:
\bea
|G_{1N}(\om)|^2&=& k^{-2}|\Delta_N(\om)|^{-2}~~{\rm with} \label{green}\\
\Delta_N(\om)&=&D_{1,N}-\Sigma' (D_{2,N}+D_{1,N-1})+{\Sigma'}^2 D_{2,N-1}\nn
\eea
where $\Delta_N(\om)$ is the determinant of $\hat{Z}$ and
the matrix elements $D_{l,m}$ are given by the following product of 
$(2\times2)$ random matrices $\hat{T}_l$:
\bea
&& \hat{D}=\left( \begin{array}{cc}
 D_{1,N} & -D_{1,N-1} \\ D_{2,N} & -D_{2,N-1} \end{array} \right)=\hat{T}_1
\hat{T}_2....\hat{T}_N \label{trans} \\ 
{\rm{where}}~~ &&\hat{T}_l=\left( \begin{array}{cc}
2-m_l \om^2/k & -1 \\ 1 & 0 \end{array}
\right) \nn
\eea
We note that the information about bath properties and boundary
conditions are now contained entirely in $\Sigma'(\om)$ while
$\hat{D}$ contains the system properties. It is known
that $|D_{l,m}|\sim e^{c N \om^2}$ \cite{matsuda70}, where $c$ is a
constant, and so we 
need to look only at the low frequency ($\om
\stackrel{<}{\sim}1/N^{1/2}$) form of $\Sigma'$ . Let us now discuss
the various cases.  
For model(a) free BC correspond to  $k'=0$ and so 
$\Sigma'=1-i\g \om/k$ while for model(b) free boundaries corresponds
to $k'=k$ and this gives, at low frequencies, $\Sigma'=1-i(m/k)^{1/2}
\om$. Other choices of $k'$ correspond to pinned boundary sites with an
onsite potential $k_o x^2/2$ where $k_o=k'$ for model(a) and
$k_o=k'-k$ for model(b). The
main difference, from the unpinned case, is  that now $Re[\Sigma'] \neq
1$. The arguments of \cite{dhar01} then immediately give $\a = 1/2$
for free BC and $\a=-1/2$ for fixed BC for both bath
models. In fact for the choice of parameters $\gamma= (mk)^{1/2}$, the
imaginary part of $\Sigma'$ is the same for both baths and hence we
expect, for large system sizes, the actual values of the current to be
the same for both bath models.  This can be seen in Fig.~(\ref{3.4:fig1})
where  the system size dependence of the current for the
various cases is shown. The current was evaluated numerically using
Eq.~(\ref{ocur1}) and averaging over many realizations.
Note that for free BC, the exponent $\alpha=1/2$ settles to its
asymptotic value at  relatively small values ($N\sim10^3$)
while, with pinning,  one needs to examine much longer chains ($N \sim
10^5$).

These results clearly show that, for both 
models(a,b) of baths, the  exponent $\alpha$ is the same and is controlled
by the presence or absence of pinning at the boundaries. The reason
that both models give the same exponents is that the imaginary part of
their self energies, given by Eq.~(\ref{selfener}), have the same
small $\om$ dependence.  If the imaginary part of the self energy,
\ie~ $\Gamma( \om)$, has a different $\om$ dependence, then one can get
different exponents for the same boundary conditions \cite{dhar01}.

For the present case, some more specific predictions can also be obtained. As mentioned
before  only modes $\om \stackrel{<}{\sim} \om_d$ are
involved in conduction.  
An observation made in \cite{dhar01a} was that {\it in this low frequency regime one can
approximate $[ \mT_N(\om) ]$ by the
transmission coefficient of the ordered chain $\mT_N^O(\om)$}. One can
then write:
\bea
[J] \sim (T_L-T_R) \int_{0}^{\om_d} \mT_N^{O}(\om)
d\om~.
\label{3.4:asyCur}
\eea 
By looking at the $N \to \infty$ limit results for the ordered
lattice, the following results are obtained for the two different
boundary conditions \cite{roydhar08b}:
\bea
{[J]}_{Fr}&=& A ~c~\f{k_B (T_L-T_R)}{\pi }\Big(\f{k m }{N \sigma^2}\Big)^{1/2}  \label{3.4:jfr}\\
{[J]}_{Fi} &=& A'~ c'~  \f{k_B (T_L-T_R) }{\pi }\Big(\f{k m
}{N \sigma^2 }\Big)^{3/2}~ \label{3.4:jfi},  
\eea
where $c=2\gamma(mk)^{1/2}/({\gamma}^2+mk),~1$ for model(a), model(b)
respectively. For fixed boundaries we have
$c'=\g (mk)^{1/2}/{k_o}^2,~ mk/{k_o}^2$ for model(a), model(b) respectively.  
$A, A'$ are constant numbers, taken to be fitting parameters. 
For model(b) this agrees with an exact
expression for 
$[J] _{Fr}$ due to Papanicolau (apart from a factor of $2 \pi$)
and with 
$A=\pi^{3/2} \int_{0}^{\infty}dt~ {[t \sinh(\pi t)]}/
{[(t^2+1/4)^{1/2} \cosh^2(\pi t)]} \approx 1.08417$ (see
[\cite{verheggen79}]).

In the case where all sites of the chain are pinned (\ie~ in the
presence of a substrate potential) it was been noted in
\cite{dhar01,likhachev06} that the current decays exponentially with
$N$ and this was proved in \cite{dharleb08}. Another interesting
result obtained in \cite{roydhar08b} is the case with a finite number
$n$ of pinned sites. It was shown, using heuristic arguments and
numerics, that $\alpha=3/2-n$ for $2 \leq n <<N$.   

A question that has been discussed in the literature is that 
of uniqueness of the steady state. This depends on the choice of heat
baths as well as the system studied. For models(a,b) of baths, the
uniqueness of the steady state of a chain has been discussed in
\cite{rubin71} and \cite{casher71}. For baths consisting of harmonic
oscillators one obvious necessary condition for uniqueness is that the
bath spectrum should include the modes of the system (see for example
\cite{dharwagh07}).  

The quantum mechanical disordered chain has been discussed in
\cite{dharshastry03} where it was argued that the asymptotic system
size dependence of the current should remain unchanged from the classical case
(unlike the low temperature behaviour for ordered case). The
temperature profile in a quantum mechanical disordered chain and  
quantum aspects such as entanglement have been numerically studied in \cite{gaul07}.

\subsubsection{Two dimensional disordered harmonic lattice}
\label{sec:disharmlat2d}
So far there has not been much progress in understanding heat
conduction in higher dimensional disordered lattices. The LEGF theory
is still applicable and provides a general expression for the steady
state current, Eq.~(\ref{3.2:SSJ}), in terms of the phonon Green's function of the harmonic
lattice. However in one dimension one could make progress by writing the
transmission coefficient  in terms of a product of random
matrices as in Eqs.~(\ref{green},\ref{trans}). This enables one to use some known mathematical theorems from
which some analytic results could be obtained. Also this
representation makes it possible to evaluate the current by  very
efficient numerical procedures.  
However in two and higher dimensions things become more complicated
and it has not been possible to make much analytic progress. 
This state of things is also reflected in the fact that while in one dimension it can be proved
exactly that all finite frequency states are localized, there is no
such proof in two dimensions, although this is the general belief. 
As far as phonon localization is concerned a renormalization  group
study by John \etal ~\cite{john83} found that things are very similar
to electronic localization. The important difference is again at low
frequencies where one gets extended states. Their study indicates the
following: in $1D$, all modes with $\om > 1/L^{1/2}$ are
localized; in $2D$, all modes with $\om > [\log (L)]^{-1/2}$
are localized; in $3D$, there is a finite band of
frequencies of non-localized states.       However this study was
unable to extract the system size dependence of heat current in a
disordered lattice in any dimension. 

There have been a few simulation studies on the $2D$
disordered harmonic lattice. Lei Yang \cite{yang02} considered a lattice with
bond-missing defects and looked at the system size dependence of
conductivity at various  defect densities. The first observation that
was made was that disorder gives rise to a temperature gradient across the
system, unlike the flat profile for the ordered case. Also it was found that at 
small densities, the conductivity diverged logarithmically, while at
larger densities, a finite heat conductivity was obtained. 
However this conclusion is probably incorrect since the paper uses
Nose-Hoover thermostats and it is known that, for harmonic systems, these have equilibration
problems \cite{posch86,frenkel02,baowenli01,dharcom01,lizhaohu01,hu03}. 

\begin{figure}
\includegraphics[width=4.5in]{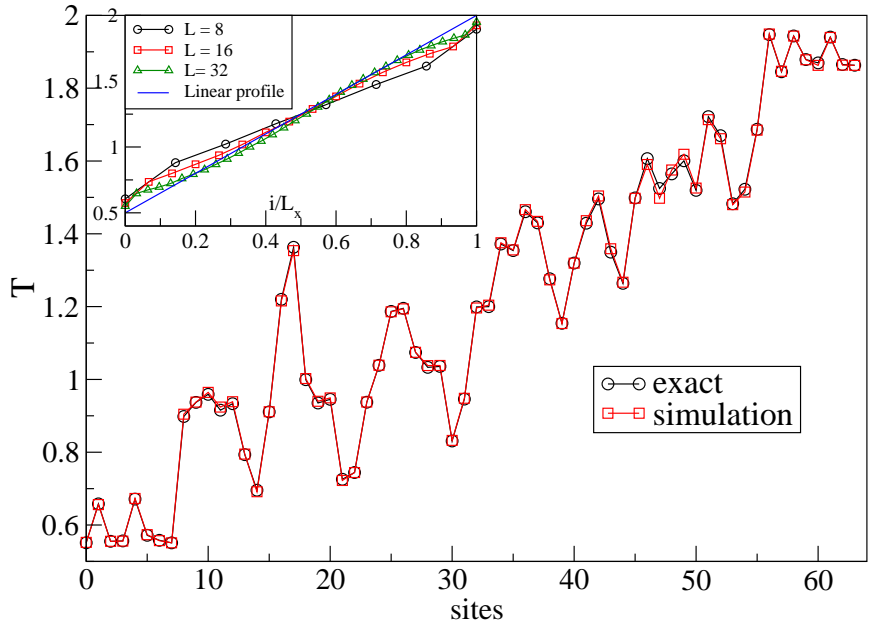}
\caption{Temperature at all the sites of a $8\times 8$ fully disordered
  lattice, from simulations and from the exact solution. Inset shows
  the disorder-averaged temperature profiles (averaged over the
  transverse direction) for different system sizes and seems to
  approach a linear form (from \cite{lee05}).}
\label{3.5:fig2}
\vspace{1cm} 
\end{figure}

\begin{figure}
\includegraphics[width=4.5in]{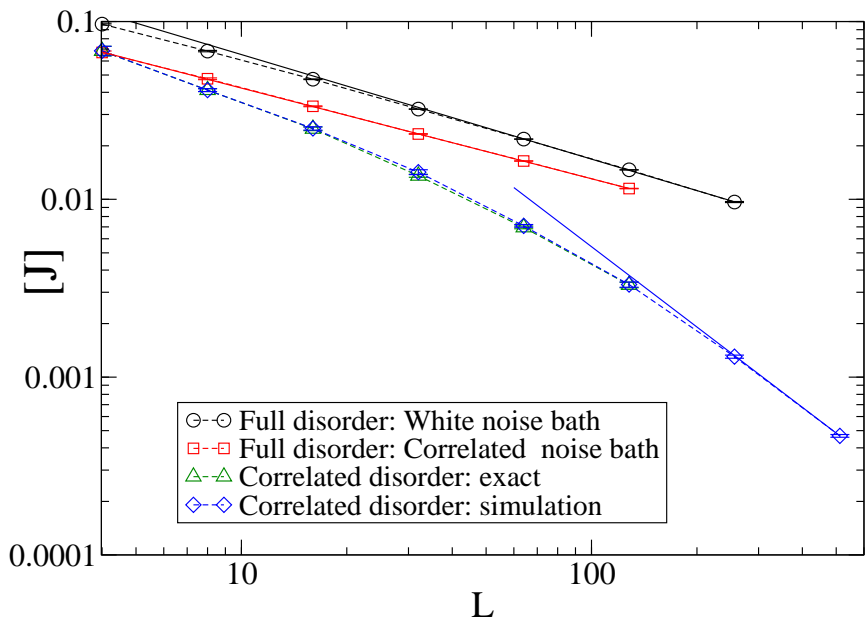}
\caption{Plot of disorder-averaged-current versus system size for a $L
  \times L$ lattice 
  two different heat baths. The case of correlated disorder
  with white noise is also shown. For the full disorder cases, the solid lines are
  fits to the last three points and have slopes $-0.59$ and $-0.51$.
  For the case of correlated disorder, the slope from exact numerics
  (and also simulations) is compared to $-1.5$ which is what one
  expects analytically (from \cite{lee05}).}
\label{3.5:fig3} 
\end{figure}
The most detailed study of the $2D$ disordered harmonic lattice is by Lee and
Dhar \cite{lee05}. They considered stochastic heat baths and looked at
the case of  mass disorder. In their model, the masses of exactly half
the particles on randomly chosen sites of a $L\times L$ square lattice were set to one and
the remaining to two.   
To see the effect of spectral properties of baths, two kinds of baths
were studied: one with uncorrelated Gaussian noise and the other with
exponentially correlated Gaussian noise. Simulations in disordered
systems have to be done with care since one can have slow 
equilibration. In \cite{lee05} the authors first checked their
simulation results by comparing them with results obtained from an
exact numerical solution of the general RLL matrix equations
Eq.~(\ref{3.1:mateq}) for a $8 \times 8$ lattice. The comparision, for the temperature at various
lattice points, obtained by the two methods is shown in
Fig.~(\ref{3.5:fig2}) and one can see excellent agreement between the
exact results and simulation. Also shown in the inset are the
disorder-averaged  temperature profiles across the system for
different sizes. One can see approach to a linear profile.  
To find the system size dependence of the current, lattices with
sizes upto $L=256$ were studied. In
Fig.~(\ref{3.5:fig3}) we show the data for the disorder averaged
current for different system sizes and for the two different bath
models. From this data the exponents $\alpha \approx 0.41$ for white
noise baths and $\alpha \approx 0.49$ for the correlated bath was
obtained. 
A special case of correlated disorder, first discussed in
\cite{oconnor74}, was also studied. Here the lattice was disordered in
the conducting direction, but ordered in the transverse
direction. Using the same methods as discussed in
sec.~(\ref{sec:ordharmlathd}), one can transform this into an
effectively $1D$ problem and then use the  numerical
techniques for evaluating the heat current as in the $1D$ case. 
In \cite{lee05}, the current was evaluated numerically  (upto $L=512$)
as well as from simulations (upto $L=128$ and with white noise
baths). Excellent agreement between the two confirmed the accuracy of
the simulations. From the transformation to an effective $1D$ problem
it is possible to argue for an exponent $\alpha = -1/2$ for the
correlated disorder case and this could be already verified at the
system size $L=512$. This is somewhat surprising considering the fact
that in the $1D$ case, one  has to go to sizes $\sim 10^5$ to see this
exponent [see Fig.~(\ref{3.4:fig1})]. This result gives some confidence that the results
obtained, for the fully disordered cases, are also close to the asymptotic
values.  An interesting observation in \cite{lee05} is that
equilibration times for local temperatures is typically much larger
than that for the current. This is expected since the temperature gets
contributions from all modes including the localized ones which are
weakly coupled to the reservoirs. On the other hand the current is
mainly carried by low frequency extended modes. This point was also noted
for the disordered $1D$ chain in \cite{likhachev06}.

\subsection{Harmonic lattices with self-consistent reservoirs}
\label{sec:selfcons}
As another application of the LEGF formalism
we consider the problem of heat transport in a harmonic
chain with each site connected to self-consistent heat
reservoirs. The classical version of this model was first studied by
Bolsteri, Rich and Visscher \cite{bolsteri70,rich75}, who introduced the self-consistent
reservoirs as a simple scattering mechanism for phonons which might ensure local
equilibration and  the validity of Fourier's law.  The extra
reservoirs connected to the system can roughly be thought of as
other degrees of freedom with which the lattice interacts. 
It is
interesting to note that the self-consistent reservoirs are 
very similar to the Buttiker probes 
\cite{butt85,butt86} which have been used to model  inelastic
scattering and phase decoherence in electron transport. In the
electron case they lead to Ohm's 
law being satisfied just as in the harmonic  chain the introduction of 
self-consistent reservoirs leads to Fourier's law being satisfied.  
In fact it has recently been shown how one can obtain both Ohm's law
and Fourier's law in an electron  model by  using
self-consistent reservoirs represented microscopically by noninteracting
electron baths \cite{roydhar07}. 

The ordered harmonic lattice with self-consistent reservoirs was
solved exactly by Bonetto et al \cite{bonetto04}, in arbitrary dimensions,  
who proved local equilibration and validity of Fourier's law, and
obtained an expression for the thermal conductivity of the system. They
also showed that the temperature profile in the wire was linear. 
The quantum version of the problem was also studied by Visscher and
Rich \cite{visscher75} who analyzed the limiting case of weak coupling to the
self-consistent reservoirs.   
We will here follow the  LEGF approach as given in \cite{dharroy06} to
obtain results
in the quantum-mechanical case. The classical results of Bonetto et al
are obtained as the high temperature limit while the quantum mechanical
results of Vischer and Rich are obtained in the weak coupling limit.
We will consider only the one-dimensional case. A generalization
to higher dimensions can be easily achieved, as in Sec.~(\ref{sec:ordharmlat}). 

Consider the quantum-mechanical Hamiltonian:
\bea
H = \sum_{l=1}^N ~[~\f{p_l^2}{2m}  +  \f{\omega_0^2 x_l^2}{2}~] +\sum_{l =1}^{N+1}
\f{m\om_c^2}{2} (x_l-x_{l-1})^2~, 
\eea
where we have chosen the boundary conditions $x_0=x_{N+1}=0$.  
All the particles are connected to heat reservoirs which are taken to
be Ohmic. The coupling strength to the reservoirs is
controlled by the 
dissipation constant $\g$. The temperatures of the first and last
reservoirs are fixed 
and taken to be $T_1=T_L$ and $T_N=T_R$. For other particles,
{\emph{i.e}}~,  $l=2,3...(N-1)$, the temperature of the attached
reservoir $T_l$ is fixed self-consistently in such a way that the net
current flowing into any of these reservoirs vanishes.  
The Langevin equations of motion for the particles on the wire are:
\bea
m \ddot{x}_l&=&-m \om_c^2 (2 x_l -x_{l-1}-x_{l+1})-m \omega_0^2
x_l-\gamma \dot{x}_l +\n_l~~~~{l=1,2...N}~,
\eea
where the noise-noise correlation, from Eq.~(\ref{3.2:nncor2}) and
with $\Gamma (\om)= \gamma \om$ for Ohmic baths, is given by:
\bea
\f{1}{2}  \la~ \n_l (\om) \n_m(\om')  + \n_l (\om') \n_m(\om)~ \ra 
&=& \f{\g \hbar \omega}{2 \pi} \coth (\f{\hbar
  \omega}{2 k_B T_l})~\d (\om+\om')~\d_{lm} ~.\label{qnnw}
\eea
From the equations of motion it is clear that the $l^{\rm th}$ particle is
connected to a bath with a self energy matrix $\hat{\Sigma}^+_l(\omega)$  whose only
non vanishing element is $(\Sigma^+_l)_{ll}=i \g \omega$.
Generalizing Eq.~(\ref{3.2:SSJ}) to the case of multiple baths,
one finds
that the heat current from the $l^{\rm th}$ reservoir into the wire is
given by: 
\bea
J_l&=& \sum_{m=1}^N ~\int_{-\infty}^\infty d \omega~ \mT_{lm}
\f{\hbar \om}{4 \pi}~[f(\om,T_l)-f(\om,T_m)]~,~~ \label{3.5:JSCR} \\
{\rm{where}}~~~\mT_{lm}&=& 4~ Tr[~
  \hat{G}^+(\omega)~  \hat{\G}_l (\omega) \hat{G}^-(\omega) ~ \hat{\G}_m (\omega)] \nn
\\ {\rm and}~~~\hat{G}^+ &=& [~-\omega^2~\hM +\hP -\sum_l \hat{\S}^+_l(\omega)~]^{-1}~,~~~\hat{\G}_l=
Im[\hat{\S}^+_l]~. \nn 
\eea
Here $\mT_{lm}$ is the transmission coefficient of phonons from the
$l^{\rm th}$ to the $m^{\rm th}$ reservoir. Using the form of
$\hat{\G}_l$ one gets, in the linear response regime: 
\bea
J_l= \g^2 ~\int_{-\infty}^\infty d
\omega~ \f{\hbar \omega^3}{\pi} \f{\p f(\omega,T)}{\p T}~
	\sum_{m=1}^N ~\mid [\hat{G}^+(\omega)]_{lm} \mid^2 ~(T_l-T_m)~. 
\eea  
For a long wire ($N >> 1$), for points far from the  boundaries of the wire ($l=y N$ where
$y=O(1),~1-y=O(1)$), one can explicitly evaluate the Green's function
and show that: 
\bea
G^+_{lm}= \f{e^{-\a |l-m|} }{2 m \om_c^2 \sinh{\a}} \label{3.5:Gasym}~,
\eea
where $e^{\alpha}=z/2\pm[(z/2)^2-1]^{1/2}$ with
$z=2+\om_0^2/\om_c^2-\om^2/\om_c^2-i \gamma \om/(m \om_c)^2$,  and we choose the root $\a$
such that $Re[\a] > 0$.  
Using this form of $G^+_{lm}$ and assuming a 
linear temperature profile given by
\bea
T_l=T_L+\f{l-1}{N-1} (T_R-T_L)~, \label{3.5:tprof}
\eea
one can see at once that, for
any point $l$  in the bulk of the wire, the zero-current condition
$J_l=0$ is satisfied since  $\sum_{m=-\infty}^\infty (l-m) |e^{-\a |l-m|}|^2 =0$. 
For points which are within distance $O(1)$ from the boundaries the
temperature profile deviates from the linear form. 
Knowing the form of the temperature profile $T_l$ and the form of
$G^+_{lm}$, one can proceed to find the net current in the wire. It is
easiest to evaluate the following quantity giving  
current  $J_{l,l+1}$ on the bond connecting sites $l$ and $(l+1)$:
\bea 
J_{l,~l+1}&=& m \om_c^2~\la x_l \dot{x}_{l+1} \ra = -\f{ m \om_c^2 \g
}{\pi}  \int_{-\infty}^\infty d \omega ~
\omega~ 
\left(\f{\hbar \om}{2 k_B T}\right)^2 {\rm{cosech}}^2 (\f{\hbar
  \om}{2k_B T})~ \nn \\
&& \times \sum_{m=1}^N k_B T_m ~Im \{ [\hat{G}^+(\omega)]_{lm}
     [\hat{G}^+(\omega)]^*_{l+1~m}\}   ~.\nn
\eea
Using  Eqs.~(\ref{3.5:Gasym},\ref{3.5:tprof}) one finally gets
the following expression for the thermal conductivity
$\kappa=JN/\Delta T$ (obtained in the large $N$ limit): 
\bea
\kappa=\f{\g k_B }{16 m \om_c^2 \pi i} \int_{-\infty}^{\infty} d \omega
~\f{\omega}{\sinh^2{\a_R}}\left(\f{\hbar \om}{2 k_B T}\right)^2 {\rm{cosech}}^2 (\f{\hbar
  \om}{2k_B T})  ~ \left( \f{1}{\sinh{\a}}-\f{1}{\sinh{\a^*}} \right),~~~~~\label{genF}
\eea
where $\a_R(\omega)=Re[\a],~\a_I(\omega)=Im[\a]$.
In the high temperature limit  $(\hbar \om/2 k_B T)^2
{\rm{cosech}}^2(\hbar \om/2k_B T)~\to~1$, this 
gives, after a change of variables from $\omega$ to $\a_I$, the
following  result for the classical thermal conductivity:
\bea
\kappa_{cl} &=&\f{2 k_B m\om_c ^2(2+\nu^2)}{\g \pi} \int_0^{\pi/2} d \a_I~ \f{\sin^2
  {(\a_I)}}{(2+\nu^2)^2 -4 \cos^2{(\a_I)}} \nn \\
&=&\f{k_B m \om_c^2}{\g~(2+\nu^2+[\nu^2(4+\nu^2)]^{1/2})}~,  
\eea
where $\nu=\om_0/\om_c$.
This agrees with the result obtained in refn.~\cite{bonetto04}.
Another  interesting limiting case is the case of weak coupling 
to the reservoirs ($\gamma \to
0$). In this case Eq.~(\ref{genF})
gives:
\bea
\kappa_{wc} &=& \left(\f{\hbar \om_c^2}{k_B T}\right)^2 \f{m k_B}{4 \gamma
  \pi} \int_0^{\pi}~ d \a_I~ \sin^2{\a_I}~
      {\rm{cosech}}^2 (\f{\hbar \om_\a}{2 k_B T}) ~,\label{wckappa}\\ 
{\rm where}~~~ \om_\a^2 &=&\om_0^2+2 \om_c^2 [1-\cos(\a_I)] \nn~.
\eea
This agrees with the result obtained in \cite{visscher75}.
In the low
temperature limit, Eq.~(\ref{wckappa}) gives $\kappa_{wc} \sim
e^{-\hbar \om_0/k_B T}/T^{1/2} $ for $\om_0 \neq 0$ and $\kappa_{wc} \sim T$ for 
$\om_0=0$.
As noted in \cite{visscher75} the expression for thermal
conductivity (in the weak scattering limit) is consistent with a
simple relaxation-time form for the thermal conductivity. The
temperature dependence  of $\kappa_{wc}$ then simply follows the
temperature dependence of the specific heat of the $1D$
chain. 

In the general case where the coupling constant has a
finite value, 
the low-temperature behaviour again depends on whether or not
there is an onsite potential. The form of the
low-temperature behaviour is very different from the case of weak
coupling. For small 
$T$ it is easy to pull out the 
temperature dependence of the integral in Eq.~(\ref{genF}) and one finds that
$\kappa \sim T^{3}$ for $\nu\neq 0$ and $\kappa \sim T^{1/2}$ for $\nu =
0$.

\begin{figure}[t]
\begin{center}
\includegraphics[width=4.5in]{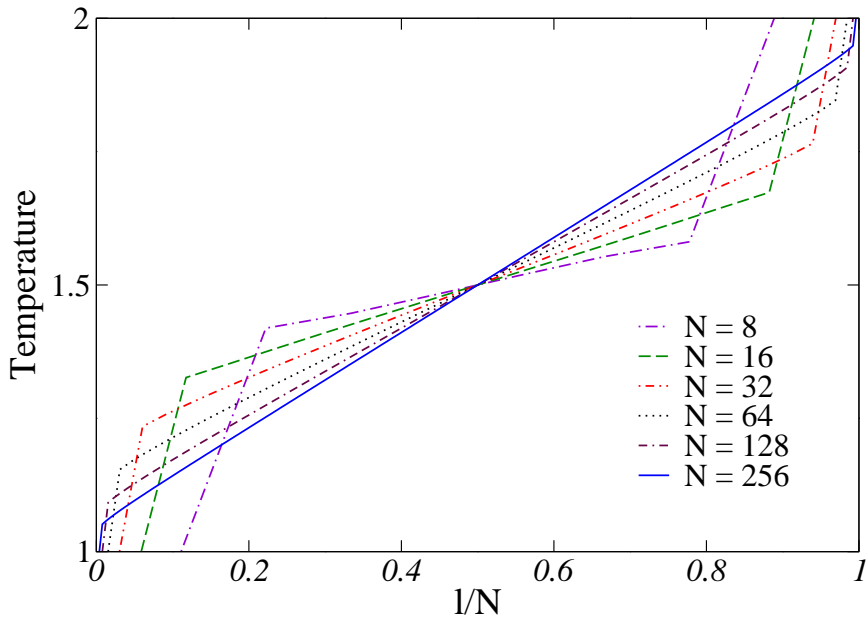}
\end{center}
\caption{ Plot of the temperature profile $\{T_l\}$ as a function of
  scaled length $l/N$ for different $N$ with $\g=1.0$ and
  $\g'=0.1$.  Here mean free path   $\ell \sim 30$ (from \cite{roy08}). } 
\label{3.5:fig1}
\end{figure}

A nice extension of the above problem has been done by Roy
\cite{roy08} who considered the case where the coupling of the bulk
lattice points to reservoirs ($\gamma'$) was taken to be different from that of the
boundary points to the end reservoirs ($\gamma$). The quantum
mechanical case with Ohmic reservoirs and the linear response regime
were studied. For small values of $\gamma'$
the transition from ballistic to diffusive transport could be seen
with increasing system size. It was shown that the current, for any
system size, could be written in the form 
\bea
J= \f{\kappa(T) \Delta T}{N +\ell}~,
\eea
where $\ell$ was an effective mean free path for phonons which
depended on $\gamma'$.  Thus for $N << \ell$ one gets ballistic
transport while for $N >> \ell$ one gets diffusive transport. 
In Fig.~(\ref{3.5:fig1}) the temperature profiles for different system
sizes is shown and one can see the transition, from a relatively flat
(in the ballistic regime), to a linear profile (in the diffusive regime). 

\section{Interacting systems in one dimension}
\label{sec:INT1D}
In the case of systems with interactions there are few analytic
results, for one-dimensional interacting particle systems, 
and these are all based on use of the Green-Kubo formula. All these theories
aim at calculating the equilibrium current-current correlation function
$C(t)=\la \mJ(0) \mJ(t)\ra $. Mainly there are
three different theoretical approaches: renormalization group theory
of hydrodynamic equations, mode coupling theory and the
Peierls-Boltzmann kinetic theory approach. We will discuss
these in Sec.~(\ref{sec:THEORY}). We will also discuss some exact
results, which have been obtained for certain models for which the
dynamics is not completely deterministic but includes some stochastic
component. All the analytic approaches predict that momentum conserving
systems in $1D$ exhibit anomalous transport with conductivity diverging as a
power law $\kappa \sim N^\alpha$. However there is disagreement on
the precise value of $\alpha$ and the number of universality classes. 
We will present the results of simulations for various momentum
conserving models in Sec.~(\ref{sec:momcons}).
In general one finds anomalous transport with $\kappa \sim
N^\alpha$ and again there is disagreement in the value of $\alpha$
obtained in simulations by different groups.  
However, the evidence seems strong that there is a single universality
class with $\alpha=1/3$. 

For momentum non-conserving interacting systems the
prediction of theory is that  Fourier's law is valid.   
Simulations also show  [Sec.~(\ref{sec:momnoncons})] that  the
presence of interactions (nonintegrable) and an 
external substrate potential are sufficient conditions to give rise to
a finite thermal conductivity.
Note that momentum non-conserving models are a bit unphysical in the
context of heat transport by molecular motion except when one is
considering wires and thin films attached to substrates. In the case
of electron transport the background ionic lattice naturally provides an
external potential for the electron and so momentum non-conservation
makes  physical sense.
We will also briefly discuss a class of rotor models that have been studied in
simulations which show Fourier behaviour inspite of absence of an
external potential \cite{giardina00,gendelman00}. These models
are somewhat special, in the sense that  
it is more natural to think of the phase space degrees of freedom
as local angle variables and thus these models should probably be thought of
as angular-momentum conserving rather than linear-momentum conserving
models. 

\subsection{Analytic results}
\label{sec:THEORY}

\subsubsection{Hydrodynamic equations and renormalization group theory}
\label{sec:RG}
This approach was first proposed by Narayan and Ramaswamy \cite{narayan02}.
Here one first notes that a one dimensional ($1D$) system of
interacting particles  will, at sufficiently large length scales,
behave like a fluid.  Suppose that the only conserved quantities  in
the system are the total number of particles, the total momentum and total
energy. One can then write hydrodynamic equations to describe the variation,
in time and space, of the density fields corresponding to these conserved quantities.
Namely we have $\rho(x,t),~g(x,t)=\rho(x,t)v(x,t),~{\rm and}~
\epsilon(x,t)$ for number, momentum and energy densities 
respectively and where $v(x,t)$ is the local average velocity field. This
basically gives the Navier-Stokes equations for a $1D$ fluid. After adding noise terms
to account for thermal fluctuations in the system the equations are
given by \cite{narayan02,mai06}: 
\bea
\partial_t \rho + \partial_x ( \rho v ) & =  & 0, \nonumber \\
\partial_t (\rho v) + \partial_x (\rho v^2) & = & -\partial_x p +
\zeta\partial_x^2v + \eta_v, \nonumber \\
\partial_t \epsilon + \partial_x [(\epsilon+p)v] & = & \kappa 
\partial_x^2 T+\zeta [(\partial_x v)^2 
+ v\partial_x^2 v] +\eta_\epsilon~,
\label{normal}
\eea
where the noise terms satisfy $\la \eta_v(x_1,t_1) \eta_v(x'_1,t'_1)
\ra \propto -k_B T \delta (x_1-x_1') \delta (t_1-t_1')$ and similarly
for $\eta_\epsilon$. 
The local temperature $T(x,t)$ and 
pressure $p(x,t)$ are implicit functions of $\rho$ and $\epsilon$.   
There are two transport coefficients, the viscosity $\zeta$
and the thermal conductivity $\kappa$. 
The above equations can be solved in the linear approximation and this
gives $C(t) \sim t^{-1/2}$ which in turn  implies a divergence of the
conductivity. However this divergence also means that the linear
approximation is not good and one needs to take into account the
nonlinear terms in the Navier-Stokes equations in order to get the
correct long time behaviour of various correlation functions. In
general this would require a RG analysis but it is argued in
\cite{narayan02} that the exponents can  be obtained from symmetry
considerations. Using the Galilean invariance of the system and the
fact that equal time correlations obey equilibrium statistical
mechanics they finally obtain:
\bea
C(t)\sim {t^{-2/3}}~.
\eea
For a finite size system, using the arguments in
Sec.~(\ref{sec:defs}), one puts a upper  cut-off $t_N \sim N$ in the
Green-Kubo integral and 
this then gives $\kappa \sim N^{1/3}$. Thus $\alpha=1/3$. Note that in
this treatment, the details of the form of the Hamiltonian are
unimportant. The only requirements are the presence of the three
conservation laws and also,  
the interactions should be such that the
nonequilibrium  state satisfies  local thermal equilibrium and 
should be describable by coarse grained hydrodynamic equations.
We note that the possibility of breakdown of hydrodynamic equations in
a one-dimensional fluid system has recently been pointed out \cite{hurtado06}.

An interesting question that arises in the context of the hydrodynamic
theory is the behaviour of the other transport coefficient in the
equations, \ie,  the bulk viscosity $\zeta$. This has not been
investigated much except in the  work in refn.~\cite{dadswell05} who,
somewhat surprisingly, find that this transport coefficient is finite.

\subsubsection{Mode coupling theory}
\label{sec:MCT}
This approach was first applied in the context of heat conduction by
Lepri, Livi and Politi \cite{lepri98a} and has subsequently been used
by several other authors  
\cite{lepri98a,lepri98b,lepri03,lepri05,delfini06,delfini07,wangli04a,wangli04b}.
We will here outline the main steps as described in
refn.~\cite{delfini07}. 
Mode coupling theory (MCT) again begins with the realization that the
divergence 
of conductivity is a result of the long time tails of the
current-current correlation
function which in turn can be attributed to the slow relaxation of
spontaneous fluctuations of long-wavelength modes in low dimensional
systems.   
For a $1D$ oscillator chain  with periodic boundary conditions one
considers the normal mode coordinates of the harmonic lattice which
are given by:
\begin{equation}
Q(q) = \frac{1}{\sqrt{N}} \sum_{n=1}^N ~ x_n ~ \exp({-iqn})~,
\label{Uk}
\end{equation}
where the wavenumber $q={2\pi k}/{N}$ with $k=-N/2+1,-N/2+2,...N/2$ (for even $N$).
The evolution of a fluctuation at wavenumber $q$ excited at $t=0$ is  
described by the  following correlation function:
\bea
G(q,t)= \frac{\langle Q^*(q,t)Q(q,0) \rangle}{\langle |Q(q)|^2~  \rangle}~.
\eea
The long time decay rate of this quantity at small $q$ is the main
object of interest and MCT is one approach to obtain this.
Basically one writes a 
set of approximate equations for $G(q,t)$ and this is then solved 
self-consistently.
Formally one can in fact write an exact equation for the time
evolution of $G(q,t)$. Using the Mori-Zwanzig projection methods
\cite{KTH91} one gets the following equation:
\bea
{\ddot G} (q,t) + 
\varepsilon \int_0^t \Gamma (q,t-s) {\dot G}(q,s) \, ds 
+ {\omega}^2(q) G(q,t)  
= 0 \quad,
\label{mct}
\eea
where the memory kernel $\Gamma(q,t)$ is proportional to $\langle
\mathcal{F}(q,t)\mathcal{F}(q,0) \rangle$, with $\mathcal{F}(q)$ being the
nonlinear part of the fluctuating force between particles. The
coupling constant $\epsilon$ and the frequency $\omega(q)$ are temperature
dependent input parameters which have to be computed independently. Equations (\ref{mct})
must be solved with the initial conditions $G(q,0)=1$  and $\dot G(q,0)=0$. 
The mode-coupling approach proceeds by replacing the exact memory
function $\Gamma (q,t)$ with an approximate one, where higher order correlators are
written in terms of $G(q,t)$.  
Consider now the  FPU  interaction potential
$U(x)=k_2x^2/2+k_3 x^3/3+k_4x^4/4$. 
In the generic case, in which $k_3$ is different from
zero, the lowest-order mode coupling approximation of the memory kernel gives:
\begin{equation}
\Gamma(q,t)= \,\omega^{2}(q)
\,\frac{2 \pi}{N} \sum_{p+p'-q=0,\pm\pi}  \,G(p,t) G(p',t) \quad .
\label{memcubic}
\end{equation}
On the other hand for the case $k_3=0$,~$ k_4\neq 0$, one  gets:
\bea
\Gamma(q,t)= \,\omega^{2}(q)
\,\Big(\frac{2 \pi}{N} \Big)^2 \sum_{p+p'+p''-q=0,\pm\pi}  
\,G(p,t) G(p',t) G(p'',t)  \quad .
\label{memquart}
\eea
Here $p~, p', p''$ range over the whole Brillouin zone  $(-\pi,\pi)$.
Using either of Eq.(\ref{memcubic}) or Eq.~(\ref{memquart}) in
Eq.~(\ref{mct}) gives a closed system of nonlinear
integro-differential equations. The coupling constant
$\varepsilon$ and the frequency $\omega(q)$ are taken as 
parameters which can be obtained from the harmonic approximation. 
The solution of these equations again involves making a number of
other approximations and the final result one obtains for $G(q,t)$ at
small values of $q$  is the following form:
\bea
G(q,t) &=& A(q,t) e^{i \omega (q) t} +  c.c \nn \\
{\rm where}~~A(q,t) &=& 
\begin{cases} 
g(\varepsilon^{1/2} t q^{3/2})~~~~{\rm for}~k_3 \neq 0 \\
 g(\varepsilon^{1/2} t q^{2})~~~~{\rm for}~k_3 =0, k_4 \neq 0. 
\end{cases}
\label{gqforms}
\eea
Finally one can relate the current-current correlation function $C(t)$
to the correlator $G(q,t)$. Again making the same approximation of
retaining only the lowest order correlation functions one gets:
\bea
C(t) \propto \sum_{q}  G^2(q,t)~, 
\label{flusso}
\eea
and plugging into this the result from Eq.~(\ref{gqforms}), one
finally obtains:
\bea
C(t) \sim 
\begin{cases}
{t^{-2/3}} ~~~~~~~{\rm for}~~ k_3 \neq 0 \\
{t^{-1/2}} ~~~~~{\rm for}~k_3 =0, k_4 \neq 0~. 
\end{cases}
\eea
Inserting this into the Green-Kubo formula (with a cut-off
proportional to $N$) then gives us $\kappa \sim N^{1/3}$ for the odd
potential and $\kappa \sim N^{1/2}$ for the even potential.

Another recent MCT study is by Wang and Li \cite{wangli04a,wangli04b}
who look at the effect of 
transverse degrees of freedom on the value of $\alpha$. They consider a one dimensional chain
where the particle positions are now two-dimensional vectors instead
of being scalar variables. The Hamiltonian they consider corresponds
to a polymer with bending rigidity and is given by:
\bea
H= \sum_l \f{{\bf{p}}_l^2}{2 m} +\f{K_r}{2}(|{\bf{r}}_{l+1}
-{\bf{r}}_{l}|-a)^2 +K_\phi \cos ( \phi_l )~, 
\eea
where $\{ {\bf{r}}_l,{\bf{p}}_l \}$, for $l=1,2...N$ denote two
dimensional vectors and $\cos (\phi_l) = -{\bf{n}}_{l-1}.{\bf{n}}_{l}$
with ${\bf{n}}_l=\Delta {\bf{r}}_{l}/|\Delta {\bf{r}}_{l}|$ and
$\Delta {\bf{r}}_{l}={\bf{r}}_{l+1}-{\bf{r}}_{l}$. Based on their MCT
analysis they suggest that the generic effect of including transverse
degrees is to give $\alpha=1/3$, while for the purely longitudinal
model one has $\alpha=2/5$. 

For momentum non-conserving systems MCT predicts a finite conductivity.
  
\subsubsection{Kinetic and Peierls-Boltzmann theory}
\label{sec:KBP}

In the  kinetic theory picture, one thinks of a gas of weakly
interacting particles, which are the heat carriers. These heat carriers
could be molcules in a gas, electrons in a metal or phonons in a
crystal. Using the idea that the
heat carriers are experiencing random collisions, and hence moving
diffusively, one can do a simple minded calculation.  This gives  us a simple expression for the 
thermal conductivity, namely $\kappa \sim c v \ell$, where $c$ is the
specific heat capacity per unit volume, $v$ the typical particle velocity and $\ell$
the mean free path of the particles between collisions. 

The Boltzmann equation approach gives a more systematic derivation of the
results of kinetic theory, and was first developed for the case of
molecular gases.  In this theory, one writes an equation of motion for
the distribution function $f(\bx,\bp,t)$, where $f(\bx,\bp,t) d^3 x d^3
p$ (in $3D$) gives the number  of particles in the volume $d^3 x d^3
p$. The presence of collisions makes the Boltzmann equation equation nonlinear, and then
one has to solve the equation under various approximations. The final
result is quite often in the form of the kinetic theory answer, with an
explicit expression for the mean free path $\ell$.   
For phonons, the Boltzmann theory of conductivity was 
developed by Peierls \cite{peierls55}. He  wrote the  Boltzmann
transport equation for the phonon gas and  pointed out the
importance of lattice momentum non-conserving processes (Umpklapp
processes) in giving rise to finite conductivity. Solving the
Boltzmann equation in the relaxation time approximation gives a simple
kinetic theory like expression for the thermal conductivity, $\kappa
\sim \int d {\bf q} c_{\bf q} v_{\bf q}^2 \tau_{\bf q}$, ~ where
$\tau_{\bf q}$ is the time between collisions, and ${\bf q}$ refers to
different phonon modes of the crystal. The relaxation time
$\tau_{\bf q}$ can get contributions from various sources, such as
phonon-phonon interactions and impurity scattering, and its calculation
from first principles is one of the main tasks.    
In three dimensional
solids, the Peierls-Boltzmann theory is well-developed \cite{ziman60,ziman72} and
probably quite accurate.  
One worry here is that the meaning of the distribution $f(\bx,{\bf q},t)$ for
phonons is not really clear, since phonons are extended objects. 
The recent work of Spohn \cite{spohn06} tries
to give a rigorous basis for the phonon Boltzmann equation for 
a crystal with a weakly anharmonic onsite potential.
We note that, as far
as making definite predictions (starting from a given Hamiltonian) on
the actual conductivity of a system 
and properties such as the temperature dependence of the
conductivity, the kinetic theory approach  probably has more chance
of success than the other two approaches  described before.

Let us now consider the application of the kinetic theory approach  to
the one dimensional case. If we look at the result $\kappa \sim \int
dq c_q v_q^2 \tau_q$, we see that a divergence with system size can
arise because the relaxation time $\tau_q$, and correspondingly the
mean free path $\ell_q=v_q \tau_q$, becomes large at small
$q$. Let us assume that  $c_q, v_q$ are constants, and that $\tau_q$
has the power-law dependence 
$\tau_q \sim q^{-a}$. Then all modes with  $q < q_m \sim L^{-1/a}$
travel ballistically and this immediately  gives $\kappa \sim
L^{1-1/a}$, and for $a > 1$ one would get a diverging conductivity.  

In a way the kinetic approach is similar to the MCT
method. Here too, one tries to 
calculate the rate of decay of long-wavelength fluctuations at small
$q$, but now using a different approximation scheme. The
current-current correlation function is then related to this decay
constant. The Kinetic theory  approach 
for the FPU chain was first considered by Pereverzev
\cite{pereverzev03} who studied 
the model with $k_3=0$ and a small non-zero value for $k_4$. It was
noted that the
approximate time evolution of a fluctuation in the average energy
$\epsilon_q$ of a
mode with wavenumber $q$ is given by the homogeneous classical
linearized Peierls equation. 
This equation is then brought to the following form, corresponding to
the relaxation time approximation: 
\bea
\f{d \la \epsilon_q(t) \ra}{dt}=-\f{1}{\tau_q} (\la \e_q \ra -k_B T)~, 
\eea
where  one has an explicit form for $\tau_q$. 
For small $q$, making some more approximations enables one to evaluate
$\tau_q$ and  one finds $\tau_q \sim q^{-5/3}$. Finally using the same
set of approximations and in the limit $N \to \infty$ one can show that:
\bea
C(t)=\f{2 k_B^2 T^2}{\pi} \int_0^\pi dq e^{-t/\tau_q} v_q^2~,
\eea
where $v_q$ is the phonon group velocity. 
At small $q$ the phonon velocity
$v_q \sim~$ const and the  above equation gives: 
\bea
C(t)\sim {t^{-3/5}}~.   
\eea
This then implies $\kappa \sim N^{2/5}$. In fact, the arguments given
at the beginning of this paragraph directly give this (putting
$a=5/3$), and one does not need to find $C(t)$.

The kinetic theory approach has been made more rigorous by the work
of Lukkarinen and Spohn \cite{spohn07}. They also work with the linearized collision
operator and make the relaxation time approximation, and for the quartic
FPU chain they confirm the result in \cite{pereverzev03}, namely
$C(t)\sim t^{-3/5}$. However they point out the possibility that the kinetic theory
approach may not be able to predict the correct long-time decay  of
the correlation function. 
Another paper using the linearized Peierls-Boltzmann equation 
for the quartic Hamiltonian also finds $\kappa \sim N^{2/5}$ \cite{nickel07}. 
Finally a quantum calculation of the phonon relaxation rate at small
$q$ has been carried out in
\cite{santhosh07,santhosh08}. They studied both the cubic and quartic
FPU chain and obtained relaxation times $\tau_q \sim
q^{-3/2},~q^{-5/3} $ for the two cases respectively.

For the case of momentum non-conserving systems, 
Lefevere and Schenkel \cite{lefevere06} and later Aoki
\etal~\cite{aoki06} have used the 
kinetic theory approach for the case of weak
anharmonicity and obtained a finite conductivity.

\subsubsection{Exactly solvable model}
\label{sec:EXACT}
A harmonic chain with a energy conserving stochastic dynamics was
considered by Kipnis \etal~ \cite{kipnis82} who could prove exactly
that the model satisfied Fourier's law. The dynamics was momentum
non-conserving and completely stochastic so it is not surprising that
Fourier's law was obtained.  Models with self-consistent
reservoirs can also be viewed as stochastic models (but with
Hamiltonian components in the dynamics) where energy is conserved on
average, while momentum is not  and again Fourier's law is satisfied.  

Recently a similar stochastic model, but  in which total momentum
conservation was also enforced, was introduced by 
Basile \etal~
\cite{basile06,basile08}. In their lattice model the dynamics
consisted of two parts. Apart from a deterministic 
Hamiltonian dynamics the system was subjected to a stochastic dynamics
which conserved both total energy and momentum exactly. The stochastic
dynamics consisted of a random exchange of momentum between three 
neighboring particles (in 
$1D$) while conserving both energy and momentum. Thus a triplet of
particles with momenta $(p_{l-1},p_l,p_{l+1})$ is chosen  and this set
performs a diffusive motion on the curve given by:
\bea
p_{l-1}+p_{l}+p_{l+1} = {\rm const.} \nn \\
 \f{p_{l-1}^2}{2 m} +\f{p_{l}^2}{2m}+\f{p_{l+1}^2}{2 m} = {\rm const.}
 \nn
\eea
The Hamiltonian of the system was taken to be that of a harmonic
system. A Fokker-Planck equation for the probability density
$P(\bx,\bp,t)$ could be
written, which in $1D$ is given by:
\bea
\f{\p P}{\p t} = (\hat{L}^H + \gamma \hat{S}) ~P~, 
\eea
where $\hat{L}^H$ is the usual Liouville operator for the given Hamiltonian
and $\hat{S}$, the generator of the stochastic perturbation of strength
$\gamma$, has the form
\bea
\hat{S} &=& \f{1}{6} \sum_l \hat{Y}_l^2, \nn \\
{\rm with}~~\hat{Y}_l&=&(p_l-p_{l+1}) \p_{p_{l-1}}+(p_{l+1}-p_{l-1})
\p_{p_{l}}+(p_{l-1}-p_{l}) \p_{p_{l+1}}~. \nn 
\eea  
The authors were able to  compute exactly an explicit form for the
current-current correlation function $C(t)$, 
 for system size $N \to \infty$, and from this
they  found the following asymptotic long-time behaviour:
\bea
C(t) \sim 
\begin{cases}
t^{-1/2}  ~~~~~~~~~~{\rm for ~no~pinning} \\
t^{-3/2}  ~~~~~~~~~~{\rm with ~pinning.} \\
\end{cases}
\eea
Plugging  this into the Green-Kubo formula 
one gets $\alpha =1/2$ in the
unpinned case, while for the pinned case   a finite conductivity is obtained.
One can argue that the stochastic dynamics in  a way mimics
anharmonicity and the problem 
considered corresponds to an even interaction potential. 
The latest prediction from MCT  also gives 
$\alpha=1/2$, which agrees with the result from this model. However
all simulation results of momentum conserving interacting Hamiltonian
models give exponents quite far from this value (between $0.3-0.4$). 
The exponent 
$\alpha=1/2$ then comes as quite a surprise.
One possibility is that the
choice of a harmonic Hamiltonian makes the model special, and thus
solvable, and at the same time makes it  a non-generic case.

Some simulations with a dynamics which is roughly similar to the above
stochastic dynamics were recently done with FPU type anharmonic terms included
in the Hamiltonian \cite{basile07}. These are equilibrium simulations
using the Green-Kubo formula. The authors have argued that their
results support the two-universality class scenario. It will be
interesting to understand in more details the role of the Hamiltonian
part of the dynamics in determining the exponent $\alpha$ in this
model. 

More recently,  a similar one-dimensional  stochastic model with random
two particle momentum exchanges has been numerically \cite{delfini08}
and  analytically \cite{lepri08} studied,  for the nonequilibrium
case with Langevin heat baths attached at the two ends. Apart from
confirming the exponent $\alpha=1/2$, these studies have also looked
at the temperature profiles. An analytic expression for the
temperature profile was obtained and it was noted that the  profiles were
very similar to those obtained for FPU chains.

\subsection{Results from simulation} 
\label{sec:SIMULATION}
\subsubsection{Momentum conserving models}
\label{sec:momcons}
{\bf Gas of elastically colliding particles of two masses}:
One of the simplest model of interacting particles that one can consider is a gas of elastically
colliding point particles where the boundary particles interact with
thermal reservoirs, usually modeled by Maxwell boundary conditions. If
all the particles have equal masses then 
this model, without reservoirs,  is the so-called Jepsen model \cite{jepsen65}. 
As far as heat conduction properties are concerned the model is
somewhat trivial. This is because
at each collision  the particles simply exchange momentum and so the
net heat transfer can be calculated by considering a single particle
that is bouncing between the hot and a cold  walls. One finds a system-
size independent heat current $J=k_B^{3/2}(2m/\pi)^{1/2}\rho (T_L^2 T_R-T_R^2
T_L)/(T_L^{1/2}T_R+T_R^{1/2}T_L)$, where $\rho$ is the particle
density,  and a flat temperature profile  
given by $T=(T_L T_R)^{1/2}$. Thus this model is somewhat like the
ordered harmonic chain. However the model  becomes 
interesting and non-trivial if one considers a dimerized model where
alternate particles have different masses say $m_1$ and $m_2$. 
In this case one finds a current which decays with system size, 
and a slowly varying temperature profile.  

The diatomic hard particle gas  model was first studied by Casati \cite{casati86} but the
numerical results were not sufficient to draw any definite conclusions.
This model, along with the diatomic Toda lattice, were later studied
by Hatano \cite{hatano99}. Using nonequilibrium
simulations and system sizes upto $N=5000$, an exponent
$\alpha\approx 0.35$ was obtained for both these models. The
current-current correlation function was also evaluated for a
periodic closed system and it was found that $C(t) \sim N^{-0.65}$
consistent with the nonequilibrium results. 
Subsequently, a number of
further studies were made using both 
nonequilibrium simulations, and also the Kubo formalism and using much larger
system sizes. Unfortunately there is not much agreement on the
numerically obtained value of the exponent. The various reported
values include: Garrido \etal~ ($\alpha=0$ implying Fourier behavior)~
\cite{garrido01,dhar02,garrido02}, Dhar~ ($\alpha \approx 0.2$)~\cite{dhar01a},
Grassberger \etal~ ($\alpha \approx 0.33$)~\cite{grass02}, Savin \etal~
($\alpha \approx 0.2 $)~\cite{savin02} and
Casati \etal~ ($\alpha \approx 0.25$)~\cite{casati03}. However, based on the
theoretical predictions, there seems reason to believe  that the value 
obtained by   Grassberger \etal~ is the correct one and here we will 
discuss their results in some detail. We also mention here the
work of Cipriani \etal~ \cite{cipriani05} who performed zero-temperature studies on
diffusion of localized pulses and using  a Levy walk interpretation concluded that 
$\alpha =0.333 \pm 0.004$. 

\begin{figure}
\begin{center}
\includegraphics[width=3in,angle=270]{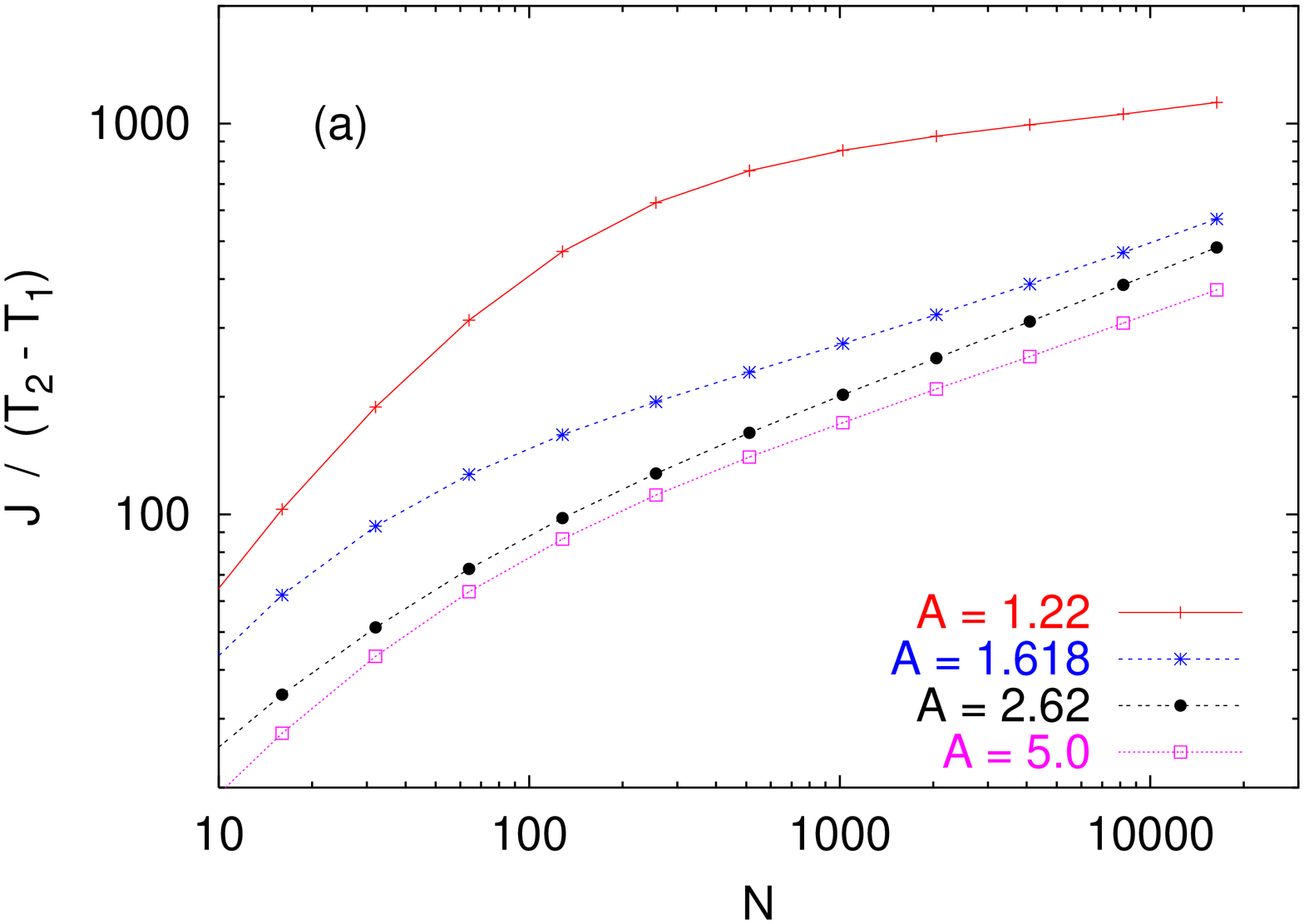}
\includegraphics[width=3in,angle=270]{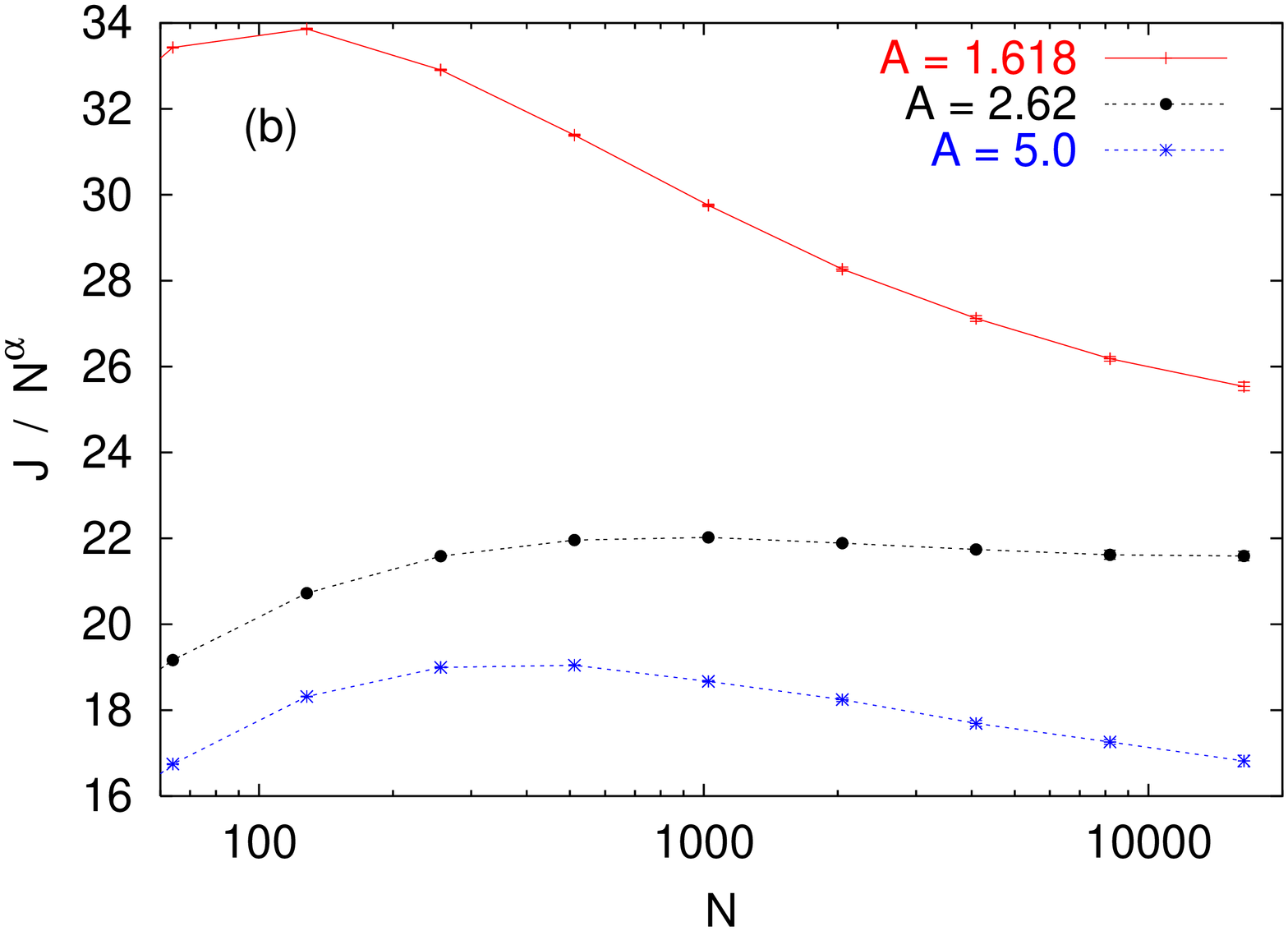}
\caption{(a): Log-log plot of $J/(T_2-T_1)$ versus $N$ for four
  values of the mass ratio $A$. (b): Part of the 
  same data divided by $N^\alpha$ with $\alpha=0.32$, so the y-axis is much 
  expanded (from \cite{grass02}).}
\label{4.2:grassfig1}
\end{center}
\end{figure}

\begin{figure}
\begin{center}
\psfig{file=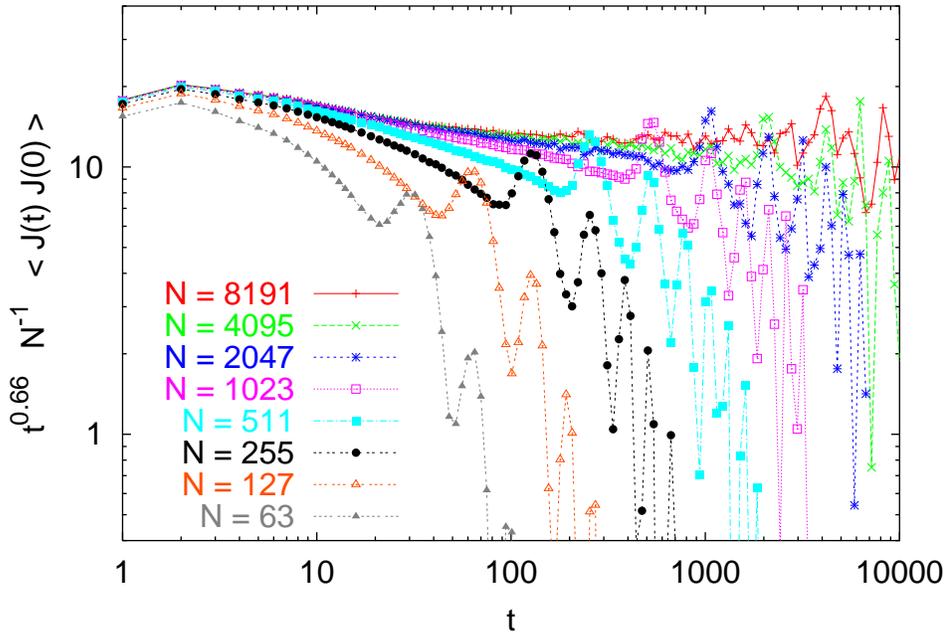,width=3.5in,angle=270} 
\caption{ Total heat current autocorrelation, $t^{0.66} N^{-1}$
  $\langle J(t)J(0)  
   \rangle$ for $A=2.2$ and $T=2$. Total momentum is $P=0$ (from \cite{grass02}). }
\label{4.2:grassfig4}
\end{center}
\end{figure}
We now present some of the results obtained by Grassberger \etal~ \cite{grass02}.
Apart from looking at much larger system sizes (upto $N=16384$),
they made the observation that the asymptotic
behaviour is easier to observe at some optimal value of the mass ratio
$A=m_2/m_1$. It was argued that $A=1$ and $A=\infty$ were special
integrable limits where one would clearly get ballistic and non-typical
behavior. If the value of $A$ was too close to $1$ or too large then one
would have to go to very large system sizes to see the correct
asymptotic form. However by choosing an appropriate value of $A$, one
can reach asymptotic behaviour much faster. This feature can be seen
in Fig.~(\ref{4.2:grassfig1})  where 
the system size dependence of the current for different values
of $A$ is given. One can see that for $A=2.62$, asymptotics is reached
faster than for $A=1.618$ and $A=5.0$. The value of the exponent
obtained from this data was $\alpha =
0.32 {+0.03\atop -0.01}$. Equilibrium simulations were also performed
and in Fig.~(\ref{4.2:grassfig4}) results are shown for the
current-current autocorrelation function obtained 
for a periodic system. For large system
sizes one can see a $t^{-0.66}$ decay with a cutoff at $t \propto N$.
This again gives $\alpha =0.34$ in agreement with the value
obtained from the nonequilibrium simulation.

\begin{figure}
\begin{center}
\psfig{file=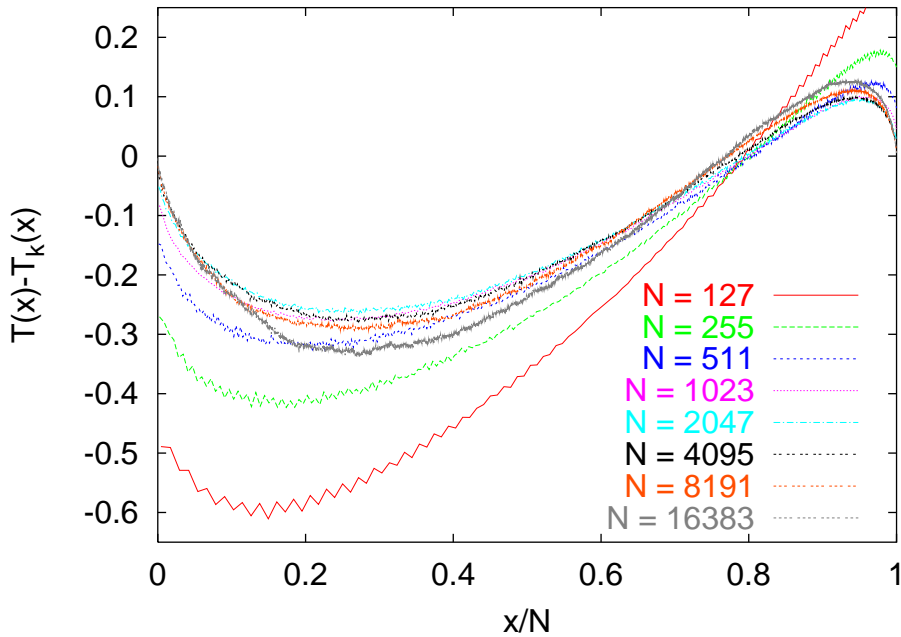,width=3.5in,angle=270}
\caption{ $T(x)-T_k(x)$ against $x/N$ for $A=1.618$. This shows
  non-convergence of the steady state temperature profile to the
profile expected from  the $\kappa \sim T^{1/2}$ result of kinetic theory
  (from \cite{grass02}). }
\label{4.2:grassfig3}
\end{center}
\end{figure}

Some interesting features were seen in the temperature profiles also
and we now discuss these.
Simple kinetic theory predicts that the thermal conductivity
of a hard particle gas should have a temperature dependence  $\kappa
\sim T^{1/2}$ (this can also be obtained from the Green-Kubo formula).  
Now if we plug this in Fourier's law then, with specified boundary
temperatures, one easily obtains the following nonlinear temperature profile:
\bea
T_k(x)=[T_L^{3/2} (1-x/N) + T_R^{3/2} x/N ]^{2/3}. 
\eea
In refn.~\cite{dhar01a}, it was noted that  a convergence, of the actual
nonequilibrium temperature profile $T(x)$, to the kinetic theory
prediction $T_k(x)$ given above, seemed to take place. However the
study in  \cite{grass02} found that this apparent convergence stops as one looks
at larger systems and in fact one finds $T(x)-T_k(x)$ attains a
non-zero profile for $N \to \infty$. This is shown in
Fig.~(\ref{4.2:grassfig3}). This result indicates that there is a
problem in writing Fourier's law in the form $J=-\kappa_N \nabla T$,
with $\kappa_N$ defined as a length dependent conductivity.  We will
see similar problems with other $1D$ models.

\begin{figure}
\begin{center}
\psfig{file=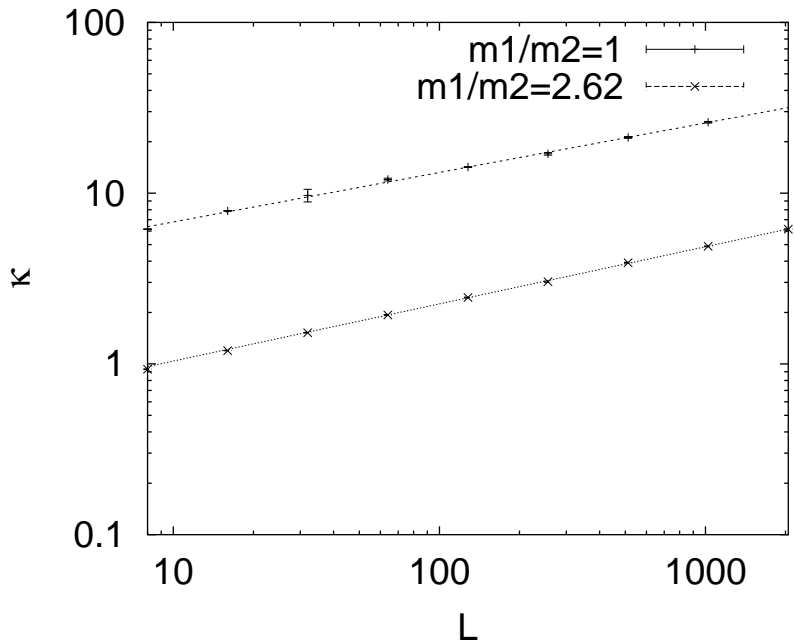,width=5in,angle=0}
\caption{
Log-log plot of the conductivity as a function of the number of particles
for the random collision model introduced in \cite{deutsch03a}. The upper plot 
is for all the particles with mass ratio $A=1$, while the lower plot is for $A=2.62$.
The slopes for the  two plots are $0.29\pm 0.01$ and $0.335\pm 0.01$
(from \cite{deutsch03a}).
}
\label{4.2:deutschfig1}
\end{center}
\end{figure}

{\bf Random collision model}: In studies on heat conduction, one often finds
that a faster convergence to the asymptotic system size limit can be
obtained by  particular choices of model and parameter
values. 
In this context one should mention a stochastic model
introduced by Deutsch and Narayan \cite{deutsch03a}. They consider a 
binary mass gas of  hard point particles where any particle's motion is strictly confined
to  one dimensions while its momentum is a two dimensional vector
$(p_x,p_y)$. During a collision between two particles their momenta
gets changed randomly while conserving total energy and both
components of  total momentum. Physically one can think of this
model as approximating a system of small particles with rough surfaces, moving in a
narrow tube. The results of nonequilibrium simulations for two different mass ratios
is shown in Fig.~(\ref{4.2:deutschfig1}). For the case $m_1/m_2=2.62$,  at
system sizes as small as $N \sim 10^3$,  one already gets $\alpha
=0.335 \pm 0.01$, which is close to the expected value $\alpha=1/3$.

{\bf Fermi-Pasta-Ulam chain}:
The Fermi-Pasta-Ulam (FPU) model consists of an oscillator chain with
harmonic as well as anharmonic nearest neighbour interparticle
interactions. This model was first studied by the authors in a
landmark paper \cite{fpu55} where they wanted to verify the common
assumption of statistical mechanics that anharmonic interactions
should lead to equilibration and equipartition. Surprisingly the
numerical experiments on the FPU chain gave a negative
result, \ie  the chain {\emph{ failed to equilibrate}}. Understanding the
FPU results  led to the development of new areas and concepts in
physics \cite{ford92}. It is probably fair to state that a complete understanding of
the problem is still lacking. For example it is believed that for high
enough energy densities and large system sizes one may achieve
equilibration but there are details which are not yet understood
precisely \cite{berman05}. 

What about heat transport across the FPU chain ? Clearly this is the
simplest model to study in order to see the effect of anharmonicity on heat
transport. One might suspect that the equilibration problem of the FPU
chain  is likely to show up in some way when one looks at transport
properties, especially so when one thinks in terms of the Green-Kubo formula.   
It turns out that a FPU chain connected to heat reservoirs is
better-behaved.  
It can in fact be proved rigorously
that a FPU chain connected to equal temperature Langevin heat baths at its two
ends will always equilibrate. At long times it will converge uniquely to the
appropriate Boltzmann-Gibbs distribution.  
Also it can be shown that even with unequal temperature baths
(stochastic or Hamiltonian) the
system reaches a unique nonequilibrium steady state \cite{eckmann99} and this is very
reassuring when one begins to consider heat transport studies in the
FPU chain.  

The first study of heat conduction in the FPU chain was by
Lepri \etal~ \cite{lepri97} who considered an interparticle potential of
the form $U(x)=k_2 x^2/2+k_4 x^4/4$ and performed nonequilibrium
simulations with  Nose-Hoover baths. Looking at
system sizes upto $N=400$ they obtained $\alpha=0.55 \pm 0.05$. In a
subsequent paper \cite{lepri98a}, by studying systems upto $N=2048$,
they obtained $\alpha \approx 0.37$.  They
also found a highly {\it nonlinear and singular} temperature profile and
noted that this was true even for relatively small temperature differences
applied to the ends. We will comment more on the temperature profile
of the FPU chain later in this section.

Since the important work  of \cite{lepri97}, a large amount of numerical  and
analytical work has been carried out on heat conduction in the FPU
chain. We first summarize the various analytic results discussed in
Sec.~(\ref{sec:THEORY}). We assume that 
the interparticle interaction is of the general FPU form  $U(x)=k_2
x^2/2+k_3 x^3/3+k_4 x^4/4$. The predictions from theory are then:

(i) Renormalization group theory of hydrodynamic equations: This
predicts that there is only one universality class with $\alpha=1/3$.

(ii) Mode-coupling theory: This predicts that there are two universality
classes depending on the parity of the leading nonlinearity in the
Hamiltonian. For the case where the leading nonlinearity is cubic, \ie
$k_3\neq 0$, the prediction is $\alpha=1/3$ while for $k_3=0, k_4 \neq
0$, the
prediction is $\alpha=1/2$.

(iii) Kinetic theory and the Peierls-Boltzmann equation approach: This
gives $\alpha=2/5$ for the quartic case.

{\it Results of simulations}: 
As we have seen in the last section  simulations of hard
particle gases~\cite{grass02,cipriani05,deutsch03a} 
seem to indicate a value $\alpha=1/3$ for the heat conduction
exponent, though even here the issue is not 
completely settled~\cite{savin02,casati03}. On the other hand, numerical
simulations of oscillator chains, including FPU chains, give various
exponents~\cite{LLP03,lepri97,lepri98a,lepri03,wangli04a} for different
systems, often slightly higher than $1/3$. This seems consistent with
early results from mode-coupling theory (MCT), which predicted 
$\alpha=2/5$~.
The most recent MCT analysis  \cite{delfini06,delfini07} predicts that
$\alpha = 1/2$ for  potentials $U(x)$ with quartic leading
nonlinearity while  for potentials with cubic nonlinearity, 
there seems to be agreement between different theories about $\alpha=1/3$.
Here we will focus on simulations for the even potential only.
We will  discuss the results of the most
recent simulations by Mai \etal~ \cite{mai07} of the even potential FPU model
and another simulation by Dhar and Saito \cite{dharsaito08} of the
alternate mass FPU chain. 

\begin{figure}
\begin{center}
\includegraphics[width=4.5in]{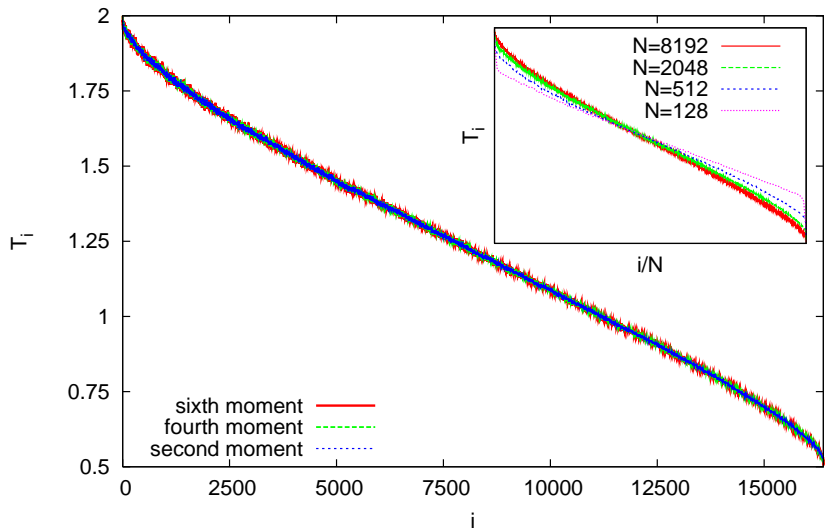}
\caption{Kinetic temperature profile for a FPU-$\beta$
chain with $N=16384$, $T_L=2.0$, $T_R=0.5$ and $\gamma=2.0$.  The
temperature evaluated from the first three even moments of the velocity are shown; their agreement
indicates a Gaussian velocity distribution.  (Inset) Normalized
temperature profiles for different  $N$ (from \cite{mai07}).}
\label{4.2:fpuTprof}
\end{center}
\end{figure}

An important aspect addressed in the simulations by Mai \etal~was
the effect of 
boundary conditions. It is well known \cite{LLP03} that the of coupling of a
system to thermal reservoirs leads to so-called contact
resistances. These show up, for example, in the jumps that one
observes in the temperature profile in such a system. It is only for
sufficiently 
large system sizes, when the resistance of the system is much larger
than the contact resistance, that one can neglect the contact
resistance. In simulations where one is interested in determining the
precise dependence of current on system size, it is 
important to ensure that one has reached the required system size where contact
resistances are negligible compared to the actual system resistance. This point has been discussed in some
detail by Aoki and Kusnezov \cite{aoki01}. 
The study by Mai \etal~ ensured
this by performing simulations with two different baths, namely,
stochastic white noise baths and the deterministic Nose-Hoover
bath. Further they did simulations for different coupling strengths
of the system to reservoir. It was found that for small
system sizes the current values were significantly different (for the
same applied 
temperature difference). This is expected since the contact resistance,
which is different for the different boundary conditions, dominates
the transport current. However       
at large system sizes,
the actual values of the currents for all the different cases tend to
converge. In this system-size regime one is thus assured that boundary
effects have been eliminated and one can then extract the correct exponent. 

\begin{figure}
\begin{center}
\includegraphics[width=4.5in]{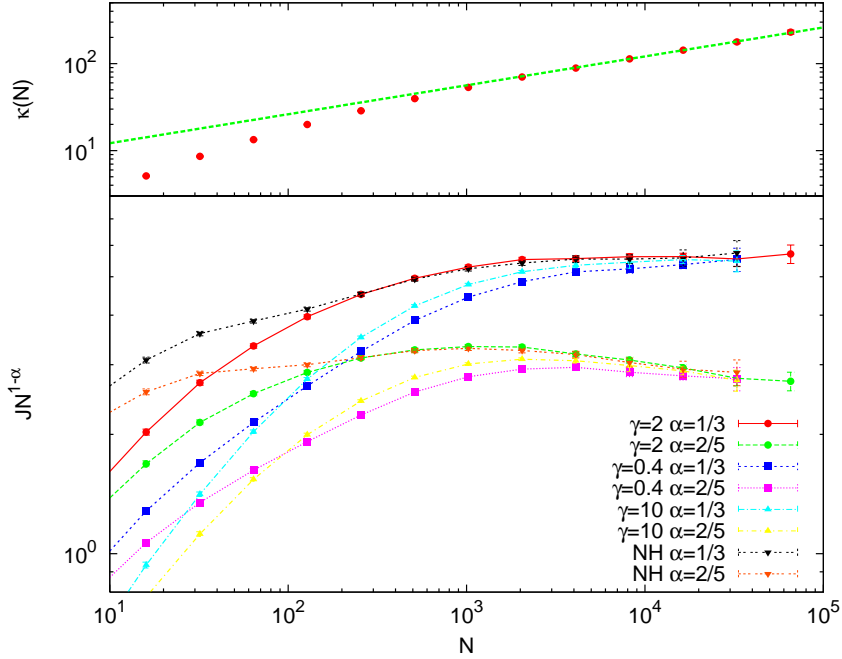}
\caption{ (Top) Conductivity versus
$N$.  The last five points fit to a slope of $\alpha =0.333\pm
0.004.$  (Bottom)
$JN^{1-\alpha}$ versus $N$ for $\alpha=1/3$ and $\alpha=2/5$.  In
the large $N$ regime, $\alpha$ is definitely less than 2/5 and
appears to agree quite well with the 1/3 prediction for all data
sets.  Langevin baths with $\gamma=0.4, 2$ and $10$, and one data
set with Nose-Hoover baths, are shown (from \cite{mai07}). }
\label{4.2:fpujvsn}
\end{center}
\end{figure}

The simulation in \cite{mai07} was done for parameter values  $k_2=1,
k_3=0, k_4=1$ and $m=1$. The temperature at the two ends were fixed at $T_L=2.0$
and $T_R=0.5$. Both white noise baths with coupling parameter $\gamma$
and Nose-Hoover baths, with coupling parameter $\theta$ were
studied. The white noise simulations were done using a velocity-Verlet
type algorithm \cite{allen87}, while the Nose-Hoover
simulations were implemented using a fourth order Runga-Kutta
integrator. Time steps of order $dt=0.0025-0.005$ were used and, for
the largest system size ($N=65536$), upto $10^9$ equilibration steps
and an equal number of data-collecting steps were used.   

The temperature profile for a chain of size $N=16384$ is
shown in Fig.~(\ref{4.2:fpuTprof}). The temperature is defined through the first three even
moments of the velocity as $T_l=\la v_l^2\ra$, $T_l=(\la v_l^4
\ra/3)^{1/2}$ and  $T_l=(\la v_l^4 \ra/2-\la v_l^6 \ra/30 )^{1/3}$ respectively.
 Their agreement indicates that local thermal
equilibrium has been achieved and the local velocity distribution is
close to Gaussian. Also we notice that the boundary jumps are almost
absent for this system size. The inset shows smaller system sizes
where the boundary jumps, arising from the contact resistance, can be
clearly seen.   As noted and discussed earlier by Lepri \etal
\cite{LLP03} the temperature profile is nonlinear and this feature 
seems to be {\it present even for small temperature differences}
and is another indication of anomalous transport. As for the hard
particle case this also indicates that one cannot find the temperature
profile using a temperature (and system size) dependent conductivity
in Fourier's law.

In Fig.~(\ref{4.2:fpujvsn}) (upper figure), the conductivity defined as
$\kappa(N)=J N/\Delta 
T$ is plotted against system size. This data gives  
\bea
\alpha = 0.333\pm 0.004~.
\eea 
The results of various simulation runs  with Langevin baths with different damping constants
$\gamma=0.4, 2$, and $10$ as well as the deterministic
Nose-Hoover thermostat is  shown   (lower figure) in
Fig.(\ref{4.2:fpujvsn}). This compares the RG prediction 
($\alpha=1/3$) and the old MCT
prediction ($\alpha=2/5$) for systems with these different baths and
bath parameters. 
As can be seen in the figure, an asymptotic exponent of $1/3$ is
attained for {\it all\/} these systems, whereas the apparent exponents
for smaller $N$ depend on system parameters.  It is
possible to understand the deviation of the apparent exponent from
$1/3$ for small system sizes.  As shown in refn.~\cite{LLP03}, if
the damping constant for the Langevin baths is very large or small,
there is a large `contact resistance' at the boundaries of the
chain. The current only depends weakly on $N$, resulting in an
apparent $\alpha > 1/3$ (similar considerations apply to Nose-Hoover
baths). This is confirmed by the plots in the figures:
the plot for $\gamma=2$ reaches the asymptotic limit fastest, whereas
$\gamma = 0.4,10$ have apparent exponents closer to $0.4$ for small
$N$.

Thus the simulations of Mai \etal~ seem to give good evidence for
$\alpha=1/3$ in the quartic  FPU model and hence gives support for
the idea of a single universality class. A discussion on these results
is contained in refns.~\cite{delfinicom08,dharrep08}. We note that the
new prediction of $\alpha=1/2$ from MCT appears to be even harder to
verify from simulations.

\begin{figure}
\begin{center}
\includegraphics[width=4.5in]{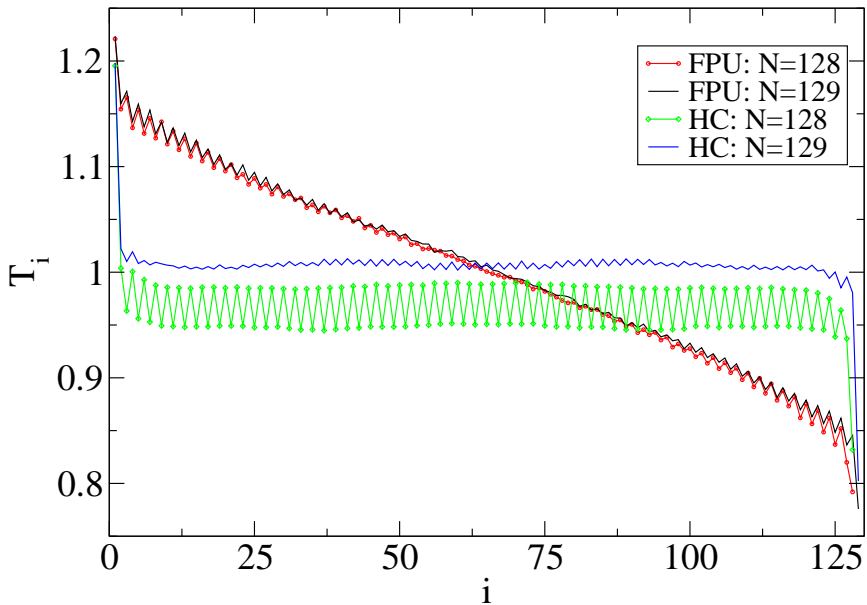}
\caption{Temperature profiles for alternate mass chains with FPU
  interactions ($k_2=1, k_4=1$) compared to those with purely harmonic
  interactions ($k_2=1$). Results for chains of even ($N=128$) and odd lengths
($N=129$)  are shown. }
\label{4.2:altmassTpr}
\end{center}
\end{figure}

\begin{figure}
\begin{center}
\includegraphics[width=4.5in]{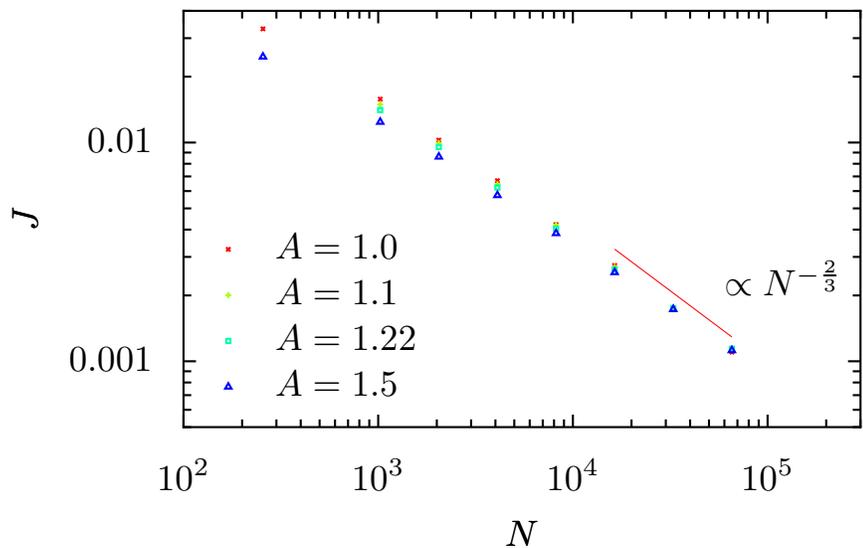}
\caption{Plot of the heat current $J$ versus
system size in the alternate mass FPU chain for different values of
the mass ratio $A=1.0,1.1,1.22$ and $1.5$ (from \cite{dharsaito08}). }
\label{4.2:fpualt_A}
\end{center}
\end{figure} 

\begin{figure}
\begin{center}
\vspace{1cm}
\includegraphics[width=4.5in]{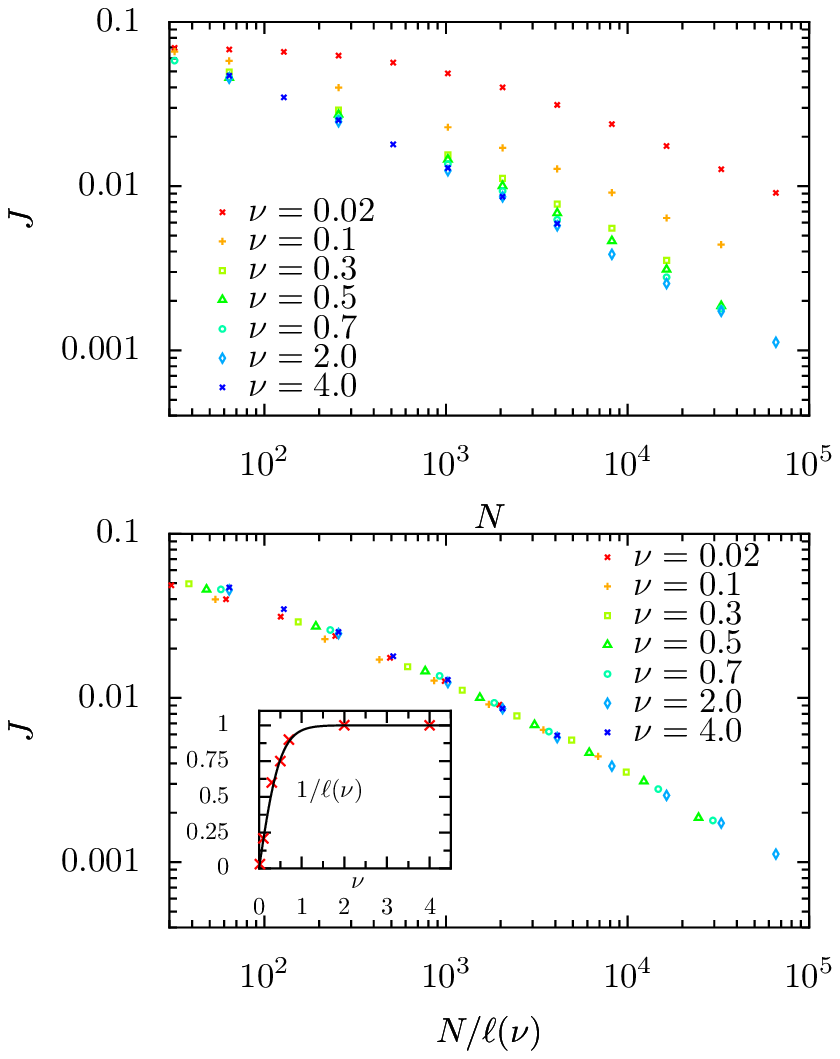}
\caption{Plot of the heat current $J$ versus
system size for the alternate mass FPU chain, for different values of
$\nu \equiv k_4$ and with the mass ratio $A=1.5$ (a). Fig (b) shows the same data
plotted with a scaled $x$-axis (from \cite{dharsaito08}).   
}
\label{4.2:fpualt_nu}
\end{center}
\end{figure}

{\bf Alternate mass FPU chain}: Further support for the value
$\alpha=1/3$ in the FPU system and its universality comes from recent
simulations of an alternate mass FPU chain \cite{dharsaito08}. In this
model one 
considers a chain with masses of particles at all even numbered sites
being $m_1$ and at odd numbered sites being $m_2$ with their ratio
being $A=m_1/m_2$. This system was first studied in \cite{mai07} where
it was noticed that the temperature profile showed peculiar
oscillations whose amplitude seemed to decay as $N^{-1/2}$ with system
size and scale linearly with the applied temperature difference. This
can be seen in Fig.~(\ref{4.2:altmassTpr}). At the hotter end the lighter
particles have a higher kinetic temperature, while at the colder end,
the heavier particles are hotter. It was pointed out in
refn.~\cite{mai07} that the temperature oscillations make it difficult
to define a local equilibrium temperature even at a coarse grained
level. Temperature oscillations can  in fact be seen even in an ordered binary
mass harmonic
chain  but there seem to be some significant 
differences. In the harmonic case, there is a big difference between    
the case where $N$ is even and that where $N$ is odd. This can be
seen in Fig.~(\ref{4.2:altmassTpr}) where we plot the temperature profiles for chains
of length $N=128$ and $N=129$ for both the FPU chain and the harmonic
chain with $m_1=0.8$ and $m_2=1.2$. For the
harmonic case the oscillations for even $N$ 
are large and do not decrease with system size while those for odd $N$ 
 decrease with system size. In the FPU case there is not much
 difference between a chain with odd or even number of particles. Also
 for the harmonic case, in
the bulk, the heavier particles are always hotter.

However, even though there appears to be a problem in defining a local
temperature, one can still measure the system size dependence of the
current in this system and this was done in refn.~
\cite{dharsaito08}. They studied the size dependence of the current for
different values of the mass ratio $A$, keeping the average mass
$(m_1+m_2)/2$ constant. Remarkably it was found that at large enough
system sizes the currents for different $A$ all tend to converge to
the same value. This can be seen in Fig.~(\ref{4.2:fpualt_A}) where one
can see that the exponent is again as that at $A=1$, \ie $\alpha
\approx 0.33$. In this paper the authors next took a fixed value of
$A=1.5$ and studied the effect of changing the interparticle
interaction strength (denoted as $\nu=k_4$ in the paper). These
results for current for different system sizes are shown in
Fig.~(\ref{4.2:fpualt_nu}). For small system sizes, 
one sees a flat region which is expected since for system sizes much 
smaller than the phonon-phonon scattering length scale, the system
will behave like a harmonic chain. The scattering length should be
larger for smaller $\nu$ and this can be seen in the plot. At larger
system sizes, all the curves tend to show the same decay
coefficient with $\alpha\approx 0.33$. A nice collapse of the data was
obtained by scaling the system size by a length factor $\ell (\nu)$
and this is shown in Fig.~(\ref{4.2:fpualt_nu}b). The $\nu$-dependence of       
the length parameter seems to be given by $\ell (\nu) =1/\tanh (2
\nu)$. A surprising point is that for any fixed system size, the value
of the current saturates to a constant non-zero value as $\nu \to \infty$.

\begin{figure}
\begin{center}
\vspace{1cm}
\includegraphics[width=4.25in]{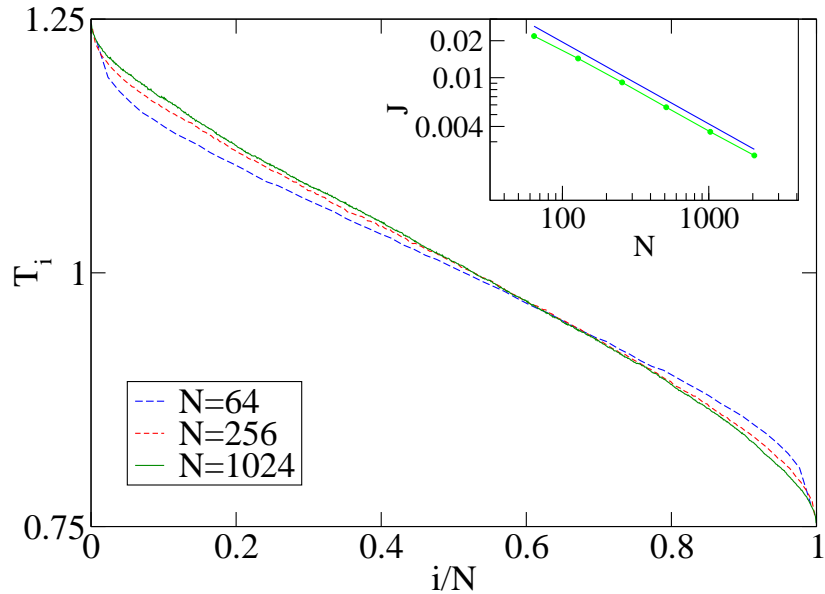}
\caption{Plot of the temperature profile in the double-well potential model for
  different system sizes. 
The inset shows the system size dependence of the current and the
  straight line drawn has slope $-2/3$. Error bars are much smaller
  than symbol sizes. The
  parameter values for the figure are $k_2=-1, k_4=1, T_L=1.25,
  T_R=0.75$ and bath coupling $\gamma=1$.
}
\label{4.2:DWfig}
\end{center}
\end{figure}
{\bf Double-well potential}: It is interesting to consider
simulation results obtained for the FPU interaction potential
$U(x)=k_2x^2/2+k_4 x^4/4$ with {\it{negative}} $k_2$ in which case we
have a double-well potential. This case was first studied in
\cite{giardina00} which had  initially reported a  finite
conductivity for this model but later it was found to have a power law
divergence \cite{LLP03}. Here we present some new simulation results for this model which
in fact show that this model exhibits a fast convergence to the
asymptotic regime  with exponent $\alpha=1/3$. The parameter values
$k_2=-1, k_4=1$ and $T_L=1.25, T_R=0.75$ were considered. 
The temperature profiles in this system for different system sizes are
shown in Fig.~(\ref{4.2:DWfig}) and are similar to the FPU profiles
[Fig.~(\ref{4.2:fpuTprof})] except that the boundary jumps are smaller. The inset of the figure 
shows a   plot of $J$ as a function of $N$. 
One can see that  that by around $N=512$ the 
curve has reached the expected asymptotic slope corresponding to $\alpha=1/3$.
Thus we again see evidence for $\alpha=1/3$.

{\bf Discussion:} In the absence of a rigorous proof it is fair to say
that the question 
of universality of the heat conduction exponent and its precise value,
in momentum conserving interacting systems in one dimensions, is still
an open problem. This is especially more so since all the analytic
methods and the exact result discussed in Sec.(\ref{sec:THEORY}) rely on use of the
Green-Kubo formula for a closed system. As pointed out  in sec~(\ref{sec:GK}) the
use of this formula and interpretation in systems with
anomalous transport is not clear. The simulation results that we have
presented in this review strongly suggest a single universality class,
with $\alpha=1/3$, 
for momentum conserving interacting systems in $1D$. However
one should 
probably bear in mind that in simulations, one can never really be sure that the 
asymptotic system size limit has been reached. It is possible for exponents
to change in unexpected ways when one goes to larger system sizes.

\subsubsection{Momentum non-conserving models}
\label{sec:momnoncons}
We will now look at heat conduction in one dimensional chains where
the particles experience, in addition to interparticle interactions,
also external potentials which physically can be thought of as arising
from interactions with a substrate.  
One of the first verification of Fourier's law in computer simulations
was obtained by Casati \etal~ in the so-called ding-a-ling model \cite{casati84,casati86}. In
this model one 
considers a system of equal mass hard point particles which interact
through  elastic 
collisions and where alternate particles are pinned by  harmonic
springs placed at fixed distances. The particles in between the pinned
ones move freely. Clearly momentum is not conserved and the authors,
by studying system sizes upto $N=20$, found evidence for 
diffusive behaviour.  They calculated the thermal conductivity using
both nonequilibrium simulations as well as using the Kubo formula and
found good agreement between the two. We note that the system
sizes studied in this paper are clearly too small to arrive at definite
conclusions. Larger system sizes with the same parameter values were
studied later by Mimnagh and Ballentine \cite{mimnagh97}. They found that in fact the
conductivity again started to increase as one went to larger
sizes. Finally though, at system sizes $N \sim 400$,  the conductivity
again saturated at a new value which is much higher than that obtained
in \cite{casati84}. This example nicely demonstrates the need for
caution in drawing conclusions from small size data (also see
discussion in \cite{LLP03} on these results).

Since the work of Casati \etal~, a number of papers have looked at
heat conduction in various momentum non-conserving models in one dimension and
have all found evidence for the validity of Fourier's law.
A model similar to the ding-a-ling is the 
ding-dong model and has all particles connected to fixed harmonic
springs. This was studied by Prosen and Robnik  and
also shows Fourier behaviour \cite{prosen92}.  
One of the first papers to recognize the fact that momentum
non-conservation is a necessary condition to get finite heat
conductivity in one-dimensional systems is that of Hu \etal~ 
\cite{hu98}. From their simulations with  various forms of Hamiltonians
including, both a harmonic  
interparticle potential $U(x)$ and a periodic onsite potential 
of the Frenkel-Kontorva form [$V(x) \sim \cos (a x)$], they found that
the presence of  an external 
potential typically led to a finite conductivity.   
The Frenkel-Kontorva model was also studied in \cite{tsironis99} who
arrived at similar conclusions. 
A study of the $\phi^4$ model [where $U(x)=k_2 x^2/2,~V(x)=\lambda
x^4/4$]  by Hu \etal~\cite{hu00}, again led to the conclusion  of a finite
conductivity. This study also 
emphasized the following point. Nonlinear integrable models usually
give a  flat temperature profile and making them non-integrable leads
to a temperature gradient. However this in itself is not a sufficient
condition to give a finite conductivity.  

\begin{figure}
\begin{center}
\epsfxsize=12.cm \epsfbox{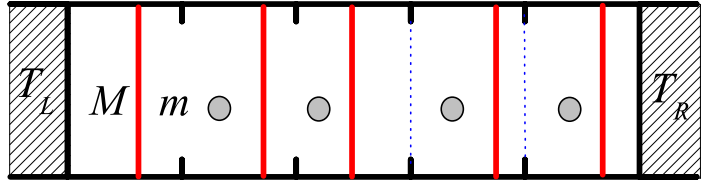}
\vspace{0cm}
\caption {The geometry of the momentum-nonconserving, alternate-mass
  hardcore model, studied in \cite{li04a}.  The elementary cell (indicated
by two dotted lines) has unit length $l=1$. The bars have mass $m_1$, and
the particles have mass $m_2$. The two heat 
baths at temperatures $T_L$ and $T_R$ are indicated. }
\label{4:lifig}
\end{center}
\end{figure}
Another nice simulation demonstrating  the role of an external
potential in giving rise to Fourier behaviour is that on the binary
mass hard particle gas \cite{li04a}. Momentum non-conservation is
ensured by confining the motion of alternate particles inside finite
cells while allowing them to interact, through elastic collisions with
neighbors [shown in Fig.~(\ref{4:lifig})]. The nonequilibrium simulations in this paper
with Maxwell heat baths (with $N \sim 512 $)  convincingly shows the validity of Fourier's
law and also the presence of local thermal equilibrium. Secondly, 
equilibrium simulations were also performed to compute current-current
correlation functions, and, using the Green-Kubo formula a value of
$\kappa$ close to the nonequilibrium result was obtained. It is worth
noting that this model   has zero Lyapunov exponent and thus is non-chaotic. 
A related study is that in refn.~\cite{gendelman04} who studied heat
conduction in a gas of hard rods placed in a periodic potential.

{\bf Results for the $\phi^4$ model}: We will describe in some more
details work on the $\phi^4$ model [$U(x)=k_2 x^2/2,~V(x)=\lambda
x^4/4$] which appears to be one of the most well-studied of the
momentum non-conserving models and where some analytic results have
also been obtained.
Heat conduction in the  $\phi^4$ model was first studied by
Aoki and Kusnezov \cite{aoki00,aoki02}  who performed both nonequilibrium measurements as
well as Green-Kubo based equilibrium measurements. Studying system
sizes upto $N=8000$ they concluded that this system had a finite
conductivity and  Fourier's law was valid. The value of $\kappa$
obtained from the nonequilibrium measurements and from the Green-Kubo
formula were again shown to be in good agreement. The authors also
numerically obtained  the temperature dependence of $\kappa$ and found
$\kappa(T) \sim T^{-1.35}$. 
A number of other papers have performed simulations of the $\phi^4$
model and studied various aspects such as the spreading of localized
disturbances \cite{hu00} and the dependence of thermal
conductivity on temperature \cite{li07}.
The model was studied analytically by
Lefevere and Schenkel \cite{lefevere06} and later by Aoki \etal~\cite{aoki06} using
a Peierls-Boltzmann kind of approach for the case of weak
anharmonicity and they too obtained a finite conductivity. They
however obtained a temperature dependence  $\kappa \sim 1/T^2$ and
this is probably the correct low temperature (corresponding to
weak anharmonicity) behaviour, since kinetic theory is expected to be
reliable in this regime. Direct nonequilibrium simulations  in
\cite{aoki06} infact found reasonable agreement with the predictions
from kinetic theory, at low temperatures. 
The study in \cite{li07} however finds a somewhat different
temperature dependence at low temperatures  ($\kappa
\sim 1/T^{1.56}$). 
We note that a scaling property of $\kappa(T,\lambda)$, to be discussed
later in Sec.~(\ref{sec:disint}), implies that $\kappa =\kappa (\lambda T)$.

\begin{figure}
\begin{center}
\vspace{1cm}
\includegraphics[width=5in]{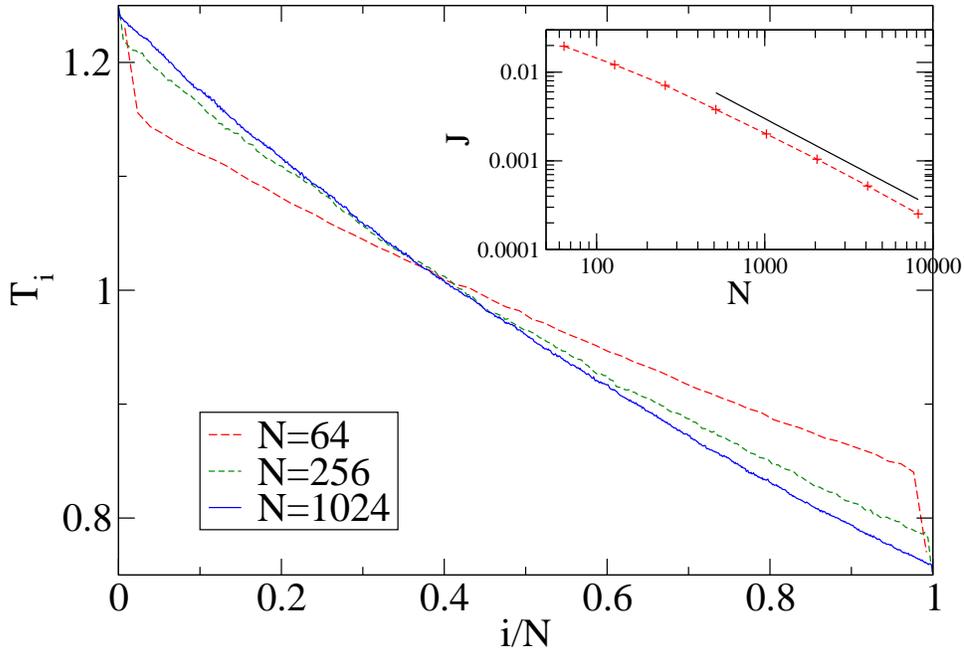}
\caption{Plot of the temperature profile in the $\phi^4$ model for
  different system sizes. 
The inset shows the system size dependence of the current and the
  straight line drawn has slope $-1$. Error bars are much smaller than
  symbol sizes. The
  parameter values for the figure are $k_2=1, \lambda=1, T_L=1.25,
  T_R=0.75$ and bath coupling $\gamma=1$.
}
\label{4.2:phi4}
\end{center}
\end{figure}

In Fig.~(\ref{4.2:phi4}) we show typical plots of the temperature
profile in the $\phi^4$ chain. The inset in the figure shows the $1/N$
dependence for the current. These simulations were performed using
white noise Langevin dynamics using the velocity-Verlet algorithm and
for the largest system size ($N=8192$) required $\sim 2 \times 10^9$
time steps with $\Delta t=0.0025$, to equilibrate.

Before concluding this section we  mention the results on a class of rotor
models studied by Giardina \etal~ \cite{giardina00} and Gendelmann
\etal~ \cite{gendelman00}. These models were originally proposed as examples of momentum
conserving systems which gave Fourier behaviour. The interparticle
potential is taken to be $U(x)=1-\cos (x)$ and the onsite term
$V(x)=0$. Nonequilibrium and equilibrium (Green-Kubo based) simulations
in \cite{giardina00,gendelman00} both indicated that this model gave a
finite conductivity.  The paper by \cite{gendelman00} also reports a
phase transition from infinite  to finite conductivity as a function
of temperature. Given that these simulations are restricted to
relatively small sizes (upto $N=2400$), one suspects that this is probably a cross-over
effect. Simulations for larger systems in \cite{yang03,yang05,gendelman05} indeed 
suggest that there may not be any true transitions and that, at all
temperatures the asymptotic beaviour is Fourier-like. An analytic study
of the rotor model using self-consistent reservoirs ( with vanishingly small coupling to
interior points ) has also claimed a transition \cite{pereira06}. 
A similar claim of possible transitions from finite to diverging
conductivity in other momentum non-conserving models such as the
Frenkel-Kontorva and $\phi^4$ model has been made in \cite{savin03}.

The fact that a momentum conserving model gives finite conductivity is
at first surprising. 
However given the form of the interparticle potential in the rotor model it is 
probably more physical to think of this model as an angular
momentum conserving model rather than linear momentum conserving
one. Thus it seems more natural to think of the position variables
$x_l$ as transverse angular degrees of freedom. In this case one
expects different hydrodynamic equations (see for example
\cite{chaikin95}) and the Fourier behaviour observed is then not surprising. 
One can also think of the rotor model as the classical limit (large
spin) of quantum lattice spin chain models which also are
momentum-nonconserving.

{\bf Quantum mechanical models}: The study of nonequilibrium steady
states of interacting quantum systems by simulations is an important and difficult problem. 
There have been a few attempts at addressing this issue, and we will
summarize these. The first set of papers were by Saito \etal~
\cite{saito96,saito03}, who used the master equation
approach to connect different temperature reservoirs to a quantum spin
chain. One interesting result was that a 
temperature gradient was formed for the case where the model was
non-integrable, while a flat profile was obtained for an integrable
model. A study of the current-curent correlator yielded a power law
decay $C(t) \sim t^{-1.5}$ implying a finite conductivity
\cite{saito03}.      

In another interesting work, Mej\'ia-Monasterio \etal~\cite{monasterio05} have
devised what they call a quantum stochastic reservoir. Using this they
have performed nonequilibrium simulations, again of a quantum spin
chain. They also observe a temperature gradient for the non-integrable
model and a flat profile for the integrable model. Further they
measured the nonequilibrium steady state current for different sytem sizes and found
a $J \sim N^{-1}$ dependence for the non-integrable case and  $J \sim
N^0$ for the integrable case.   

\section{Systems with disorder and interactions}
\label{sec:disint}

As discussed in sec.~(\ref{sec:disharmlat})  localization of
eigenfunctions or  of normal modes 
strongly affects transport in materials containing random impurities. 
In electronic systems localization
has its strongest effect in one dimensions where 
any finite disorder makes all eigenstates localized 
and one has an insulator. The presence of
inelastic scattering, such as  is caused by electron-phonon interactions, 
leads to hopping of electrons between localized states and
gives rise to a finite conductivity. The question as to whether
electron-electron interactions  lead to a similar effect
has attracted much attention recently
 but is still not fully understood
 \cite{LLA78,GMP05,NGD91,basko06,vadim07,pikovsky08,kopidakis08}.  The   
main interest is to understand the transition, from an insulating state
governed by the physics of Anderson localization, to a conducting state
as one increases interactions.  
One can ask the same question in the context of heat
conduction by phonons
and consider the effect that phonon-phonon interactions have on localization.  
Here we will mainly discuss the  effect of anharmonicities on the steady state 
transport of heat through a chain of oscillators with random masses .
The effect of interactions between phonons on localization 
caused by disorder has also been investigated by looking at the
spreading of wave packets
\cite{bourbonnais90,snyder06,pikovsky08,kopidakis08} and we will
briefly discuss these results at the end of this section. 

An early work on steady state heat conduction in
disordered anharmonic systems  is that of
Payton, Rich and Visscher \cite{payton67} who studied mass-disordered
lattices in the presence of cubic and quartic interparticle anharmonicities. They performed
nonequilibrium simulations with stochastic baths in one and two
dimensions. Their main conclusion was that
in most cases interactions (interparticle anharmonicity) seemed to greatly enhance the conductivity of
the system (except for the case of very weak disorder). We note that,
at that time simulations were restricted to small sizes and it was
wrongly assumed by the authors that the disordered harmonic lattices in one
and two dimensions, as well as the anharmonic ones, had 
finite thermal conductivities. The system size dependence was not
studied systematically. Similarly a study by Poetzsch and
Bottger \cite{poetzsch94} for a two dimensional lattice system found
that, while quartic anharmonicity enhances the 
conductivity of the disordered system, cubic anharmonicity reduces
it. Again this study was restricted to small system sizes and 
assumed that the conductivity is finite.

The first systematic study of the joint effects of anharmoncity and
disorder on the system-size dependence of heat current was by Li \etal~
\cite{baowenli01}.  They studied the mass-disordered FPU chain using
Nose-Hoover nonequilibrium simulations. 
Their  conclusion was  that this model
showed a  transition, from a 
Fourier like scaling $J \sim N^{-1}$ at low temperatures, to a pure FPU
like behaviour  with $J \sim N^{-0.57}$ at high temperatures.
A more recent simulation of the same model by Dhar and Saito
\cite{dharsaito08} suggests that this
conclusion may be incorrect. What Li \etal observe is probably a
cross-over effect and  there is really no true transition
in transport properties. It appears that {\it disorder becomes irrelevant}
as far as the value of the exponent $\alpha$ is concerned. 
We will
here present the latest simulation results of the disordered FPU
chain as well as results from the study of the disordered $\phi^4$ lattice,
where similar conclusions are reached \cite{dharleb08}. 
Temperature driven phase transitions in one dimensional heat
conduction have also been reported in some other models. However as has
been pointed out in refn.~\cite{yang03} these are probably cross-over effects
and there is no true transition in these models.

The general form of the  Hamiltonian that has been  studied by various
people is, in the one-dimensional case, given by:     
\bea
H &=& \sum_{l=1,N} [ \f{p_l^2}{2 m_l} + k_o \f{x_l^2}{2} + 
\lambda \f{x_l^4}{4} ]\nn \\
&+&  \sum_{l=1,N+1}
[ ~ k \f{(x_l-x_{l-1})^2}{2} +\nu \f{(x_l-x_{l-1})^4}{4} ~] \label{disintham}
\eea
with fixed boundary conditions  $x_0=x_{N+1}=0$.
The masses $\{m_l\}$ are chosen independently from some 
distribution $p(m)$, {\emph{e.g.}} one uniform in the interval
$(\bar{m}-\Delta,\bar{m}+\Delta)$ or a binary distribution given by
$P(m)=\delta [ m-(\bar{m}-\Delta)]/2   +\delta [ m-(\bar{m}+\Delta)]/2$.   
The chain is connected at its ends to two heat baths at temperatures 
$T_L$ and $T_R$ 
respectively. Here we will mostly consider white noise reservoirs, but
will also give some results with Nose-Hoover baths.
The equations of motion of the chain are then given by:
\bea
\label{eq: 2}
m_l \ddot{x}_l&=&- k_o x_l - \l x_l^3-k (2 x_l-x_{l-1}-x_{l+1}) \nn \\
&-& \nu [ (x_l-x_{l-1})^3 + (x_l-x_{l+1})^3 ]-\g_l \dot{x}_l 
+\n_l~,  
\eea
with $\n_l=\n_L \delta_{l,1}+\n_R \delta_{l,N},~\g_l=\g (\delta_{l,1}+
\delta_{l,N})$, 
 and where the Gaussian noise terms satisfy the  fluctuation dissipation 
relations $\la \eta_L(t) \eta_L(t') \ra = 2 \g k_B T_L \delta(t-t')$,
$\la \eta_R(t) \eta_R(t') \ra = 2 \g k_B T_R \delta(t-t')$.

Note that Eq.(\ref{eq: 2}) is invariant under the transformation $T_{L,R} \to s T_{L,R}$,
$\{ x_l\} \to \{ s^{1/2} x_l \}$ and $(\l,\nu) \to  (\l,\nu)/s$.  This implies
the scaling relation $ J (sT_L,sT_R,\l,\nu)\ra = s 
J (T_L,T_R,s\l,s \nu)$. For the conductivity $\kappa$ this implies
$\kappa= \kappa(\nu T, \l T)$. Thus the effect of changing temperatures 
can be equivalently studied by changing anharmonicity. 
We will first discuss the unpinned (momentum conserving) case and then
the pinned (momentum non-conserving) case.

{\bf Disordered FPU chain}: This corresponds to  taking 
$k_o=\lambda=0$ in the  Hamiltonian in Eq.~(\ref{disintham}), and is the case studied by
\cite{baowenli01} and in \cite{dharsaito08}. There are two important
parameters in the problem, namely the disorder strength given by
$\Delta$ and the anharmonicity given by $\nu$.   
Let us consider the two limiting cases, of the disordered harmonic
chain ($\nu=0, \Delta \neq 0$),  and of the ordered FPU chain ($\nu \neq 0,
\Delta=0$). For the former, with fixed
boundary conditions it is expected that $\alpha = -1/2$, while for the
ordered FPU chain one expects $\alpha = 1/3$. In the presence of both
disorder and interactions a possible scenario is that for strong
disorder one gets $\alpha = -1/2$ while with strong interactions, one
gets $\alpha=1/3$ and there is a phase transition between the two
behaviours as we change parameters. The numerical results that we will
discuss, suggest that there is no such transition.
Note that in both the limiting cases, the low frequency long
wavelength  modes are believed to play an important  role in transport.  

\begin{figure}
\begin{center}
\includegraphics[width=4.5in]{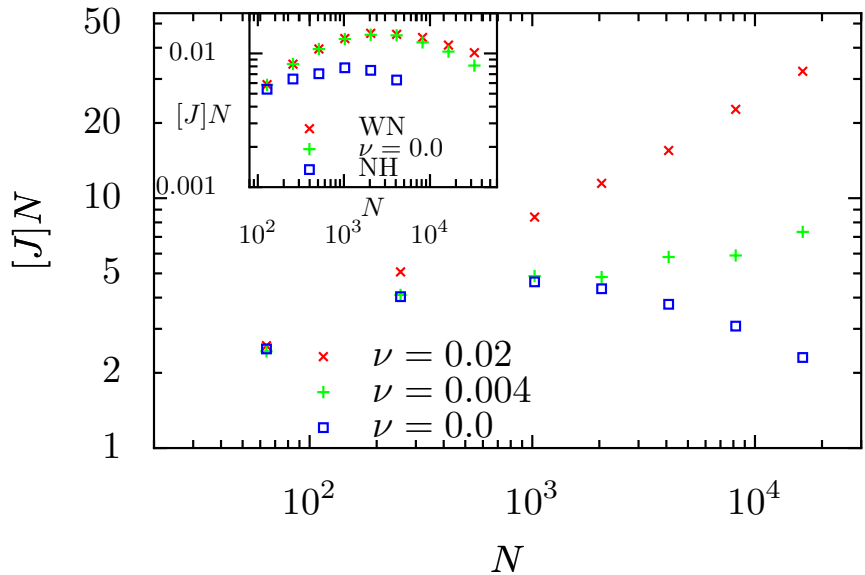}
\caption{Plot of heat current versus system size, for the disordered
FPU  chain, for different values of $\nu$. The data in the inset 
corresponds to parameters as in \cite{baowenli01}, namely $(T_L, T_R )=(0.001,0.0005)$ with 
Gaussian white noise bath for $\nu=1$ (WN) and $\nu=0$, and 
Nose-Hoover bath (NH) for $\nu=1$ (from \cite{dharsaito08}).}
\label{5:disfpu_jvsn}
\end{center}
\end{figure}

\begin{figure}
\begin{center}
\includegraphics[width=4.5in]{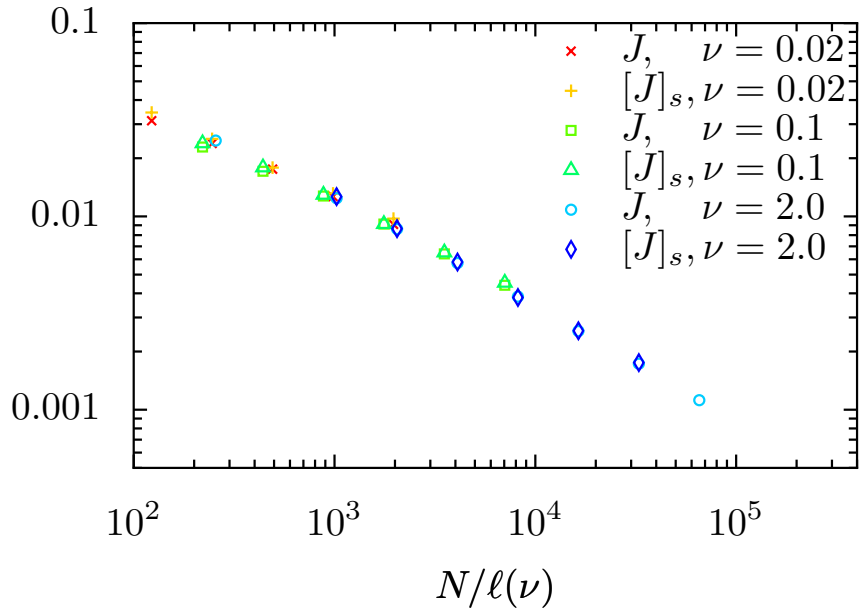}
\caption{Plot of scaled heat current $[J]_s$ for the disordered FPU
  chain and the current $J$ for  the ordered  chain, for
  different values of $\nu$.
 The $x-$axis is  scaled as in Fig.~(\ref{4.2:fpualt_nu})  (from \cite{dharsaito08}).
}
\label{5:disfpu_scal}
\end{center}
\end{figure}

The simulations in \cite{dharsaito08} looked at the case of  binary
mass distribution  with $\bar{m}=1, \Delta = 0.2$ and different values
of the interaction strength $\nu=0.004, 0.02, 0.1, 2.0$. 
Averages were taken over $50-100$ samples  for $N < 1024$, $10$ samples for
$N=1024-16384$, and $2$ samples for $N=32768$ and $65536$. 
In Fig.~(\ref{5:disfpu_jvsn}) the results of simulations for the
disorder averaged current $[J]$ 
for $\nu=0.004$, $0.02$ and $\nu=0.0$ are shown. For small values of
$\nu$ one sees  that, at small system 
sizes the current value is close to the $\nu=0$ value. As expected
one has to go to large system sizes to see the effect of the weak
anharmonicity. At sufficiently large $N$  the same system size
dependence of $J$ is obtained as that for the ordered FPU chain,
namely with $\alpha=1/3$. The authors in \cite{dharsaito08} then show
that by scaling the current by appropriate factors,
the data for the disordered case can be made to collapse
on to the binary-mass ordered case. This is shown in
Fig.~(\ref{5:disfpu_scal}) (for $\nu=0.02,0.1,2.0$).
Thus these results show that the asymptotic
power law dependence of the current is always dominated by
anharmonicity and there seems to be no transition. Disorder only
decreases the overall conductance of a sample. 

The authors of \cite{dharsaito08} have also investigated the parameter
range studied in \cite{baowenli01} and explained 
the reasons which led to the 
erroneous conclusions in \cite{baowenli01}, of a transition in conducting
properties at low temperatures (or equivalently small anharmonicity).
In fact this can be understood even from the data for  $[J]N$ in
Fig.~(\ref{5:disfpu_jvsn}) for 
$\nu=0.004$. We see that at around $N\sim 1000-2000$ the data seems to
flatten and if one had just looked at data in this
range, as was done by \cite{baowenli01}, one would conclude that 
Fourier's law is valid. However 
the behaviour changes drastically when one looks at larger system
sizes and one again gets the usual FPU behaviour. 
The inset of Fig.~(\ref{5:disfpu_jvsn}) shows results for parameters as in
\cite{baowenli01} but for much larger system sizes. This case
corresponds to a much smaller value of 
$\nu$ and so it is expected that it will follow the $\nu=0$ curve till
very long length scales and this is clearly seen. However, at around $N=16384$, 
there is a tendency
for the curve to turn up and it can be expected that the same asymptotic
behaviour to eventually show up.  While a transition cannot be ruled
out at even lower temperatures and with stronger disorder, this seems unlikely. Also, if there is
such a transition, it should probably be to a disordered phase with $[J]\sim
N^{-3/2}$.

It is interesting to consider the temperature dependence of
conductivity in the disordered FPU chain.  
The scaling property of the current, mentioned
earlier [after Eq.~(\ref{eq: 2})], implies that the thermal
conductivity  has the form $\kappa=\kappa(\nu T)$.  For small
anharmonicity ($\nu << 1$), the earlier results for the ordered
alternate mass FPU chain imply that at large
system sizes $\kappa \sim N^{1/3}/\nu^{2/3}$ and from the scaling property
this immediately gives $\kappa \sim 1/T^{2/3}$ at low
temperatures. However at small system sizes  [$ N << \ell (\nu)$], we
expect the system to behave like a harmonic system with $\kappa
\sim T^0$. At high temperatures the conductivity will saturate to a
constant value. 
Experimentally, the temperature
dependence of the thermal conductivity may be easier to measure
and one can verify if this is unaffected by disorder [see, for example
sec.~(\ref{sec:expts})].

{\bf Disordered $\phi^4$ chain}: Let us now look at the case where the
particles are subjected to an external pinning 
potential in addition to nearest neighbor harmonic interactions. We
will consider the anharmonicity to be an onsite quartic term (thus
$\lambda > 0, \nu=0$, also $k_o,k > 0$ ) in which
case this corresponds to the discrete $\phi^4$ model. Pinning greatly 
enhances the  difference between heat transport in a random chain 
with and without anharmonicity 
and thus is a good testing ground for the effect of anharmonicity on 
localization.  This model is also closer in spirit to charge transport by hopping in
random media.
Again let us look at the two limiting cases. In the case with a
pinning potential at all sites the disordered case 
($\lambda=0, \Delta \neq0$) gives $J \sim e^{-c N}$. For $\Delta = 0$ and
$\lambda \neq 0$ we have seen from sec.~(\ref{sec:momnoncons}) that one expects Fourier's
law to be valid and so $\alpha=0$. 

\begin{figure}
\begin{center}
\includegraphics[width=4.25in]{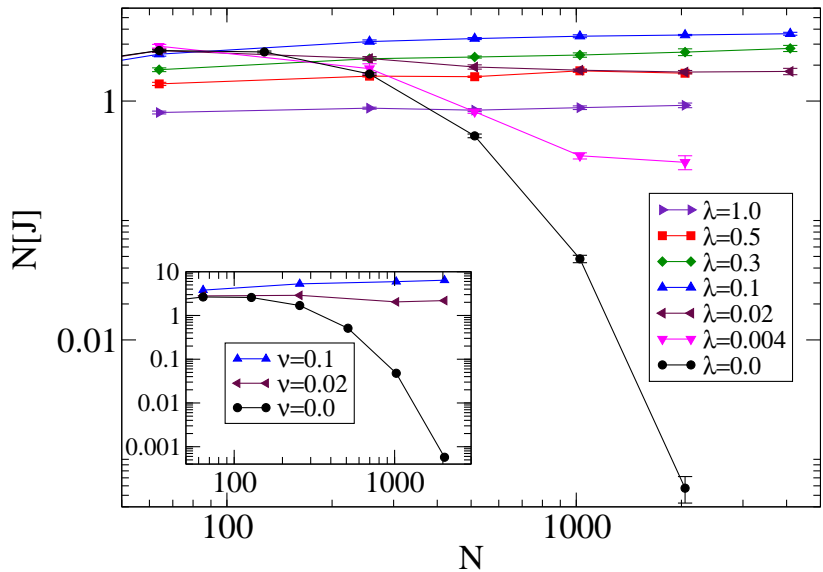}
\caption{Plot of $[J] N$
versus $N$ for the disordered $\phi^4$ chain, for different values of
$\lambda$. The inset shows results   obtained for the case with
interparticle anharmonicity and onsite harmonic pinning (from \cite{dharleb08}). 
}   
\label{5:disphi4_jvsn}
\end{center}
\end{figure}

The case with parameters $k= k_o=1, \lambda > 0$  and a uniform
mass distribution with $\bar{m}=1.0$ and $\Delta=0.2$ was studied in
\cite{dharleb08}.   
In Fig.~(\ref{5:disphi4_jvsn}) the result of simulations for  
different values of anharmonicity  $\lambda=0.004-1.0$ is given. 
As can be seen from the data, there is a 
dramatic {\it{increase}} in the heat current on introduction of a small amount
of anharmonicity and the system-size dependence goes from exponential decay to
a $1/N$ dependence implying diffusive transport. For smaller $\lambda$
the diffusive regime sets in at larger length scales but as in the FPU
case, here too one finds that anharmonicity determines the system size
scaling and no transition is observed. 
A measure of the relative 
strengths of anharmonicity and disorder is obtained by looking at the 
ratio of the energy scales  $E_a= \l \la x^4 \ra/4$ and $E_d=
T \Delta/m $. For the given parameters one finds 
$\epsilon= E_a/E_d \approx 0.008$ for $\lambda =0.004$.
Unlike the FPU case, in this model, it does not seem that any simple
scaling of the data is possible.

Thus this study shows  that  introduction of a
small amount of 
 phonon-phonon interactions in the disordered harmonic chain  leads to
 diffusive energy transfer, {\emph {i.e.}}, the insulating chain becomes a normal heat 
conductor. How exactly this occurs is not clear. It is possible that
anharmonicity gives rise to extended states or leads to hopping of
energy between states which are now approximately localized
({\emph{i.e}} they are no longer exact normal modes, but have a small rate of energy
leakage to nearby modes) . There is no  evidence  of the existence of a finite
critical value of anharmonicity  required for this transition. 
For small values of anharmonicity 
it is necessary to go to larger system sizes to see the transition 
from insulating to diffusive. 
As in the FPU case, a transition to a localized phase
at a very small value of anharmonicity is possible and would be
difficult to observe in simulations, because 
equilibration times increase rapidly with decreasing $\lambda$.

An interesting question in this model is 
the limiting behavior of $\kappa (\lambda
T)$ for $(\lambda T)\to 0$. It turns out that the temperature profiles for
the disordered $\phi^4$ chain are qualitatively different from the
ordered chain and this means that the temperature
dependence of conductivity is different for the two cases. For the
ordered case, from kinetic theory one gets $\kappa \sim 1/(\lambda^2
T^2)$ for small $\lambda T$ \cite{aoki06}, while for the disordered
case \cite{dharleb08} found $\kappa \approx (\lambda T)^{1/2}$. For
the FPU chain on other hand, the ordered and disordered cases give
similar temperature profiles \cite{dharsaito08}.

There have been a number of studies on
disordered anharmonic chains which
have investigated the spreading of localized pulses of energy injected
into a system at zero temperature. 
The study by Bourbonnais and Maynard \cite{bourbonnais90} looked at FPU type of systems in
one and two dimensions and observed that anharmonicity destabilizes
the localized modes and the diffusion of pulses was found to be
anomalous. This seems to be consistent with the heat conduction
results on the disordered FPU chain. A similar zero temperature study
of the mass disordered FPU system was carried out by Snyder and
Kirkpatrick \cite{snyder06} who however found evidence for diffusive transport at
sufficiently strong anharmonicity. 
In the case of the $\phi^4$ and related models
there have been some extensive recent numerical studies and here the
conclusions are somewhat contradictory to the heat conduction results.
The  spreading of localized energy pulses has been reported to be
sub-diffusive  in \cite{pikovsky08} while  \cite{kopidakis08} reports  absence of
diffusion. The study in refn.~\cite{kopidakis08} offers a picture of
spreading of an initially  localized energy wavepacket to a limiting profile as
taking place through nonlinearity induced coupling between the
localized modes.   
All these studies suggests that  the  behaviour of a heat pulse at zero
temperature and that at finite temperature can be very different. 
Indeed as pointed out  nicely in \cite{zhao06}, it is necessary to
look at appropriate spatiotemporal correlation functions of closed
systems at finite temperatures in order to understand diffusion in the open system.
This of course is also what one effectively does in the Green-Kubo approach.

Finally we note an interesting related problem that was studied by Rich
and Visscher \cite{rich75}.
They considered  heat conduction in a disordered Harmonic chain with
self-consistent reservoirs. Since self-consistent reservoirs can be
roughly though of as some sort of nonlinearity leading to incoherent  scattering
of phonons, this problem has some similarity with that considered in
this section. Based on exact numerical calculations on small chains,
their main conclusion was that the presence of self-consistent
reservoirs leads to a finite conductivity for  chains with both
free and fixed boundary conditions (and no bulk pinning). The
self-consistent reservoirs makes the model momentum non-conserving so
this is consistent with the results of the disordered $\phi^4$ chain
presented here. A very interesting conjecture made in this paper is
that a finite conductivity will be obtained if the limits $N\to \infty$
first, and then coupling to self-consistent reservoirs $\to 0$ are taken. 
A recent paper \cite{bernardin08} has studied heat conduction in
a disordered harmonic chain with an energy conserving stochastic
dynamics and has obtained  rigorous results which indicate a finite thermal
conductivity of the system.

\section{Interacting systems in two dimensions}
\label{sec:intsys2D}
We have seen that, in the one-dimensional case, it is usually quite difficult to
obtain the asymptotic system size dependence of the current. In order
to get the correct exponent requires one to go to large system sizes
and at some point the sizes required are beyond current computational
capabilities. Of course a combination of simulations and results from
analytic work gives one some confidence about the results obtained so
far. In the case of higher dimensional  systems naturally one can expect
the same computational difficulties and in fact here they become more
pronounced  since the number of particles is now $L^d$ where $L$ is
the linear size and $d$ the dimensionality. The good news is that there is
general agreement on the system-size dependence of conductivity from
different analytical methods. Both MCT \cite{LLP03} and the
hydrodynamics approach \cite{narayan02} predict that for a
momentum conserving system, the thermal conductivity  diverges
logarithmically with system size in $2D$ and is finite in $3D$.
In the presence of pinning all theories predict a finite
conductivity in all dimensions. 

There are  few simulation studies in higher dimensional systems and we
summarize the main results obtained so far. 
Early studies of heat conduction were mainly interested in finding the
temperature dependence of thermal conductivity and {\it assumed } that
this was finite \cite{payton67,mountain83,michalski92}. One of the first paper to study system size
dependence was probably that by Jackson and Mistriotis
\cite{jackson89}. They studied the diatomic Toda lattice and concluded
that the  thermal conductivity was finite for mass ratio greater than
a critical value and diverged otherwise.

More extensive studies on the system size dependence were
made by Lippi and Livi \cite{lippi00} for an 
oscillator system in two dimensions with vector displacements and
interparticle interactions. 
 They considered a $L_x \times L_y$
lattice with the following Hamiltonian:
\bea
H= \sum_{i=1}^{L_x} \sum_{j=1}^{L_y}
   \f{|{\bf{p}}_{ij}|^2}{2 m}  +U(|{\bf{x}}_{i+1,j}-{\bf{
    x}}_{i j}|) + U(|{\bf{x}}_{i,j+1}-{\bf{x}}_{i j}|)~, \label{6:ham2d} 
\eea
 where ${\bf x}_{ij}$ denotes the vector displacement (taken to
 be two-dimensional vectors) of a particle at
 lattice site $(i,j)$ where $i=1,2..L_x$ and $j=1,2...L_y$ and ${\bf
   p}_{ij}$ denotes the corresponding momentum vector. Two kinds of
 interparticle potentials were studied, namely a FPU type potential given
 by $U(x)=x^2/2+k_4 x^4/4$ and  a Lennard-Jones potential given by
 $U(x) = A/x^{12}-B/x^6$. Both models gave  similar
 results. Nonequilibrium simulations using
 Nose-Hoover baths, as well as  equilibrium simulations based on the Kubo
 formula, were performed.  Nonequilibrium simulations were first performed
 on  strips of width $L_y$ with aspect ratio $L_y/L_x < 1 $ and with
 heat conduction in the
 $x$-direction. It was  observed that for fixed $L_x$, as one
 increased $L_y$, the current seemed to saturate to a constant value 
for quite small values of $L_y/L_x$. Subsequently, to save on computational time, the
authors considered the value $L_y/L_x=1/2$ in all their simulations. 
Studying system sizes upto $L_x=128$ they obtained a  
logarithmic divergence, with system size, of the  conductivity \ie
$\kappa \sim \ln (L_x)$. The equilibrium simulations, performed over
similar system sizes, and using a microcanonical ensemble gave a $t^{-1}$
dependence for the current-current correlation function. Using the
Green-Kubo formula this implies again a logarithmic
divergence, with system size,  of the conductivity.     

\begin{figure}
\psfig{file=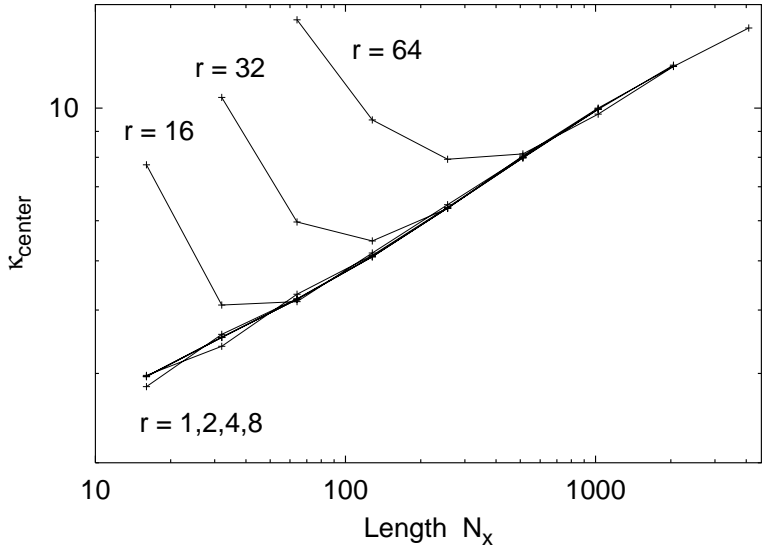,width=3.0in,angle=270}
\caption{Plot of conductivity, defined by Eq.~(\ref{eq:kapcent}),
  versus system size $L_x$ for $2D$ scalar FPU model.
 Curves for different  aspect ratios $r=L_x/L_y$ are
  shown. The asymptotic slope was $\approx 0.22$ (data from \cite{grassyang02}).
 }
\label{6:grassfig1}
\end{figure}

\begin{figure}
\psfig{file=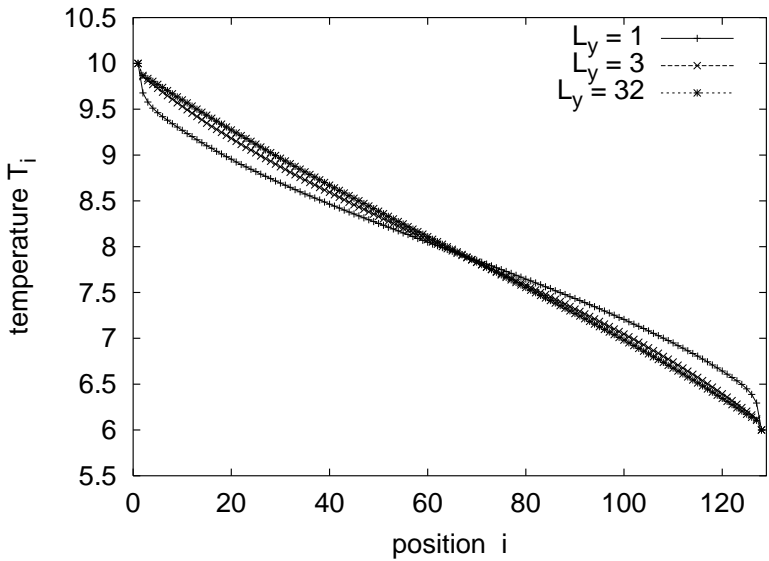,width=3.0in,angle=270}
\caption{Temperature profiles for scalar $2D$ FPU lattices 
 with $L_x=128$ and $L_y=1, 
  3$, and $32$. The temperatures of the heat baths at both ends were
  fixed at $(T_L, T_R)=(10.0, 6.0)$ (data from \cite{grassyang02}). } 
\label{6:grassfig2}
\end{figure}

The same Hamiltonian as in Eq.~(\ref{6:ham2d})
with FPU interactions but 
with a scalar displacement field 
was studied later by Yang and Grassberger \cite{grassyang02}. This paper
looked at somewhat bigger system sizes than \cite{lippi00} but were unable to
verify the logarithmic divergence and instead obtained a power law
dependence with an exponent $ \alpha \approx 0.22$. A careful
investigation of the value of $r=L_x/L_y$, at which a dimensional cross-over from $1D$
to $2D$ behaviour occured was carried out.  
Their conclusion was  that 
at large values of $r$, the conductivity  $\kappa$ diverged as a power
law with $\alpha =0.37 \pm 0.01$ while for small $r$ they obtained
$\alpha \approx 0.2$. The data for conductivity versus system size for
different values of $r$ is shown in Fig.~(\ref{6:grassfig1}). The
conductivity plotted in the figure was defined as
\bea
\kappa_{center}=\f{J}{(dT/dx)_{center}}~, \label{eq:kapcent}
\eea
where $(dT/dx)_{center}$ is the
temperature gradient evaluated numerically at the center. This
definition was used to take care of the boundary temperature jumps
that are usually present for small system sizes [see
Fig.~(\ref{6:grassfig2})].    
  Based on
the data in Fig.~(\ref{6:grassfig1})  the
authors also made the interesting suggestion that the cross-over from
$1D$ to $2D$ behaviour 
takes place at $r \to \infty$ in the limit  $L_x \to \infty$. This has
obvious implications for experimental tests on the dependence of conductivity
on length, for systems such as nanotubes and nanowires. 
Another  paper \cite{yang06} again studying the vector model for even
larger system sizes 
(upto $64 \times 65536$) has claimed observing a logarithmic
divergence. However one of the authors of the paper has expressed
doubts about whether convergence has been attained at these sizes
\cite{grassnote} and this seems very likely to be the case. 

\begin{figure}
{\includegraphics[width=4.5in]{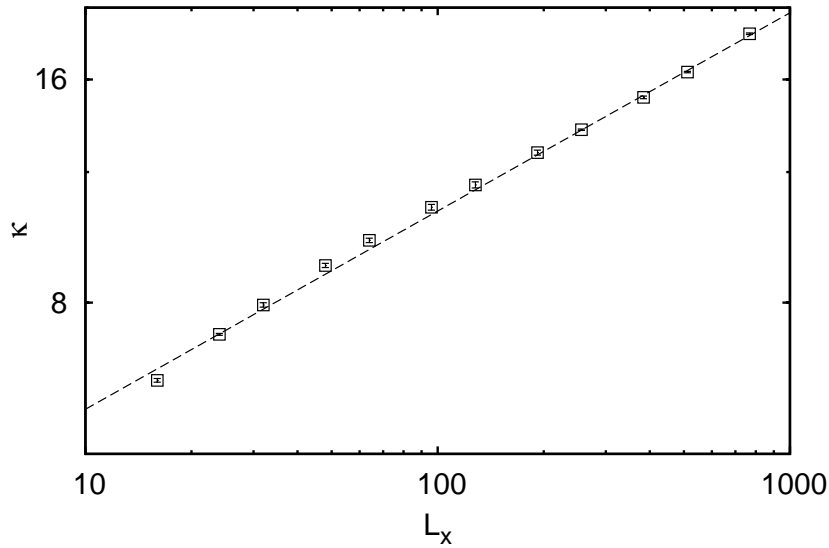}}
\caption{
System size dependence of the thermal conductivity for 2D
FPU lattice with $L_y:L_x=1:2$, plotted on a $\log-\log$ scale. The dashed line
represents the result of a power-law fitting in the region $L_x\ge 128$,
yielding the result $\kappa (L_x)\sim L_x^{0.267(5)}$  (data from \cite{shiba08})}.
\label{6:shibafig1} 
\end{figure}

\begin{figure}
 {\includegraphics[width=4.5in] {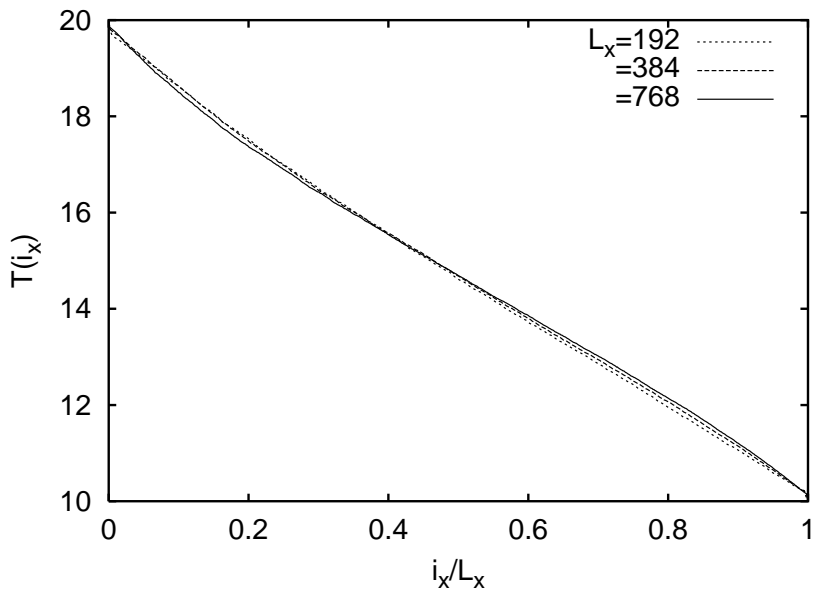}}
\caption{ Temperature profiles for the 2D FPU lattice  with $L_y:L_x=1:2$. 
The sequences represent the results for the sizes $L_x= 192, 384$, and $768$. 
The temperatures of the heat baths at both ends were fixed to $(T_L, T_R)=(20.0, 10.0)$. 
The horizontal axis represents the position in the $x$-direction, scaled
by the system size $L_x$, and the vertical axis represents the local
temperature  (data from \cite{shiba08}). }
\label{6:shibafig2} 
\end{figure}

The most recent simulations by Shiba
and Ito \cite{shiba08} considered the same Hamiltonian as in
Eq.~(\ref{6:ham2d}) and used the same parameter set as \cite{lippi00},
namely $k_4=0.1, T_L=20, T_R=10$.
They performed nonequilibrium Nose-Hoover simulations and studied
system sizes upto $384 \times 768$. 
Their data for conductivity versus system size is plotted in
Fig.~(\ref{6:shibafig1}).  
The exponent  $\alpha
\approx 0.268$ obtained by them appears to be significantly different from
logarithmic behaviour. We also show the temperature profiles for
different system sizes [Fig.~(\ref{6:shibafig2})] and it appears that
the boundary jumps  are quite negligible. 

Unlike in $1D$, where the hard particle gas has been intensely
studied, there have been very few studies on hard disc systems. 
For a hard disk fluid system, Shimada \etal~ \cite{shimada00} reported
$\alpha$ to be less than $0.2$. 
Thus for momentum-conserving systems in $2D$ it is fair to say that
simulations have not been able to unambiguously establish   the
logarithmic divergence of the conductivity predicted from 
theory. Further work is clearly needed here.

As far as momentum non-conserving
interacting systems are concerned one would naturally expect Fourier
behaviour, given that this is the case even in one dimension. 
In the next section we will discuss a number of momentum
non-conserving models of {\it non-interacting} particle systems which can be
shown (including some rigorously) to satisfy Fourier's law. In these
models noninteracting particles are scattered from fixed scatterers. These
models however suffer from the drawback that there is no mechanism for local
thermal equilibration and so the meaning of temperature and Fourier's
law is somewhat artificial. There have been a few papers which have
introduced particle interactions in these kind (hard particle
scattering) of models. Here we discuss two such models.  

\begin{figure}[!t]
\centerline{\epsfxsize=0.9\columnwidth \epsffile{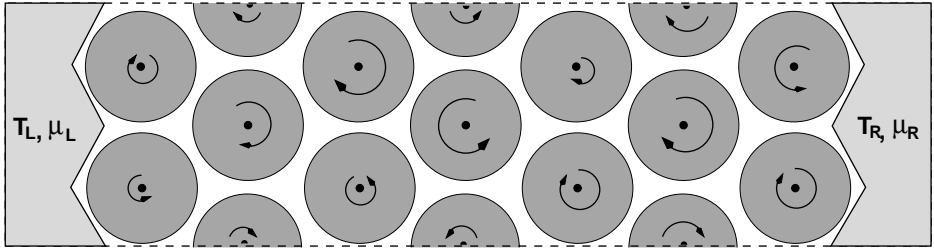}}
\vspace{0.5cm}
\caption{Schematic illustration of the model studied in \cite{monasterio01}. The scatterers are
on a triangular array arranged such as to avoid infinitely
long trajectories. The scatterers can perform rotational motion after
collisions with the moving point particles. 
Periodic boundary conditions are used in the vertical
direction (from \cite{monasterio01}).}
\label{6:onsagfig}
\end{figure}

The first model, introduced by Mej\'ia-Monasterio \etal~
\cite{monasterio01,larralde03}, is one in which noninteracting
particles move among a periodic array of circular scatterers [see
  Fig.~(\ref{6:onsagfig})]. The dynamics is specified as
follows. Consider dimensionless units such that the mass of the moving particles is
$1$ and the moment of inertia of the scatterers is $\eta$. Then, if
$(v_n,v_t)$ are the normal and tangential components of velocity of the particle
at the time of collision, and $\om$ is the angular velocity of the
discs, then after the collision they are transformed to
$(v_n',v_t',\om')$ which are given by the linear transformation:
\bea
v_n'&=&-v_n,\nn \\
v_t'&=& v_t-\f{2 \eta}{1+\eta} (v_t-\om), \nn \\
\om'&=&\om +\f{2 }{1+\eta} (v_t-\om)~. 
\eea
The dynamics  conserves total energy
$v_n^2/2+v_t^2/2+\eta \om^2/2$ and angular momentum and is
time-reversal invariant (however, the transformation is
non-syplectic). This system was then connected to 
two reservoirs of both heat and particle and which are specified by
temperature and chemical potentials $(T_L,\mu_L)$ and $(T_R,\mu_R)$
respectively. Thus both heat and particle currents were
generated. Performing detailed simulations on this system, some of the
main conclusions of the paper were:
(i) the system satisfied local thermal equilibrium, (ii) both heat and
particle currents satisfied usual linear response relations with
finite transport coefficients, (iii) Onsager reciprocity relations were satisfied. 
The largest system studied had about $100$ discs in the conducting
direction (and two discs in the vertical direction).   
Note that in this model interactions between particles is introduced
indirectly.  Motivated by this model, refn.~\cite{eckmann04} studied
an idealized model with noninteracting tracer particles moving between
fixed energy storing centres and exchanging energy with these. Local
thermal equilibration and temperature profiles were analytically
studied in this work. 
Another model where an explicit verification, of linear response
relations for heat and particle transport were obtained, as well as Onsager
reciprocity relations, is a $1D$ electronic system with 
self-consistent reservoirs \cite{roydhar07}.

\begin{figure}[htp]
  \centering
  \includegraphics[width = .44\textwidth]{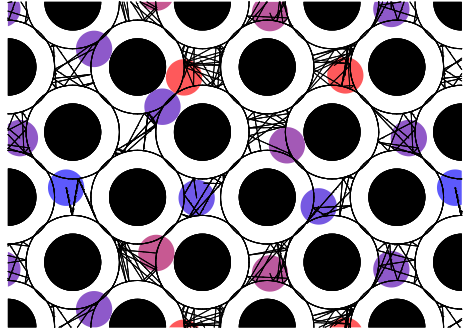}
  \caption{ Model studied in \cite{gaspard08}.
    The small colored discs move among a periodic array of 
    fixed black discs. Each small disc is confined to move in a cell
    bounded by four fixed discs. Most of the time the small disc moves
    with constant energy and undergoing elastic collisions with the
    fixed discs. Once in a while there is a collision between
    particles in two neighboring cells and there is exchange of energy.
 The solid lines 
    show the trajectories of the centers of the moving particles about their
    respective cells. The colors are coded according to the particles
    kinetic temperatures (from blue to red with increasing
    temperature) (from \cite{gaspard08}).} 
  \label{6:gaspardfig}
\end{figure}

Another recent study by Gaspard and Gilbert \cite{gaspard08,gaspard08b,gaspard08c} has considered a system where hard disc particles
are confined within periodic array of cells formed by fixed
scatterers. The model is explained in Fig.~(\ref{6:gaspardfig}). The
main idea of the authors has been to introduce a three time-scale mechanism
in generating the heat conduction state: (i) a short time scale $\tau_{wall}$
corresponding to particles motion within a cell with negligible energy
transfers, (ii) an intermediate time scale $\tau_{binary}$
corresponding to binary collisions which lead to local equilibrium and
(iii) a long time scale $\tau_{macro}$ of the macroscopic relaxation 
of Fourier modes. Based on a master equation approach the authors are
able to demonstrate the validity of Fourier's law and find an explicit
expression for the thermal conductivity of the system.
Note the similarity of the model with the one dimensional model
described in Fig.~(\ref{4:lifig}), though the mechanism leading to
Fourier behaviour there is possibly different. 

As far as simulations of oscillator systems is concerned,
a two dimensional momentum non-conserving lattice model was studied
by Barik \cite{barik06,barik07} who studied a scalar displacement model with
harmonic interparticle interactions and an onsite potential $V(x)$.
Several different forms of $V(x)$ were studied. 
With a Frenkel-Kontorva interaction given by $V_a(x)=\cos(x)$ the author
reports a power law divergence of the conductivity \cite{barik07}. For
two other models, of the form $V_b(x)=-\cos(x)- \sin (2x)/2$, a logarithmic
divergence of the conductivity was found \cite{barik06}. However the
reason why a 
finite conductivity has not been obtained in these momentum
non-conserving systems is probably because the system sizes studied
are too small (upto $240 \times 240$). This is evident from the quite        
large boundary temperature jumps that can be seen in the temperature
profiles reported in these papers \cite{barik06,barik07}. This means
that the contact resistances are contributing significantly to the
measured resistance. As we have discussed earlier in sec.~(\ref{sec:momcons}) this
will give a higher apparent divergence of the conductivity than is
actually the case. In this case looking at $\kappa_{center}$, defined
in Eq.~(\ref{eq:kapcent}), may be a good idea. 
Another point is that specific
forms of interaction strengths also might lead to a faster
convergence. Thus in \cite{barik06} it can be seen that for the same
applied temperature difference and same system size, the potential
$V_a(x)$ gives a flat 
temperature profile while with $V_b(x)$ one gets a significant
gradient. This feature has also been observed earlier in
sec.~(\ref{sec:momcons}) where we saw that the random collision model
and the $1D$ double-well potential gave fast convergence to the
asymptotic limit.

It appears that the simulation results in two dimensions  are
quite inconclusive for momentum conserving systems as
regards the system size dependence of conductivity. 
More extensive studies with larger system sizes
and different forms of interaction potentials are needed  to
confirm the theoretical predictions.   
For momentum non-conserving lattice systems it is likely that simulations on larger system
sizes will verify validity of Fourier's law.

\section{Non-interacting non-integrable systems}
\label{sec:nonint}

Probably the first  rigorous demonstration of Fourier-like dependence of the current
in a Hamiltonian system was by Lebowitz and Spohn \cite{lebowitz78} in
the Lorenz gas model. This model consists of a gas of classical point
particles in dimension $d \geq 2$ which  undergo elastic collisions
with fixed randomly placed spherical scatterers. The authors studied
this system with  stochastic boundary conditions on two bounding
walls corresponding to temperatures $T_L$ and $T_R$. They could prove
rigorously that in the Boltzmann-Grad limit of  a large number of 
small scatterers, one gets $J \sim (T_L-T_R)/L$, where $L$ is the
length of the box. 

A number of quasi-$1D$ Lorenz-gas-like systems have been numerically
investigated recently and have provided some more insight. Alonso
\etal~ \cite{alonso99} studied a channel with an array of periodically placed
semicircular scatterers. From their nonequilibrium simulations with
Maxwell baths, they verified Fourier's law and showed that
the nonlinear temperature profile could be understood by a simple
model of diffusing particles with an energy dependent diffusion
constant $D(E) \sim E^{1/2}$.    

The particle dynamics of the Lorenz-gas model with convex
scatterers is known to be  chaotic.
Li \etal \cite{li02} explored the question as to whether a positive Lyapunov
exponent, \ie, a chaotic dynamics, is a necessary condition in order to
get Fourier's law. They considered a system of non-interacting
particles   moving in a quasi $1D$ channel and suffering elastic collisions
with fixed triangular shaped scatterers placed throughout the
channel. The cases of a regular, and a random,  array of scatterers was
considered. It can be shown that in both these cases, the dynamics has
zero Lyapunov exponent unlike the case with convex scatterers. The
authors found from their nonequilibrium simulations that while the
regular array gave a diverging thermal conductivity, the random array
gave a finite conductivity. In both cases  a  temperature gradient was
obtained, though they had very different forms, highly nonlinear in
the periodic case and linear in the random case. Thus the random array
of non-chaotic scatterers  gave rise to normal Fourier heat transport
in the channel. Correspondingly it was shown that the particle motion
was superdiffusive for the periodic case and diffusive for the random
case. Thus this simulation shows {\it that chaos is not a necessary 
condition for diffusive transport}. In other  studies 
\cite{alonso02,li03} it has been seen that, even with  a periodic array of triangular
scatterers, one gets Fourier transport whenever  the internal angles
are irrational multiples of $\pi$. Finally Li and  Wang
\cite{liwang03} and Denisov \etal~\cite{denisov03} have  given
analytic arguments to relate the diffusion 
exponent of the heat carriers to the conductivity exponent $\alpha$.  
That chaos is not a necessary condition for normal transport 
was also seen  earlier in sec.~(\ref{sec:INT1D}). There we saw that
the interacting hard particle dimer gas (which is non-chaotic)  gives
normal transport in the presence of an external potential.

The validity of Fourier's law in Lorenz-gas models basically arises
due to the diffusive motion of the heat carriers. 
However the absence of interactions makes these models somewhat
ill-behaved from the thermodynamic point of view. As pointed out in
\cite{dhar99} these models lack local thermal equilibrium and so the
meaning of temperature and Fourier's law in  these systems is somewhat
different from that one usually has in nonequilibrium thermodynamics.  
The models studied in \cite{monasterio01,larralde03,gaspard08}, and
discussed in 
sec.~(\ref{sec:intsys2D}), are examples of similar momentum nonconserving models of particles with
collisional  dynamics where however the introduction of interactions
leads to local thermal equilibrium.   

\section{Experiments}
\label{sec:expts}
The experimental measurement of thermal conductivity of a system is
much more difficult than, for example, its electrical
conductivity. One realizes this from the simple fact that it is easy
to construct an ammeter to measure electrical  current but that is not so in
the case for heat current. Thus measurements of thermal conductivity 
require special methods and often the interpretation of experimental
data themselves require involved theoretical modeling.   
This is perhaps one reason as to why there has, until recently, not
been much experimental studies which 
have addressed the precise question of the system size dependence of
thermal conductivity and its expected divergence in  low dimensional systems.
The situation has changed recently with the advent of
nanophysics. Understanding heat transfer in systems such as nanowires
and nanotubes is not only a question of basic interest but of
technological importance too. With amazing advances in nano-technology
it has now become actually possible to measure the thermal conductivity of
a  nanotube suspended between two thermal
reservoirs. Here we briefly discuss some of the experiments on
nanowires and nanotubes. We will
try to explain the present understanding and also try to emphasize the
relevance, of the knowledge that has been obtained from studies of
simple models discussed in this review. In all the experiments
that we will describe here the heat current is believed to be
mainly due to phonons. 

The most common approach to understanding experimental data on heat
conduction is perhaps through simple kinetic theory picture which
says that the conductivity in a phonon system is proportional to $c v \ell$ where $c$ is
the specific heat per unit volume, $v$ the sound speed and $\ell$ the
phonon mean free path. For system sizes smaller than $\ell$, one
expects ballistic transport and roughly one can replace $\ell$ by $L$
in the conductivity formula, and hence get $\kappa \sim L$. On the
other hand for $\ell >> L$ it is normally expected that a finite
conductivity will be obtained. 
However, from the results presented in the previous sections, it is
clear that this picture cannot be correct. At sufficiently large
length scales, for low dimensional systems such as nanowires and
nanotubes, we expect the conductivity to diverge as a power law 
$\kappa \sim L^\alpha$.

 In the ballistic limit (defined as one where
anharmonicity can be neglected) one can use the Landauer or NEGF
formula and here there are examples where  good agreement between theory and
experiments can be seen.   
For nanotubes and
nanowires with low impurity level it turns out that phonon mean free
paths can be quite long and so transport is ballistic over fairly long length scales.   

{\bf Experiments on nanowires}: One of the first measurements of
phonon thermal conductance of a nanosystem 
was that by Tighe \etal~ \cite{tighe97}.  In a  beautiful
experiment they measured the conductance of insulating GaAs wires of
length $\approx 5.5 \mu m$ and cross-section $\approx 200 nm \times
300 nm$. At low temperatures ($1.5-5 K$) they obtained conductances of
order $\sim 10^{-9} W/K$. From their data and using  kinetic theory
arguments they estimated the phonon mean free path to be of order $\sim 1 \mu m$.   
Now one can ask the question: what is the thermal conductance of a
perfectly transmitting $1D$ wire? 
This can be easily obtained from Eq.~(\ref{3.2:SSJLR})  by setting
$\mT(\om)=1$ and one gets:
\bea
G=\f{J}{\Delta T} = \f{k_B}{2 \pi} \int_0^{\om_m} d \om \left( \f{\hbar
  \om}{k_B T} \right)^2 \f{ e^{\hbar \om /(k_B T)}}{[e^{\hbar \om /(k_B
	T)}-1]^2}~,
\eea
where we assume a phonon dispersion between $0-\om_m$. At temperatures
$T << \hbar \om/ k_B$ this gives:
\bea
G_{th}=g_0= \pi^2 k_B^2 T/(3 h) \label{8:quantum}
\eea
where $g_0=(9.456 \times 10^{-13} W /K^2)T$ has been proposed as the
quantum of thermal conductance \cite{rego98,schwab00} and is the
maximum value of energy transported per phonon mode. 
In a wire that is not strictly $1D$, as is the case of a wire of
diameter $\sim 200 nm$, other modes would contribute also to the
current. In another nice  and difficult experiment, Schwab \etal~
\cite{schwab00} were able to
measure the quantum of thermal conductance. They used a
silicon nitride wire which had four lowest massless modes and other
massive modes corresponding to the finite width of the wire. By going
to sufficiently low temperatures ($T < 1K$)  they were able to suppress
transport by the massive modes. Also one had to ensure very good
contacts between the wires and reservoirs. The authors were able to
verify that the 
resulting conductance corresponded to the value $g_0$. The agreement
is in fact quite impressive.     
At high temperatures all modes would contribute and so it will be
difficult to verify the classical $1D$ result with this system. 
Note that theoretically $\mT(\om)=1$  can probably be achieved only 
with Rubin baths. For other baths (for example Ohmic) this cannot be
obtained even for ordered chains and as a result, the  
temperature dependence of conductivity can be quite different [for
example see sec.(\ref{sec:ordharmlat1d})] from the
linear dependence in Eq.~(\ref{8:quantum}).

Another experiment \cite{liwu03} reported measurements of thermal
conductivity of 
single crystalline Si nanowires with wires several microns long and with
varying diameters between $22 nm$ to $115 nm$. They found that the
thermal conductivity increased rapidly with diameter and was almost
two orders of magnitude smaller ($\sim 20 W/mK$) than the bulk
value. These results have 
not been clearly understood. So far there has been no experiments
measuring the dependence of conductivity on  length in  nanowires.

{\bf Experiments on nanotubes}: Apart from nanowires, there have also been a  number of measurements of
heat current in nanotubes. One of the first measurements of
conductance in  individual samples was by Kim \etal~\cite{kim01}, on a $2.4 \mu m$
long and $14 nm$ diameter multiwalled carbon nanotube (MWCNT). They found a
very high thermal conductivity of $\approx 3000 W/mK$ (at room
temperature) and noted that
this corresponded to a phonon mean free of $\sim 500
nm$. Somewhat surprisingly this was close to a theoretically predicted value
by Berber \etal~ \cite{berber00} who had 
performed classical molecular dynamics simulations (using
the Green-Kubo formula) for a $(10,10)$ carbon
nanotube.  
Using realistic potentials they reported a high 
thermal conductivity of $\approx 6000 W/mK$, at room temperature. 
From our expectations of diverging conductivity we expect that these
reported values, both in experiments and simulations, will increase with
increasing length of the wire. In fact  simulations by Maruyama
\cite{maruyama02}, Zhang and Li \cite{zhang05}, and by Yao \etal~\cite{yao05}, again
with realistic potentials, do find such a increase. A more recent
simulation \cite{donadio07} however finds a converging conductivity. 

As far as
theoretical work on heat conduction in carbon nanotubes is concerned,
we mention the insightful paper by Mingo and Broido \cite{mingo05},
who point out that it is necessary to perform quantum mechanical
calculations in 
order to understand the experimental results and that classical
calculations can be misleading. They mostly consider the ballistic
conductance using the Landauer formula [ Eq.~(\ref{3.2:SSJLR})] and
also the Boltzmann-Peierls equation at longer lengths. However they do
not comment on the system size dependence of conductivity.   
It is likely that as far as the question of system size dependence
of conductivity is concerned, the answer should probably be
independent of whether one is doing a classical or a quantum
calculation. Hence it will be useful to settle this issue through
classical molecular dynamics simulations.  From our experience with the
difficulty in reaching asymptotic system sizes for $1D$ and $2D$
systems, it is clear that one has to be careful before coming to quick conclusions.

The experimental results of Fujii \etal~ \cite{fujii05} on individual MWCNT
 give some hints of anomalous behaviour. They also obtain
large values of thermal conductivity but find that it decreases with
the diameter of the tubes. At room temperatures the thermal
conductivity of a  $3.7 \mu m$ long nanotube with diameter $\approx 10
nm$ was about $2500 W/mK$ while that of a $3.6 \mu m$ long, 
diameter $\approx 30 nm$ nanotube was about $500 W/mK$.  Note that the
dependence of $\kappa$ on diameter is opposite to that found for nanowires
mentioned earlier.

Measurements on individual single-wall carbon nanotubes (SWCNT) also have now
been done. Yu \etal~ \cite{yu05} observed that the thermal conductance of a $2.76
\mu m $ long suspended tube was very close to the calculated ballistic
thermal conductance (calculated using the Landauer formula) of a $1 nm
$ diameter tube.  In the temperature range of $100-300 K$ they found increasing
conductance and no signs of significant phonon-phonon
scattering. Another measurement on a single-walled carbon nanotube by
Pop \etal~ \cite{pop06} measured the conductance to temperatures upto
$800 K$ and they found a $1/T$ decay at large temperatures. They also
report measurements on various lengths, ranging from $0.5 \mu m$ to
$10 \mu m$ and diameter $1.5 nm$, and  curiously, they  found increasing
conductance with length. This the authors explain can be understood
to be a result of the large phonon mean free path $\sim 0.5 \mu m$ and
phonon boundary scattering.   

An experimental proof of the divergence of thermal conductivity with
system size is probably the dream of many theorists. There seems to be
rapid progress in the direction of making this possible. The first  indication of length
dependence was reported by Wang \etal~ \cite{zlwang07}, for the case of a SWCNT placed
on a silicon substrate. They measured samples of lengths between
$0.5-7 \mu m$ at room temperature and found a slow increase of the conductance. 
The most recent experiments by Chang \etal~\cite{chang08} makes a
detailed investigation of the length dependence of conductivity in multiwalled
nanotubes, of carbon and boron-nitride, and claim to have found convincing
evidence for violation of Fourier's law. These room temperature measurements were on
suspended tubes of effective length between $\approx 3.7-7 \mu m$ and
their estimate of phonon mean free path is $\sim 20-50 nm$. From their
measurements (over the rather limited length scale) the authors
conclude that $\alpha \approx 0.6$  for the carbon nanotube
$\alpha \approx 0.5$ for the boron nitride sample (which is isotopically
disordered).  It is interesting to note that the two samples in this
experiment, approximately correspond to the ordered and disordered FPU models.

Experiments on suspended  single layer graphene sheets have also been
made recently \cite{balandin08} and so interesting experimental results from two dimensional
systems can also be expected in the near future. It is of course too early
to make definite conclusions from these  experiments.

Finally we briefly discuss one other area, that of thermal rectifiers,
where  an experiment was motivated by theoretical work on simple
models of heat conduction. In a paper by Terraneo \etal~
\cite{terraneo02}, an inhomogeneous nonlinear lattice model of heat
conduction was proposed. This model had the interesting property that
by changing a single parameter 
on a part of the chain one could cause a transition from insulating to conducting
behaviour. A related observation was that 
the absolute value of the heat current depended 
on the sign of $\Delta T= T_L-T_R$. Thus one basically had a model for
a thermal rectifier. Note that it can be proved rigorously, for 
both harmonic systems (with inhomogeneity but without self-consistent
reservoirs) as well as homogeneous anharmonic 
systems, that $J(\Delta T) = -J (-\Delta T)$ and so these systems cannot work as
rectifiers.  For harmonic systems this follows immediately from the general
expression for current given in sec.~(\ref{sec:legf}).
 Clearly one needs both inhomogeneity as well as
anharmonicity to get rectification and Terraneo \etal~ gave a simple
explicit demonstration of how this could be achieved. Physically their
results can be understood easily by thinking of the anharmonicity as
giving rise to effective phonon bands which can be moved up and down
by increasing or decreasing  local temperatures. Phonon flow from the
reservoirs into the system can thus be controlled. 

Since the work in \cite{terraneo02}, a number of papers have observed
this effect in a number of models
\cite{li04b,segal05,li05,li06,saito06,hu06,eckmann06,casati07,yang07,hu08,segal08}.
Based on the model of thermal recifier, Yang and Li have proposed a
design of a thermal logic gate \cite{wang07}.  
An experimental observation of thermal rectification was made recently by Chang
\etal~ \cite{changrecti07,casati07b}. They made measurements of the heat current
in a boron-nitride nanotube which was 
mass-loaded externally in an inhomogeneous way, and were able to
obtain a small rectification.

We conclude this section with the note that the situation looks
hopeful for vigorous interactions between theory and experiments.

\section{Concluding remarks}
\label{sec:conclusions}

The fact that Fourier's law is not valid in $1D$ and $2D$ systems is a
surprising result and probably the most important knowledge gained
from the large number of studies on heat conduction in low dimensional systems.
Even in the limit of the system length being much larger than typical
scattering lengths in a system, one 
finds that it is not possible to define a thermal conductivity as an
intrinsic size-independent property of the system.
This discovery is not only of academic interest but also important
from the point of view of understanding real experiments. For example
this tells us that it does not make sense to talk about the thermal
conductivity of a carbon nanotube since this will keep changing with the
length of the nanotube.

To  summarize, the main conclusions of this review are:

(i) { Fourier's law is not valid in momentum-conserving systems in 
one and two dimensions}. 

 For disordered harmonic systems, $\kappa \sim
N^\alpha$, where $\alpha$ depends on boundary conditions and
spectral properties of heat baths. 

For nonlinearly interacting systems without disorder, simulation results on a number of
models indicate that in $1D$, $\alpha=1/3$, and that there is only one
universality class. There is disagreement between predictions from
different theoretical approaches. In $2D$, the theoretical prediction
of $\kappa \sim \log(N) $ has not been verified in the latest
simulations.  

(ii) { Fourier's law, as far as the scaling $J \sim 1/N$ is concerned,
  is valid in momentum-non-conserving non-integrable systems in all dimensions}.
Both theory and most simulations agree on this.  

(iii) { In $1D$ oscillator systems with both disorder and anharmonicity, the
    asymptotic system size dependence of current is determined by
    anharmonicity alone, and localization becomes irrelevant.}

(iii) { Chaos is neither a necessary nor a sufficient condition for
    validity of Fourier's law}.
This result follows from the observation that Fourier's law is valid
in billiard-like systems with polygonal scatterers which have
    zero-Lyapunov coefficient and hence are non-chaotic. 
    On the other hand Fourier's law is not valid for the FPU system in
    any parameter regime even though it has positive Lyapunov exponents. 

(iv) { For momentum-conserving  $1D$ systems, it is not possible to
write Fourier's law in the form $J = -\kappa_N \nabla T$ with $\kappa_N$
defined as a size dependent conductivity.}  This follows from the
anomalous steady state temperature profiles that 
seem to be invariably obtained in such systems.

(v) { For harmonic lattice systems, the Langevin equation Green's
  function (LEGF) formalism provides a very useful theoretical framework for
understanding heat transport, in  both classical and quantum systems.}
\bigskip

Some interesting  problems  that need to be
addressed in the future are the following:

(a) Exact determination of the exponent $\alpha$ in any one dimensional
momentum conserving model with purely Hamiltonian dynamics and without
use of the usual Green-Kubo formula. 

(b) Simulations on nonlinearly interacting systems in  $1D$, $2D$ and
 $3D$ for larger system sizes and 
 different models, in order to establish the exponents convincingly. 
It will be nice to have more results on systems such as hard
 discs and spheres. 

We point out that  the understanding of heat conduction even in  
three dimensional macroscopic systems is incomplete. A nice
example of this  can be seen from the discussion given in
\cite{wei93}, in the context of understanding experimental results on
 heat conductivity of a highly purified single crystal diamond.
A related point: in $3D$ it is a belief (see for example
   \cite{ziman72}) that at low temperatures, where Umpklapp processes
   become exponentially rare, normal processes along with impurity
   scattering lead to a finite conductivity for the system. Can this
   be given some more rigorous justification, or, verified in simulations ?

(c) Finding $\alpha$ for  two and three dimensional
   disordered harmonic systems analytically. Further simulations are
   also necessary here. What are the connections with localization
   theory ?

(d) For disordered anharmonic systems, for disorder strength and
   anharmonicity strengths denoted by $\Delta$ and $\lambda$
   respectively,  what is the phase diagram in the $\Delta-\lambda$
   plane ?  

(e)  For open systems there is a rigorous derivation of a 
    linear response result which is valid for finite systems. Is
   it possible to prove the equivalence of this  with the  usual
    Green-Kubo formula for closed systems, 
   in some example ? 
This is probably true for systems with normal transport
and probably {\it not true} for systems with anomalous transport.

(f) Proof of non-existence (or existence) of phase transitions in one
   dimensional models of heat conduction with short range interactions. 

(g) One needs studies for quantum interacting systems since
most experimental work seems to be in this domain. We have seen
that the Green's function approach has been  successful in
understanding harmonic systems (\ie~ballistic transport). An extension
of this approach to the anharmonic case 
would be very useful. 
Apart from the  Green's function formalism, the approach used by
Chen \etal~ \cite{chen89} may be  a useful method for this
problem.  

(h) How valid is the hydrodynamic description for systems with
   anomalous transport? If they are valid, what are the correct
   hydrodynamic equations? For example we have  seen that one cannot
   use the equation $J = -\kappa_N (T) \nabla T$ to describe the
   steady state.

(i) Non steady state properties: most of the studies on heat conduction
   have been on measurement of current and temperature in the
   nonequilibrium steady state. In general of course one is interested
   also in time dependent properties. In fact the diffusion equation
   following from Fourier's  law is itself a time-dependent  equation.  
Also many experiments make measurements in non-steady state
   conditions, such as by studying heat pulses and frequency dependent
   studies.  
Thus it is necessary to have more theoretical studies on heat flow in
   non steady state  situations. Interesting questions can be asked
   here, {\emph{e.g.}}, can one talk of a frequency  dependent thermal 
   conductivity  \cite{shastry06} ?

(j) Scalar versus vector models: for lattice models one question is whether
   the dimensionality of the displacement vectors matters as  far as
   exponents are concerned. While it is usually assumed that
   dimensionality does not matter, it will be nice to have a proof of
   this. 

(k) Temperature dependence of conductivity or conductance: Apart from
   the system size dependence it will be useful to get more results on
   the temperature dependence of the linear response heat current,
   since this is one of the things that experimentalists are
   interested in.

\vspace{2cm}
{\bf Acknowledgements:} I thank A. Kundu and R. Marathe for
reading the manuscript. I  thank D. Roy for
reading the manuscript as well as making valuable suggestions. I am
grateful to P. Grassberger, W. Nadler and L. Yang for permission in
using data from two of their papers \cite{grass02,grassyang02}. I am also
grateful to J. Deutsche and O. Narayan for permission in using data
from their paper \cite{deutsch03a} and H. Shiba and N. Ito for using
data from their paper \cite{shiba08}. Finally, I thank a large number
of colleagues who read the first draft of this review and made
valuable suggestions towards improving it.

\end{document}